\documentclass[11pt]{article}

\usepackage[utf8]{inputenc}

\usepackage{amstext,amsmath,amssymb,amsfonts,bbm}
\usepackage{hyperref}
\usepackage{authblk}
\usepackage{caption} 
\usepackage{epsfig} 
\usepackage{color} 
\usepackage{graphicx} 
\usepackage{braket} 
\usepackage{slashed} 
\usepackage{dsfont} 
\usepackage{amsthm}  
\usepackage{physics}

\usepackage{cite}

\allowdisplaybreaks[4]

\usepackage{psfrag}
\usepackage{tikz} 
\usetikzlibrary{arrows}
\usetikzlibrary{plotmarks}
\usetikzlibrary{patterns}
\usetikzlibrary{external}
\usepackage{pgfplots}
\pgfplotsset{compat=1.9}
\usetikzlibrary{patterns}
\usetikzlibrary{calc}
\usepackage[compat=1.1.0]{tikz-feynman}
\usetikzlibrary{automata,positioning}

\usepackage{float}  
\usepackage{subfig}
\usepackage[lmargin=50pt,rmargin=60pt,tmargin=60pt,bmargin=65pt]{geometry}

\newcommand{\be}{\begin{equation}}
\newcommand{\ee}{\end{equation}} 
\newcommand{\nn}{\nonumber}
\newcommand{\f}{\frac}
\newcommand{\p}{\partial}

\newcommand{\la}{\langle}
\newcommand{\ra}{\rangle}

\let\a=\alpha \let\b=\beta    \let\d=\delta
\let\z=\zeta       \let\k=\kappa \let\l=\lambda
\let\m=\mu    \let\n=\nu          \let\r=\rho 
 \let\t=\tau     
    
\let\G=\Gamma \let\D=\Delta    \let\X=F
         
  \let\eps=\epsilon

\newcommand{\phib}{\bar{\phi}}

\newcommand{\cB}{\mathcal{B}}

\newcommand{\cF}{\mathcal{F}}
\newcommand{\cG}{\mathcal{G}}

\newcommand{\cN}{\mathcal{N}}
\newcommand{\cO}{\mathcal{O}}

\newcommand{\cZ}{\mathcal{Z}}

\DeclareMathOperator{\im}{\mathrm{i}}

\newcommand{\mba}{\mathbf{a}}
\newcommand{\mbb}{\mathbf{b}}
\newcommand{\mbc}{\mathbf{c}}
\newcommand{\mbd}{\mathbf{d}}
\newcommand{\mbe}{\mathbf{e}}
\newcommand{\mbf}{\mathbf{f}}
\newcommand{\mbg}{\mathbf{g}}
\newcommand{\mbh}{\mathbf{h}}
\newcommand{\mbj}{\mathbf{j}}
\newcommand{\mbk}{\mathbf{k}}
\newcommand{\mbm}{\mathbf{m}}
\newcommand{\mbn}{\mathbf{n}}

\newcommand{\mbG}{\pmb{G}}

\allowdisplaybreaks[4]

\theoremstyle{remark}

\definecolor{orange}{rgb}{0.88,0.39,0.12} 
\definecolor{rouge}{rgb}{0.8, 0.0, 0.0}
\definecolor{vert}{rgb}{0.4, 0.69, 0.2}
\definecolor{bleu}{rgb}{0.19, 0.55, 0.91}
\definecolor{lavenderpurple}{rgb}{0.59, 0.48, 0.71}

\begin{document}

\title{\bf Sextic tensor field theories in rank $3$ and $5$}

\author[1]{Dario Benedetti}
\author[1,2]{Nicolas Delporte}
\author[1]{Sabine Harribey}
\author[3,4]{Ritam Sinha}

\affil[1]{\normalsize \it 
 CPHT, CNRS, Ecole Polytechnique, Institut Polytechnique de Paris, Route de Saclay, \authorcr 91128 PALAISEAU, 
 France \authorcr
\hfill }

\affil[2]{\normalsize\it Laboratoire de Physique Th\'eorique (UMR 8627), CNRS, Univ.Paris-Sud, \authorcr
\it Universit\'e Paris-Saclay, 91405 Orsay, France \authorcr
\hfill }

\affil[3]{\normalsize \it Instituto de Fisica Teorica IFT-UAM/CSIC, Cantoblanco 28049, Madrid, Spain \authorcr
\hfill }

\affil[4]{\normalsize \it The Racah Institute of Physics, The Hebrew University of Jerusalem,
Jerusalem 91904, Israel \authorcr
\hfill 

\bigskip
Emails: dario.benedetti@polytechnique.edu, nicolas.delporte@th.u-psud.fr, sabine.harribey@polytechnique.edu, ritam.sinha@mail.huji.ac.il\authorcr \hfill}

\date{}

\maketitle

\hrule\bigskip

\begin{abstract}
We study bosonic tensor field theories with sextic interactions in $d<3$ dimensions. We consider two models, with rank-3 and rank-5 tensors, and $U(N)^3$ and $O(N)^5$ symmetry, respectively. For both of them we consider two variations: one with standard short-range free propagator, and one with critical long-range propagator, such that the sextic interactions are marginal in any $d<3$.
We derive the set of beta functions at large $N$, compute them explicitly at four loops, and identify the respective fixed points. 
We find that only the rank-3 models admit melonic interacting fixed points, with real couplings and critical exponents: for the short-range model, we have a Wilson-Fisher fixed point with couplings of order $\sqrt{\epsilon}$, in $d=3-\epsilon$; for the long-range model, instead we have for any $d<3$ a line of fixed points, parametrized by a real coupling $g_1$ (associated to the so-called wheel interaction).
By standard conformal field theory methods, we then study the spectrum of bilinear operators  associated to such interacting  fixed points, and we find a real spectrum for small $\epsilon$ or small $g_1$.

\end{abstract}
\bigskip
\hrule\bigskip

\tableofcontents

\section{Introduction}

Tensor field theories are models of $N^r$ bosonic or fermionic fields with a Lagrangian invariant under 
$U(N)^r$ \cite{Bonzom:2012hw},  $O(N)^r$ \cite{Carrozza:2015adg}, or $Sp(N)^r$ \cite{Carrozza:2018psc} transformations, and with $r>2$.\footnote{Different types of field content and symmetry groups are possible, with similar features to what we describe below, but they are typically more complicated to describe or analyze \cite{Bonzom:2011zz,Witten:2016iux,Klebanov:2017nlk,Benedetti:2017qxl,Carrozza:2018ewt}.} Their study was initially undertaken as a generalization of matrix models in the tentative to build models of higher-dimensional quantum gravity from a geometric interpretation of their Feynman diagrams \cite{ambj3dqg,sasa1,review}, but it has recently developed into new directions. 

An important property that makes tensor models particularly interesting for field theory is that, like vector and matrix models, they admit a $1/N$ expansion \cite{Gurau:2011xq,Gurau:2011aq}, which however is typically different from that of vector and matrix models (see Ref.~\cite{Klebanov:2018fzb} for a comparative review). In particular, the leading order lies somewhat in between that of their lower-rank cousins, as their large-$N$ limit is richer than that of vectors, but more manageable than the planar limit of matrix models.
The leading order diagrams are the so-called \emph{melonic} diagrams \cite{Bonzom:2011zz}, for which one can write closed Schwinger-Dyson and Bethe-Salpeter equations. They are the same type of diagrams encountered in the Sachdev-Ye-Kitaev model as well \cite{Sachdev:1992fk,Kitaev, Maldacena:2016hyu,Polchinski:2016xgd,Gross:2016kjj}, and to which the model owes its solution. In the case of the SYK model, the fields are in a vector representation, but they interact through a random tensor, whose quenched average leads to the melonic dominance, for the same combinatorial reasons as for tensor models \cite{Bonzom:2017pqs}. Tensor models have thus been extensively studied in one dimension as an alternative to the SYK model without quenched disorder \cite{Witten:2016iux,Gurau:2016lzk,Peng:2016mxj,Krishnan:2016bvg,Krishnan:2017lra,Bulycheva:2017ilt,Choudhury:2017tax,Halmagyi:2017leq} (see also Ref.~\cite{Klebanov:2018fzb,Delporte:2018iyf} for reviews).
Such developments have also prompted interest in tensor models as a novel class of quantum field theories for which we can hope to control non-perturbative aspects via the large-$N$ limit, and in particular discover new interacting conformal field theories \cite{Giombi:2017dtl,Prakash:2017hwq,Benedetti:2017fmp,Giombi:2018qgp,Benedetti:2018ghn,Benedetti:2019eyl,Benedetti:2019ikb,Popov:2019nja}.

Not all tensor models lead automatically to the interesting melonic limit mentioned above. We can of course build tensor models which are actually vector or matrix models in disguise, or we can also consider proper tensor models whose Feynman diagrams are cactus diagrams as in vector models (e.g.\ Ref.~\cite{Benedetti:2018ghn}), and so on.
Several papers worked out conditions under which a tensor model possesses a melonic limit. Starting with an $O(N)^3$ quartic model, Ref.~\cite{Carrozza:2015adg} introduced an \emph{optimal rescaling} of the couplings which leads to melonic dominance.
For interactions of higher order, including the sextic ones we consider below in rank 3, Ref.~\cite{Lionni:2017yvi} identified the structure of diagrams at first leading orders.
Ref.~\cite{Ferrari:2017jgw} proved that, with tensors of prime rank and for a particular class of complete interactions, the dominant Feynman diagrams are melonic.
Ref.~\cite{Gubser:2018yec} classified the different structures of complete interactions for tensors of odd rank. 
Ref.~\cite{Prakash:2019zia} showed melonic dominance in sextic subchromatic models (rank-3 and a particular rank-4 model). 

Those works have mostly concentrated on the combinatorial aspects of the large-$N$ limit.
The next important question is then to identify more precisely the field theoretic content of those melonic theories, and look for dependence on the dimension, rank and interaction of those models. 
In dimension 1, Ref.~\cite{Klebanov:2019jup} obtained the spectrum of a rank-5 generalization of the CTKT model \cite{Carrozza:2015adg,Klebanov:2016xxf} and generalized to ranks greater than 3 the conformal spectrum of bilinears.
In higher dimensions, very quickly it seemed difficult to find a non-trivial RG fixed point \cite{Benedetti:2017fmp} or a real spectrum of conformal dimensions at integer dimensions \cite{Prakash:2017hwq,Giombi:2017dtl, Giombi:2018qgp}. However, using a long-range free propagator such that the quartic interactions are marginal in $d<4$, Ref.~\cite{Benedetti:2019eyl} managed to construct the renormalization group flow of a quartic model and find a non-trivial infrared-attractive fixed point with a purely imaginary coupling for the (unbounded) ``tetrahedral'' coupling, but with real critical exponents. Later, the dimensions and OPE coefficients of bilinear operators were computed in Ref.~\cite{Benedetti:2019ikb}, and found to be real, thus providing an important step in order to assert unitarity of the conformal field theory (CFT).

One of the aims of this paper is to understand how general are some of those findings, such as the need for an imaginary coupling in combination with the long-range propagator. In particular, we would like to understand how they depend on the rank of the tensors and on the order of the interactions.
For this purpose, we chose to study models with sextic interactions in rank 3 and 5, and with either short or (critical) long-range propagators.
Short-range sextic models have been considered before, but either without actually studying the existence of fixed points \cite{Giombi:2017dtl} (and only for rank 5), or for a different scaling in $N$ of the couplings than the optimal one \cite{Giombi:2018qgp}.
Here, we will compute beta functions for our models, at leading order in the $1/N$ expansion, and at four-loop order. In the presence of a small parameter, such as $\epsilon=3-d$ in the short-range case, or an exactly marginal coupling in the long-range case, the four-loop expansion is sufficient in order to identify interacting fixed points.

Our main results are: in rank 3, we find two non-trivial infrared fixed points for the short-range model, and a line of infrared fixed points for the long-range model, for real couplings. In both cases, we find a window with real spectrum of bilinear operators. Surprisingly, in rank 5, the only fixed point is non-interacting.

The rest of the paper is organised as follows. In Sec.~\ref{sec:models}, we start by setting the scene with definitions of our models in rank 3 and 5, long- and short-range, and a description of the leading order diagrams. We continue in Sec.~\ref{sec:SDeq} and \ref{sec:kernels} by computing the two- and four-point functions. Sec.~\ref{sec:betas} contains a detailed derivation of the $\beta$-functions of our sextic couplings, as well as their fixed points. Before concluding, we compute in Sec.~\ref{sec:BSeq} the spectrum of bilinears (including spin dependence) through the now standard eigenvalue equation. In three appendices, we spell out details on our conventions and on the main loop integrals.

\section{The models}
\label{sec:models}

Both models we are going to consider can be viewed as symmetry-breaking perturbations of a free $O(\cN)$-invariant action for $\cN$ scalar fields $\phi_{\bf a}(x)$, with ${\bf a}= 1,\ldots,\cN$, $x\in \mathbf{R}^d$:\footnote{As usual, a summation is implied for repeated indices.}
\be
S_{\rm free}[\phi,\phib] = \int d^d x  \, \phib_{\bf a}(x) (   - \p_\m\p^\m)^{\zeta} \phi_{\bf a}(x) \, .
\ee
The scalar fields will be either complex or real (in the latter case $\phib_{\bf a}=\phi_{\bf a}$ and we multiply the action by a factor $1/2$).
 $\zeta$ is a free parameter, which must be positive in order to have a well-defined thermodynamic limit, and it must be bounded above by one in order to satisfy reflection positivity.
We will later fix it to be either $\zeta=1$, as in Ref.~\cite{Klebanov:2016xxf,Giombi:2017dtl,Giombi:2018qgp}, or $\zeta=d/3$, as in Ref.~\cite{Benedetti:2019eyl,Benedetti:2019ikb}.\footnote{For $\zeta<1$, the fractional Laplacian can be defined in several ways \cite{Kwasnicki_2017}. In Fourier space, with the convention that $f(x) = \int\,\f{d^d p }{(2\pi)^d} e^{-i p\cdot x}\,f(p)$, we simply have:
\begin{equation*}
S_{\rm free}[\phi,\phib] = \int \f{d^d p}{(2\pi)^d}  \, \phib_{\bf a}(p) (p^2)^{\zeta} \phi_{\bf a}(p) \, .
\end{equation*}
In direct space we can instead write it as a kernel:
\begin{equation*}
S_{\rm free}[\phi,\phib] =  c(d,\zeta) \int d^d x\, d^d y  \, \f{\phib_{\bf a}(x)  \phi_{\bf a}(y)}{|x-y|^{d+2\z}} \, ,
\end{equation*}
with $c(d,\zeta) = \f{2^{2\z}\G\left(\f{d+2\z}{2}\right)}{\pi^{d/2}|\G(-\z)|}$. Notice that often in the literature on the long-range Ising model (e.g.\ \cite{Behan:2017emf,Paulos:2015jfa}) one finds the free action to be defined as above, but with $c(d,\zeta) = 1$.
}

The free propagator is
\be
C(p) = \f{1}{p^{2\zeta}}\,, \;\;\;\; C(x,y) = \f{\G\left(\Delta_{\phi}\right)}{2^{2\z}\pi^{d/2}\G(\z)} \f{1}{|x-y|^{2\D_{\phi}}}\,,
\ee
with $\Delta_{\phi}= \f{d-2\zeta}{2}$.

Perturbing the free action above by a quartic $O(\cN)$-invariant potential leads to the usual short-range ($\zeta=1$, e.g.\ Ref.~\cite{Moshe:2003xn}) or long-range ($\zeta<1$, e.g.\ Ref.~\cite{Brezin_2014,Defenu_2015}) $O(\cN)$ model.

The general type of tensor field theories we have in mind will have $\cN=N^r$, and a potential explicitly breaking the $O(\cN)$ symmetry group down to $\cG^r$, with either $\cG=O(N)$ (for real fields) or $\cG=U(N)$ (for complex fields).
For example, for $r=3$, we will write the field label as a triplet, ${\bf a}=(abc)$, and impose invariance of the action under the following transformation rule:
\be
\phi_{abc}(x) \to R^{(1)}_{aa'}\, R^{(2)}_{bb'}\,  R^{(3)}_{cc'}\,  \phi_{abc}(x)\, , \;\;\;\; R^{(i)}\in \cG \,.
\ee
Proper tensor field theories have $r>2$, otherwise we talk of vector ($r=1$) or matrix ($r=2$) field theories.
We will explicitly consider two models with sextic interactions, for $r=3$ and $r=5$. For $r=4$ we could write a model qualitatively very similar to $r=5$, but we would not learn much more, so we will not present it.

\subsection{Rank $3$}

\paragraph{Action.}

We first consider a rank-3 bosonic tensor model in $d\leq 3$ dimensions, with $U(N)^3$ symmetry and sextic interactions.
The bare action is
\begin{align} \label{eq:action}
S[\phi,\phib] &= \int d^d x  \, \phib_{abc} (   - \p_\m\p^\m)^{\zeta} \phi_{abc} + S_{\rm int}[\phi,\phib] \, ,\\
 \label{eq:int-action}
S_{\rm int}[\phi,\phib] &= \int d^d x  \sum_{b=1}^5 \f{\l_b}{6 N^{3+\r(I_b)}} I_b \,.
\end{align}
The $U(N)^3$ invariants $I_b$ are all those that can be constructed with six fields, and their respective parameter $\r(I_b)$ will be chosen according to the optimal scaling defined in Ref.~\cite{Carrozza:2015adg}:
\be
\r(I_b)=\frac{F(I_b)-3}{2} \,,
\ee
with $F(I_b)$ counting the total number of cycles of alternating colors $i$ and $j$ with $i,j ~\in \lbrace 1,2,3\rbrace$, and the colors being introduced in the following paragraph.

It is customary to represent the tensor invariants as {\it colored graphs} \cite{RTM}. To that end, we represent every tensor field as a node (black and white  for $\phi$ and $\phib$, respectively) and every contraction of two indices as an edge. Each edge is assigned a color red, blue, or green (or a label $1$, $2$, or $3$) corresponding to the positions of the indices in the tensor. We call the resulting graphs $3$-colored graphs. As a consequence of the $U(N)^3$ symmetry, such graphs are bipartite, that is, edges always go from a white to a black node.
With the aid of such representation we can write the interacting part of the action as:
\be
\begin{split} \label{eq:int-action-graph}
S_{\rm int}[\phi,\phib] = & \int d^d x  
\left( \f{\l_1}{6 N^{3}} \vcenter{\hbox{\tikzsetnextfilename{wheel2}
\begin{tikzpicture}[line cap=round,line join=round,>=triangle 45,x=1.0cm,y=1.0cm,scale=0.25]
\draw [line width=1.pt,color=bleu] (-2.,3.9282032302755114)-- (2.,3.92820323027551);
\draw [line width=1.pt,color=rouge] (2.,3.92820323027551)-- (4.,0.46410161513775483);
\draw [line width=1.pt,color=bleu] (4.,0.46410161513775483)-- (2.,-3.);
\draw [line width=1.pt,color=rouge] (2.,-3.)-- (-2.,-3.);
\draw [line width=1.pt,color=bleu] (-2.,-3.)-- (-4.,0.46410161513775794);
\draw [line width=1.pt,color=rouge] (-4.,0.46410161513775794)-- (-2.,3.9282032302755114);
\draw [line width=1.pt,color=vert] (-4.,0.46410161513775794)-- (4.,0.46410161513775483);
\draw [line width=1.pt,color=vert] (-2.,-3.)-- (2.,3.92820323027551);
\draw [line width=1.pt,color=vert] (-2.,3.9282032302755114)-- (2.,-3.);
\begin{scriptsize}
\draw [color=black,fill=white] (-2.,-3.) circle (5pt);
\draw [fill=black] (2.,-3.) circle (5pt);
\draw [color=black,fill=white] (4.,0.46410161513775483) circle (5pt);
\draw [fill=black] (2.,3.92820323027551) circle (5pt);
\draw [color=black,fill=white] (-2.,3.9282032302755114) circle (5pt);
\draw [fill=black] (-4.,0.46410161513775794) circle (5pt);
\end{scriptsize}
\end{tikzpicture}}}
+ \f{\l_2}{6 N^{4}} \vcenter{\hbox{\tikzsetnextfilename{long_pillow_rank32}
\begin{tikzpicture}[line cap=round,line join=round,>=triangle 45,x=1.0cm,y=1.0cm,scale=0.25]
\draw [line width=1.pt,color=bleu] (2.,3.92820323027551)-- (4.,0.46410161513775483);
\draw [line width=1.pt,color=rouge] (4.,0.46410161513775483)-- (2.,-3.);
\draw [line width=1.pt,color=rouge] (-2.,-3.)-- (-4.,0.46410161513775794);
\draw [line width=1.pt,color=bleu] (-4.,0.46410161513775794)-- (-2.,3.9282032302755114);
\draw [line width=1.pt,color=vert] (-4.,0.46410161513775794)-- (4.,0.46410161513775483);
\draw [shift={(0.,0.7163553774951986)},line width=1.pt,color=rouge]  plot[domain=1.0138566155959863:2.1277360379938064,variable=\t]({1.*3.783644622504801*cos(\t r)+0.*3.783644622504801*sin(\t r)},{0.*3.783644622504801*cos(\t r)+1.*3.783644622504801*sin(\t r)});
\draw [shift={(0.,8.384780999581391)},line width=1.pt,color=vert]  plot[domain=4.2905543607066985:5.134223600062681,variable=\t]({1.*4.884780999581392*cos(\t r)+0.*4.884780999581392*sin(\t r)},{0.*4.884780999581392*cos(\t r)+1.*4.884780999581392*sin(\t r)});
\draw [shift={(0.,0.21184785278031246)},line width=1.pt,color=bleu]  plot[domain=4.1554492691857785:5.269328691583599,variable=\t]({1.*3.7836446225048*cos(\t r)+0.*3.7836446225048*sin(\t r)},{0.*3.7836446225048*cos(\t r)+1.*3.7836446225048*sin(\t r)});
\draw [shift={(0.,-7.456577769305882)},line width=1.pt,color=vert]  plot[domain=1.1489617071169056:1.9926309464728869,variable=\t]({1.*4.884780999581395*cos(\t r)+0.*4.884780999581395*sin(\t r)},{0.*4.884780999581395*cos(\t r)+1.*4.884780999581395*sin(\t r)});
\begin{scriptsize}
\draw [color=black,fill=white] (-2.,-3.) circle (5pt);
\draw [fill=black] (2.,-3.) circle (5pt);
\draw [color=black,fill=white] (4.,0.46410161513775483) circle (5pt);
\draw [fill=black] (2.,3.92820323027551) circle (5pt);
\draw [color=black,fill=white] (-2.,3.9282032302755114) circle (5pt);
\draw [fill=black] (-4.,0.46410161513775794) circle (5pt);
\end{scriptsize}
\end{tikzpicture}}}
+ \f{\l_3}{6 N^{4}} \vcenter{\hbox{\tikzsetnextfilename{circle2}
\begin{tikzpicture}[line cap=round,line join=round,>=triangle 45,x=1.0cm,y=1.0cm,scale=0.25]
\draw [line width=1.pt,color=bleu] (-2.,3.9282032302755114)-- (2.,3.92820323027551);
\draw [line width=1.pt,color=bleu] (4.,0.46410161513775483)-- (2.,-3.);
\draw [line width=1.pt,color=bleu] (-2.,-3.)-- (-4.,0.46410161513775794);
\draw [shift={(0.2475952641916737,0.3211524227066167)},line width=1.pt,color=rouge]  plot[domain=2.12803655173777:3.1079512042452184,variable=\t]({1.*4.25*cos(\t r)+0.*4.25*sin(\t r)},{0.*4.25*cos(\t r)+1.*4.25*sin(\t r)});
\draw [shift={(-6.247595264191662,4.071152422706646)},line width=1.pt,color=vert]  plot[domain=5.269629205327561:6.249543857835013,variable=\t]({1.*4.25*cos(\t r)+0.*4.25*sin(\t r)},{0.*4.25*cos(\t r)+1.*4.25*sin(\t r)});
\draw [shift={(-0.2475952641916725,0.3211524227066176)},line width=1.pt,color=rouge]  plot[domain=0.03364144934457438:1.0135561018520238,variable=\t]({1.*4.25*cos(\t r)+0.*4.25*sin(\t r)},{0.*4.25*cos(\t r)+1.*4.25*sin(\t r)});
\draw [shift={(6.2475952641916646,4.071152422706645)},line width=1.pt,color=vert]  plot[domain=3.175234102934366:4.155148755441817,variable=\t]({1.*4.25*cos(\t r)+0.*4.25*sin(\t r)},{0.*4.25*cos(\t r)+1.*4.25*sin(\t r)});
\draw [shift={(0.,-6.75)},line width=1.pt,color=vert]  plot[domain=1.0808390005411683:2.060753653048625,variable=\t]({1.*4.25*cos(\t r)+0.*4.25*sin(\t r)},{0.*4.25*cos(\t r)+1.*4.25*sin(\t r)});
\draw [shift={(0.,0.75)},line width=1.pt,color=rouge]  plot[domain=4.222431654130961:5.202346306638418,variable=\t]({1.*4.25*cos(\t r)+0.*4.25*sin(\t r)},{0.*4.25*cos(\t r)+1.*4.25*sin(\t r)});
\begin{scriptsize}
\draw [color=black,fill=white] (-2.,-3.) circle (5pt);
\draw [fill=black] (2.,-3.) circle (5pt);
\draw [color=black,fill=white] (4.,0.46410161513775483) circle (5pt);
\draw [fill=black] (2.,3.92820323027551) circle (5pt);
\draw [color=black,fill=white] (-2.,3.9282032302755114) circle (5pt);
\draw [fill=black] (-4.,0.46410161513775794) circle (5pt);
\end{scriptsize}
\end{tikzpicture}}}
 \right. \\
 & \left.
 + \f{\l_4}{6 N^{5}} \vcenter{\hbox{\tikzsetnextfilename{pillow_trace_rank32}
\begin{tikzpicture}[line cap=round,line join=round,>=triangle 45,x=1.0cm,y=1.0cm,scale=0.25]
\draw [line width=1.pt,color=rouge] (-2.,1.5)-- (-2.,-1.5);
\draw [line width=1.pt,color=rouge] (2.,1.5)-- (2.,-1.5);
\draw [shift={(0.,-1.75-0.5)},line width=1.pt,color=vert]  plot[domain=1.0808390005411683:2.060753653048625,variable=\t]({1.*4.25*cos(\t r)+0.*4.25*sin(\t r)},{0.*4.25*cos(\t r)+1.*4.25*sin(\t r)});
\draw [shift={(0.,5.75-0.5)},line width=1.pt,color=bleu]  plot[domain=4.222431654130961:5.202346306638418,variable=\t]({1.*4.25*cos(\t r)+0.*4.25*sin(\t r)},{0.*4.25*cos(\t r)+1.*4.25*sin(\t r)});
\draw [shift={(0.,1.75+0.5)},line width=1.pt,color=vert]  plot[domain=4.222431654130961:5.202346306638418,variable=\t]({1.*4.25*cos(\t r)+0.*4.25*sin(\t r)},{0.*4.25*cos(\t r)+1.*4.25*sin(\t r)});
\draw [shift={(0.,-5.75+0.5)},line width=1.pt,color=bleu]  plot[domain=1.0808390005411683:2.060753653048625,variable=\t]({1.*4.25*cos(\t r)+0.*4.25*sin(\t r)},{0.*4.25*cos(\t r)+1.*4.25*sin(\t r)});
\draw [fill=white] (-2.,2.-0.5) circle (5pt);
\draw [fill=black] (-2.,-2.+0.5) circle (5pt);
\draw [fill=black] (2.,2.-0.5) circle (5pt);
\draw [fill=white] (2.,-2.+0.5) circle (5pt);

\draw [line width=1.pt,color=rouge] (-2.,0.+4.5)-- (2.,0.+4.5);
\draw [shift={(0.,-3.75+4.5)},line width=1.pt,color=bleu]  plot[domain=1.0808390005411683:2.060753653048625,variable=\t]({1.*4.25*cos(\t r)+0.*4.25*sin(\t r)},{0.*4.25*cos(\t r)+1.*4.25*sin(\t r)});
\draw [shift={(0.,3.75+4.5)},line width=1.pt,color=vert]  plot[domain=4.222431654130961:5.202346306638418,variable=\t]({1.*4.25*cos(\t r)+0.*4.25*sin(\t r)},{0.*4.25*cos(\t r)+1.*4.25*sin(\t r)});
\draw [fill=white] (-2.,0.+4.5) circle (5pt);
\draw [fill=black] (2.,0.+4.5) circle (5pt);
\end{tikzpicture}}}
  + \f{\l_5}{6 N^{6}} \vcenter{\hbox{\tikzsetnextfilename{triple_trace_rank32}
\begin{tikzpicture}[line cap=round,line join=round,>=triangle 45,x=1.0cm,y=1.0cm,scale=0.25]
\draw [line width=1.pt,color=rouge] (-2.,0.)-- (2.,0.);
\draw [shift={(0.,-3.75)},line width=1.pt,color=bleu]  plot[domain=1.0808390005411683:2.060753653048625,variable=\t]({1.*4.25*cos(\t r)+0.*4.25*sin(\t r)},{0.*4.25*cos(\t r)+1.*4.25*sin(\t r)});
\draw [shift={(0.,3.75)},line width=1.pt,color=vert]  plot[domain=4.222431654130961:5.202346306638418,variable=\t]({1.*4.25*cos(\t r)+0.*4.25*sin(\t r)},{0.*4.25*cos(\t r)+1.*4.25*sin(\t r)});
\draw [fill=white] (-2.,0.) circle (5pt);
\draw [fill=black] (2.,0.) circle (5pt);

\draw [line width=1.pt,color=rouge] (-2.,0.+3)-- (2.,0.+3);
\draw [shift={(0.,-3.75+3)},line width=1.pt,color=bleu]  plot[domain=1.0808390005411683:2.060753653048625,variable=\t]({1.*4.25*cos(\t r)+0.*4.25*sin(\t r)},{0.*4.25*cos(\t r)+1.*4.25*sin(\t r)});
\draw [shift={(0.,3.75+3)},line width=1.pt,color=vert]  plot[domain=4.222431654130961:5.202346306638418,variable=\t]({1.*4.25*cos(\t r)+0.*4.25*sin(\t r)},{0.*4.25*cos(\t r)+1.*4.25*sin(\t r)});
\draw [fill=white] (-2.,0.+3) circle (5pt);
\draw [fill=black] (2.,0.+3) circle (5pt);

\draw [line width=1.pt,color=rouge] (-2.,0.-3)-- (2.,0.-3);
\draw [shift={(0.,-3.75-3)},line width=1.pt,color=bleu]  plot[domain=1.0808390005411683:2.060753653048625,variable=\t]({1.*4.25*cos(\t r)+0.*4.25*sin(\t r)},{0.*4.25*cos(\t r)+1.*4.25*sin(\t r)});
\draw [shift={(0.,3.75-3)},line width=1.pt,color=vert]  plot[domain=4.222431654130961:5.202346306638418,variable=\t]({1.*4.25*cos(\t r)+0.*4.25*sin(\t r)},{0.*4.25*cos(\t r)+1.*4.25*sin(\t r)});
\draw [fill=white] (-2.,0.-3) circle (5pt);
\draw [fill=black] (2.,0.-3) circle (5pt);
\end{tikzpicture}}}
  \right)\,,
\end{split}
\ee
where a (normalized) sum over color permutations should be understood, whenever it is non-trivial (see App.~\ref{ap:conventions} for more details on our conventions).
The graphs representing the tensor invariants are also called \emph{bubbles}. Bubbles which are composed of one, two, or three connected components are referred to as single-trace, double-trace, or triple-trace, respectively, for analogy with the matrix case, and bubbles $I_b$ for which $\rho(I_b)=0$ are called \emph{maximally single trace}  (MST), as each of their 2-colored subgraphs are single trace.
The $I_1$ invariant is the only MST bubble in our action.

\paragraph{Colored graphs and Feynman diagrams.} 

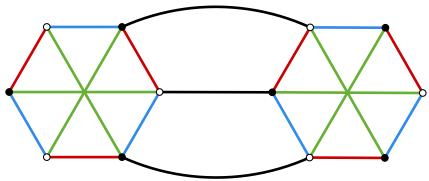
\begin{figure}[htbp]
\centering
\tikzsetnextfilename{4colored2}
\begin{tikzpicture}[line cap=round,line join=round,>=triangle 45,x=1.0cm,y=1.0cm,scale=0.25]
\draw [line width=1.pt,color=bleu] (-2.,3.9282032302755114)-- (2.,3.92820323027551);
\draw [line width=1.pt,color=rouge] (2.,3.92820323027551)-- (4.,0.46410161513775483);
\draw [line width=1.pt,color=bleu] (4.,0.46410161513775483)-- (2.,-3.);
\draw [line width=1.pt,color=rouge] (2.,-3.)-- (-2.,-3.);
\draw [line width=1.pt,color=bleu] (-2.,-3.)-- (-4.,0.46410161513775794);
\draw [line width=1.pt,color=rouge] (-4.,0.46410161513775794)-- (-2.,3.9282032302755114);
\draw [line width=1.pt,color=vert] (-4.,0.46410161513775794)-- (4.,0.46410161513775483);
\draw [line width=1.pt,color=vert] (-2.,-3.)-- (2.,3.92820323027551);
\draw [line width=1.pt,color=vert] (-2.,3.9282032302755114)-- (2.,-3.);
\draw [line width=1.pt,color=bleu] (16.00889143984295,3.8862244623504685)-- (12.008934908181699,3.9048723796736873);
\draw [line width=1.pt,color=rouge] (12.008934908181699,3.9048723796736873)-- (9.992807072221494,0.4501323678831591);
\draw [line width=1.pt,color=bleu] (9.992807072221494,0.4501323678831591)-- (11.976635767922538,-3.023255561230589);
\draw [line width=1.pt,color=rouge] (11.976635767922538,-3.023255561230589)-- (15.976592299583789,-3.0419034785538086);
\draw [line width=1.pt,color=bleu] (15.976592299583789,-3.0419034785538086)-- (17.992720135543994,0.4128365332367223);
\draw [line width=1.pt,color=rouge] (17.992720135543994,0.4128365332367223)-- (16.00889143984295,3.8862244623504685);
\draw [line width=1.pt,color=vert] (17.992720135543994,0.4128365332367223)-- (9.992807072221494,0.4501323678831591);
\draw [line width=1.pt,color=vert] (15.976592299583789,-3.0419034785538086)-- (12.008934908181699,3.9048723796736873);
\draw [line width=1.pt,color=vert] (16.00889143984295,3.8862244623504685)-- (11.976635767922538,-3.023255561230589);
\draw [shift={(6.978788798239646,-7.099605555190955)},line width=1.pt]  plot[domain=1.142053805626341:1.994876851745199,variable=\t]({1.*12.099624147314989*cos(\t r)+0.*12.099624147314989*sin(\t r)},{0.*12.099624147314989*cos(\t r)+1.*12.099624147314989*sin(\t r)});
\draw [line width=1.pt] (4.,0.46410161513775483)-- (9.992807072221494,0.4501323678831591);
\draw [shift={(7.014018273981845,8.013839538211876)},line width=1.pt]  plot[domain=4.2836464592161345:5.136469505334992,variable=\t]({1.*12.099624147314993*cos(\t r)+0.*12.099624147314993*sin(\t r)},{0.*12.099624147314993*cos(\t r)+1.*12.099624147314993*sin(\t r)});
\begin{scriptsize}
\draw [color=black,fill=white] (-2.,-3.) circle (5pt);
\draw [fill=black] (2.,-3.) circle (5pt);
\draw [color=black,fill=white] (4.,0.46410161513775483) circle (5pt);
\draw [fill=black] (2.,3.92820323027551) circle (5pt);
\draw [color=black,fill=white] (-2.,3.9282032302755114) circle (5pt);
\draw [fill=black] (-4.,0.46410161513775794) circle (5pt);

\draw [fill=black] (16.00889143984295,3.8862244623504685) circle (5pt);
\draw [color=black,fill=white] (12.008934908181699,3.9048723796736873) circle (5pt);
\draw [fill=black] (9.992807072221494,0.4501323678831591) circle (5pt);
\draw [color=black,fill=white] (17.992720135543994,0.4128365332367223) circle (5pt);
\draw [fill=black] (15.976592299583789,-3.0419034785538086) circle (5pt);
\draw [color=black,fill=white] (11.976635767922538,-3.023255561230589) circle (5pt);
\end{scriptsize}
\end{tikzpicture}
\caption{$4$-colored graph corresponding to a two-loop Feynman diagram with external tensor contractions equivalent to $I_2$.}
\label{fig:4colored}
\end{figure}

We introduce some (mostly standard) notation for the perturbative expansion of the free energy (and the connected $n$-point functions) of the theory \cite{Bonzom:2012hw}. Each interaction invariant is represented as a $3$-colored graph as above.
Expanding around the free theory, the Gaussian average leads to the usual Wick contraction rules, for which we represent the propagators as edges of a new color, connecting a white and black node. We choose the black color for such propagators, or equivalently, the label $0$. We give an example of the resulting \emph{$4$-colored graphs} in Fig.~\ref{fig:4colored}.

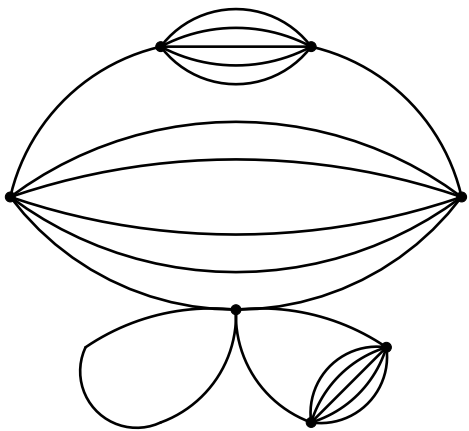
\begin{figure}[htbp]
\centering
\tikzsetnextfilename{melon_dtadpole2}
\begin{tikzpicture}[line cap=round,line join=round,>=triangle 45,x=1.0cm,y=1.0cm,scale=0.5]
\draw [shift={(0.,-17.5)},line width=1.pt]  plot[domain=1.240498971965643:1.9010936816241504,variable=\t]({1.*18.5*cos(\t r)+0.*18.5*sin(\t r)},{0.*18.5*cos(\t r)+1.*18.5*sin(\t r)});
\draw [shift={(0.,-8.)},line width=1.pt]  plot[domain=0.9272952180016122:2.214297435588181,variable=\t]({1.*10.*cos(\t r)+0.*10.*sin(\t r)},{0.*10.*cos(\t r)+1.*10.*sin(\t r)});
\draw [shift={(0.,8.)},line width=1.pt]  plot[domain=4.068887871591405:5.355890089177974,variable=\t]({1.*10.*cos(\t r)+0.*10.*sin(\t r)},{0.*10.*cos(\t r)+1.*10.*sin(\t r)});
\draw [shift={(0.,17.5)},line width=1.pt]  plot[domain=4.382091625555436:5.0426863352139435,variable=\t]({1.*18.5*cos(\t r)+0.*18.5*sin(\t r)},{0.*18.5*cos(\t r)+1.*18.5*sin(\t r)});
\draw [line width=1.pt] (-2.,4.)-- (2.,4.);
\draw [shift={(0.,0.25)},line width=1.pt]  plot[domain=1.0808390005411683:2.060753653048625,variable=\t]({1.*4.25*cos(\t r)+0.*4.25*sin(\t r)},{0.*4.25*cos(\t r)+1.*4.25*sin(\t r)});
\draw [shift={(0.,2.5)},line width=1.pt]  plot[domain=0.6435011087932844:2.498091544796509,variable=\t]({1.*2.5*cos(\t r)+0.*2.5*sin(\t r)},{0.*2.5*cos(\t r)+1.*2.5*sin(\t r)});
\draw [shift={(0.,7.75)},line width=1.pt]  plot[domain=4.222431654130961:5.202346306638418,variable=\t]({1.*4.25*cos(\t r)+0.*4.25*sin(\t r)},{0.*4.25*cos(\t r)+1.*4.25*sin(\t r)});
\draw [shift={(0.,5.5)},line width=1.pt]  plot[domain=3.7850937623830774:5.639684198386302,variable=\t]({1.*2.5*cos(\t r)+0.*2.5*sin(\t r)},{0.*2.5*cos(\t r)+1.*2.5*sin(\t r)});
\draw [shift={(-0.8078947368421046,-1.1921052631578954)},line width=1.pt]  plot[domain=1.7964843933151022:2.915904587069588,variable=\t]({1.*5.327201143392342*cos(\t r)+0.*5.327201143392342*sin(\t r)},{0.*5.327201143392342*cos(\t r)+1.*5.327201143392342*sin(\t r)});
\draw [shift={(0.8078947368421041,-1.192105263157896)},line width=1.pt]  plot[domain=0.2256880665202054:1.3451082602746909,variable=\t]({1.*5.327201143392343*cos(\t r)+0.*5.327201143392343*sin(\t r)},{0.*5.327201143392343*cos(\t r)+1.*5.327201143392343*sin(\t r)});
\draw [shift={(0.,4.5)},line width=1.pt]  plot[domain=3.7850937623830774:5.639684198386302,variable=\t]({1.*7.5*cos(\t r)+0.*7.5*sin(\t r)},{0.*7.5*cos(\t r)+1.*7.5*sin(\t r)});
\draw [line width=1.pt] (2.,-6.)-- (4.,-4.);
\draw [shift={(4.875,-6.875)},line width=1.pt]  plot[domain=1.8662371639386166:2.8461518164460733,variable=\t]({1.*3.005203820042827*cos(\t r)+0.*3.005203820042827*sin(\t r)},{0.*3.005203820042827*cos(\t r)+1.*3.005203820042827*sin(\t r)});
\draw [shift={(3.75,-5.75)},line width=1.pt]  plot[domain=1.4288992721907328:3.283489708193957,variable=\t]({1.*1.7677669529663689*cos(\t r)+0.*1.7677669529663689*sin(\t r)},{0.*1.7677669529663689*cos(\t r)+1.*1.7677669529663689*sin(\t r)});
\draw [shift={(2.25,-4.25)},line width=1.pt]  plot[domain=-1.7126933813990606:0.1418970546041639,variable=\t]({1.*1.7677669529663689*cos(\t r)+0.*1.7677669529663689*sin(\t r)},{0.*1.7677669529663689*cos(\t r)+1.*1.7677669529663689*sin(\t r)});
\draw [shift={(1.125,-3.125)},line width=1.pt]  plot[domain=5.0078298175284095:5.987744470035866,variable=\t]({1.*3.005203820042827*cos(\t r)+0.*3.005203820042827*sin(\t r)},{0.*3.005203820042827*cos(\t r)+1.*3.005203820042827*sin(\t r)});
\draw [shift={(3.044848484848485,-3.1367676767676764)},line width=1.pt]  plot[domain=3.096705097539215:4.362485416735506,variable=\t]({1.*3.047918583737522*cos(\t r)+0.*3.047918583737522*sin(\t r)},{0.*3.047918583737522*cos(\t r)+1.*3.047918583737522*sin(\t r)});
\draw [shift={(0.6467105263157892,-8.913157894736843)},line width=1.pt]  plot[domain=0.9719039150693753:1.6797314122666895,variable=\t]({1.*5.948417503247086*cos(\t r)+0.*5.948417503247086*sin(\t r)},{0.*5.948417503247086*cos(\t r)+1.*5.948417503247086*sin(\t r)});
\draw [shift={(-3.0448484848484862,-3.136767676767675)},line width=1.pt]  plot[domain=-1.2208927631457138:0.04488755605057786,variable=\t]({1.*3.047918583737523*cos(\t r)+0.*3.047918583737523*sin(\t r)},{0.*3.047918583737523*cos(\t r)+1.*3.047918583737523*sin(\t r)});
\draw [shift={(-0.6467105263157911,-8.913157894736841)},line width=1.pt]  plot[domain=1.4618612413231034:2.169688738520418,variable=\t]({1.*5.948417503247084*cos(\t r)+0.*5.948417503247084*sin(\t r)},{0.*5.948417503247084*cos(\t r)+1.*5.948417503247084*sin(\t r)});
\draw [shift={(-2.6395384615384616,-4.639538461538462)},line width=1.pt]  plot[domain=2.702158596409465:5.151823037565018,variable=\t]({1.*1.5032847506111804*cos(\t r)+0.*1.5032847506111804*sin(\t r)},{0.*1.5032847506111804*cos(\t r)+1.*1.5032847506111804*sin(\t r)});
\begin{scriptsize}
\draw [fill=black] (-6.,0.) circle (3.75pt);
\draw [fill=black] (6.,0.) circle (3.75pt);
\draw [fill=black] (-2.,4.) circle (3.75pt);
\draw [fill=black] (2.,4.) circle (3.75pt);
\draw [fill=black] (0.,-3.) circle (3.75pt);
\draw [fill=black] (4.,-4.) circle (3.75pt);
\draw [fill=black] (2.,-6.) circle (3.75pt);
\end{scriptsize}
\end{tikzpicture}
\caption{An example of melon-tadpole Feynman diagram. Double tadpoles are based on the $I_b$ ($b \in [1,5]$) vertices and melons are based on $I_1$ vertices.}
\label{fig:melon_dtadpole}
\end{figure}

Ordinary \emph{Feynman diagrams}, the only objects that we will actually call by such name here, are obtained by shrinking each interaction bubble to a point, which we will call an  interaction vertex, or just vertex.
We give an example of such a Feynman diagram in Fig.~\ref{fig:melon_dtadpole}. While Feynman diagrams are sufficient for representing Feynman integrals, the $4$-colored graphs are necessary in order to identify the scaling in $N$. 
Indeed, in a $4$-colored graph, each propagator identifies all three indices on its two end tensors whereas each edge of color $i$ identifies only one pair of indices between its end tensors. The indices will then circulate along the cycles of color $0i$, which we call \emph{faces}, hence each face gives rise to a free sum, that is, a factor $N$. 
The amplitude of a Feynman diagram $\mathcal{G}$ thus scales as $A(\mathcal{G})\sim N^{F-3n_{1}-4n_2-4n_3-5n_4-6n_5}$, with $F$ the total number of faces in the associated $4$-colored graph and $n_i$ the number of bubbles of the interaction $i$. 
The existence of the large-$N$ limit relies on the fact that the power of $N$ is bounded from above \cite{RTM,Carrozza:2015adg}.

\paragraph{Melonic graphs and melonic diagrams.} 

Melonic $k$-valent graphs are defined constructively starting from the fundamental melon, i.e.\ the unique graph built out of two $k$-valent vertices without forming self-loops (or tadpoles), and then iteratively inserting on any edge a melonic 2-point function, i.e.\ the graph obtained from the fundamental melon by cutting one edge in the middle. Notice that melonic $k$-valent graphs are always bipartite, and edge colorable with $k$ colors. 

An important result in rank-$r$ tensor models is that if one only allows for interaction bubbles which are melonic $r$-valent graphs, then in the perturbative expansion the leading order vacuum graphs at large $N$ are melonic $(r+1)$-valent graphs \cite{Bonzom:2012hw}.
However, it is important to notice that melonic $(r+1)$-valent graphs do not correspond to melonic Feynman diagrams, i.e.\  they do not remain melonic after shrinking the colors from 1 to $r$.
From the point of view of the Feynman diagrams, melonic $(r+1)$-valent graphs reduce to the same type of cactus diagrams appearing in the large-$N$ limit of vector models, and therefore field theories based on such interaction are not expected to lead to very different results than vector models.\footnote{They can nevertheless lead to new phases with patterns of spontaneous symmetry breaking which are impossible in the vector case \cite{Benedetti:2018ghn}.} 

Adding non-melonic bubbles, things get more complicated, and possibly more interesting. In particular, it was found in Ref.~\cite{Carrozza:2015adg} that non-melonic interaction bubbles can be scaled in such a way that they also contribute at leading order in the $1/N$ expansion, and that for some interactions (in that specific example, the quartic tetrahedron interaction) their leading-order Feynman diagrams are melonic. The possibility of restricting the spacetime Feynman diagrams to the melonic type by means of a large-$N$ limit has been a main reason for studying tensor field theories in dimension $d\geq 1$, starting from \cite{Klebanov:2016xxf}.

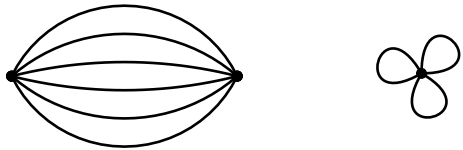
\begin{figure}[htbp]
\centering
\captionsetup[subfigure]{labelformat=empty}
\subfloat[]{\tikzsetnextfilename{melon_freeenergy3}
\begin{tikzpicture}[line cap=round,line join=round,>=triangle 45,x=1.0cm,y=1.0cm,scale=0.75]
\draw [shift={(0.,-7.875)},line width=1.pt]  plot[domain=1.3220863377013738:1.8195063158884195,variable=\t]({1.*8.125*cos(\t r)+0.*8.125*sin(\t r)},{0.*8.125*cos(\t r)+1.*8.125*sin(\t r)});
\draw [shift={(0.,-2.2916666666666665)},line width=1.pt]  plot[domain=0.8532549862537522:2.288337667336041,variable=\t]({1.*3.0416666666666665*cos(\t r)+0.*3.0416666666666665*sin(\t r)},{0.*3.0416666666666665*cos(\t r)+1.*3.0416666666666665*sin(\t r)});
\draw [shift={(0.,-0.975)},line width=1.pt]  plot[domain=0.45359769610777173:2.6879949574820214,variable=\t]({1.*2.225*cos(\t r)+0.*2.225*sin(\t r)},{0.*2.225*cos(\t r)+1.*2.225*sin(\t r)});
\draw [shift={(0.,2.2916666666666665)},line width=1.pt]  plot[domain=3.9948476398435453:5.429930320925834,variable=\t]({1.*3.0416666666666665*cos(\t r)+0.*3.0416666666666665*sin(\t r)},{0.*3.0416666666666665*cos(\t r)+1.*3.0416666666666665*sin(\t r)});
\draw [shift={(0.,7.875)},line width=1.pt]  plot[domain=4.463678991291166:4.961098969478212,variable=\t]({1.*8.125*cos(\t r)+0.*8.125*sin(\t r)},{0.*8.125*cos(\t r)+1.*8.125*sin(\t r)});
\draw [shift={(0.,0.975)},line width=1.pt]  plot[domain=3.595190349697565:5.829587611071815,variable=\t]({1.*2.225*cos(\t r)+0.*2.225*sin(\t r)},{0.*2.225*cos(\t r)+1.*2.225*sin(\t r)});
\begin{scriptsize}
\draw [fill=black] (-2.,0.) circle (2.5pt);
\draw [fill=black] (2.,0.) circle (2.5pt);
\draw [fill=black] (2.,0.) circle (2.5pt);
\draw [fill=black] (-2.,0.) circle (2.5pt);
\draw [fill=black] (2.,0.) circle (2.5pt);
\draw [fill=black] (-2.,0.) circle (2.5pt);
\draw [fill=black] (-2.,0.) circle (2.5pt);
\draw [fill=black] (2.,0.) circle (2.5pt);
\end{scriptsize}
\end{tikzpicture}}
\hspace{1cm}
\subfloat[]{\tikzsetnextfilename{ttadpole_freeenergy2}
\begin{tikzpicture}[line cap=round,line join=round,>=triangle 45,x=1.0cm,y=1.0cm,scale=.75]
\draw [fill=black] (0,0) circle (2.5pt);

\begin{feynman}
\diagram*{
 a -- [out=90, in=0, loop, min distance=1.5cm,line width=1.pt] a ;
 a -- [out=210, in=120, loop, min distance=1.5cm,line width=1.pt] a ;
 a -- [out=330, in=240, loop, min distance=1.5cm,line width=1.pt] a 
 };
\end{feynman}

\end{tikzpicture}}
\caption{The two Feynman diagrams (the fundamental melon on the left, and the triple-tadpole, or trefoil, on the right) starting from which all the vacuum melon-tadpole diagrams can be built. The melon is based on the wheel vertices and the triple-tadpole is based on any of the interactions $I_i$ (for explicit examples of the corresponding colored graphs in rank 5, see Figure \ref{fig:fund_vacuum_rank5}).}
\label{fig:fund_vacuum}
\end{figure}

\begin{figure}[htbp]
\centering
\captionsetup[subfigure]{labelformat=empty}
\subfloat[]{\tikzsetnextfilename{melonSDE2}
\begin{tikzpicture}[line cap=round,line join=round,>=triangle 45,x=1.0cm,y=1.0cm,scale=0.75]
\draw [line width=1.pt] (-2.,0.)-- (2.,0.);
\draw [shift={(0.,-3.75)},line width=1.pt]  plot[domain=1.0808390005411683:2.060753653048625,variable=\t]({1.*4.25*cos(\t r)+0.*4.25*sin(\t r)},{0.*4.25*cos(\t r)+1.*4.25*sin(\t r)});
\draw [shift={(0.,-1.5)},line width=1.pt]  plot[domain=0.6435011087932844:2.498091544796509,variable=\t]({1.*2.5*cos(\t r)+0.*2.5*sin(\t r)},{0.*2.5*cos(\t r)+1.*2.5*sin(\t r)});
\draw [shift={(0.,3.75)},line width=1.pt]  plot[domain=4.222431654130961:5.202346306638418,variable=\t]({1.*4.25*cos(\t r)+0.*4.25*sin(\t r)},{0.*4.25*cos(\t r)+1.*4.25*sin(\t r)});
\draw [shift={(0.,1.5)},line width=1.pt]  plot[domain=3.7850937623830774:5.639684198386302,variable=\t]({1.*2.5*cos(\t r)+0.*2.5*sin(\t r)},{0.*2.5*cos(\t r)+1.*2.5*sin(\t r)});
\draw [line width=1.pt,dash pattern=on 5pt off 5pt] (-2.,0.)-- (-3.,0.);
\draw [line width=1.pt,dash pattern=on 5pt off 5pt] (2.,0.)-- (3.,0.);
\begin{scriptsize}
\draw [fill=black] (-2.,0.) circle (2.5pt);
\draw [fill=black] (2.,0.) circle (2.5pt);
\end{scriptsize}
\end{tikzpicture}}
\hspace{1cm}
\subfloat[]{\tikzsetnextfilename{dtadpoleSDE2}
\begin{tikzpicture}[line cap=round,line join=round,>=triangle 45,x=1.0cm,y=1.0cm,scale=0.75]
\draw [fill=black] (0,0) circle (2.5pt);
\draw [line width=1.pt,dash pattern=on 5pt off 5pt] (0,0) -- (-0.5,-0.87);
\draw [line width=1.pt,dash pattern=on 5pt off 5pt] (0,0) -- (0.5,-0.87);
\draw[white] (-2,-1)--(2,-1);

\begin{feynman}
\diagram*{
 a -- [out=150, in=90, loop, min distance=1.5cm,line width=1.pt] a ;
 a -- [out=90, in=30, loop, min distance=1.5cm,line width=1.pt] a };
\end{feynman}

\end{tikzpicture}}
\caption{The two minimal two-point function Feynman diagrams used in the iterative construction of melon-tadpole diagrams. The melon is based on wheel vertices and the double tadpole is based on any of the interactions $I_i$.}
\label{fig:SDE}
\end{figure}
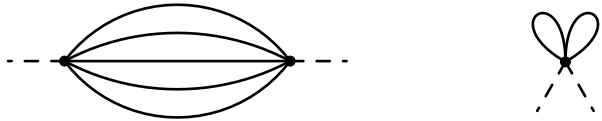

\paragraph{The large-$N$ limit.}  The $I_1$ invariant in \eqref{eq:int-action} (i.e.\ the first bubble in \eqref{eq:int-action-graph}, which we call the {\it wheel} graph, and which is also known as the complete bipartite graph $K_{3,3}$) stands out as the only non-melonic bubble in our action, and as a consequence, as the only interaction that does not lead only to tadpole corrections to the propagator at large $N$. It leads instead to Feynman diagrams which are of \textit{melon-tadpole} type \cite{Lionni:2017xvn,Lionni:2017yvi,Prakash:2019zia} (see Fig.~\ref{fig:melon_dtadpole}), i.e. diagrams obtained by repeated insertions of either melon or tadpole two-point functions (Fig.~\ref{fig:SDE}) on the propagators of either one of the two fundamental vacuum graphs in Fig.~\ref{fig:fund_vacuum}.
The 4-colored graph corresponding to the fundamental melon is built from two mirror wheel graphs (i.e.\ completing in a straightforward way Fig.~\ref{fig:4colored}), while the triple-tadpole is built on any of the interactions.\footnote{Notice that the leading 4-colored graph of the trefoil is unique for the melonic bubbles (essentially tadpoles like to be based on multilines), while there are three leading-order trefoils that can be built on the wheel. \label{foot:trefoil}}

As tadpole corrections just renormalize the mass, the effect of $I_2$ to $I_5$, and of the $I_1$ tadpoles, will be ignored in the discussion of the Schwinger-Dyson equations for  the two-point function, assuming that we are tuning the bare mass to exactly set the effective mass to zero.
Along the same line of thoughts, we have not included quartic interactions in our action, assuming that they can be tuned to zero. 
In fact, we will be using dimensional regularization, which for massless theories results in the tadpoles (and other power-divergent integrals) being regularized to zero (e.g.\  \cite{zinnjustin}); thus we will actually need no non-trivial tuning of bare parameters, and we will be able to keep mass and quartic couplings identically zero.

\subsection{Rank $5$}
\label{sec:rank5}
\paragraph{Action.}

The second sextic model we will consider in this paper is a $O(N)^5$ bosonic tensor model in $d$ dimensions. 
We consider a real tensor field of rank $5$, $\phi_{abcde}$ transforming under $O(N)^5$ with indices distinguished by their position. The action of the model is\footnote{The optimal scaling is now defined as $\r(J_b)=\frac{F(J_b)-10}{4}$ with a straightforward generalisation of Ref.~\cite{Carrozza:2015adg}.}:
\begin{align}
S[\phi] & = \f12 \int d^d x  \, \phi_{abcde} (   - \p_\m\p^\m)^{\zeta} \phi_{abcde} + S_{\rm int}[\phi] \, ,\\
 \label{eq:int-action-rank5}
S_{\rm int}[\phi] &= \int d^d x  \sum_{b=1}^6 \f{\k_b}{6 N^{5+\r(J_b)}} J_b \,.
\end{align}

The interaction part of the action can be written with the same graphical representation as for the previous model. However, because we are now considering a rank-5 model, the graphs representing the interactions will be $5$-colored graphs, and because we have real fields with $O(N)^5$ symmetry, the graphs will not be bipartite and the nodes will have all the same color (black). An action containing all the $O(N)^5$ invariants would be rather long,\footnote{Using the code provided in Ref.~\cite{Avohou:2019qrl} (built on a generalization of the methods of Ref.~\cite{BenGeloun:2013kta,BenGeloun:2017vwn}), we can count the total number of sextic invariants, with their different coloring choices, to be 1439.} and difficult to handle. We will restrict the potential by exploiting the large-$N$ limit: we start from the interaction whose bubble is a complete graph (i.e.\ in which for every pair of nodes there is an edge connecting them), and then include only the other interactions which are generated as radiative corrections, until  we obtain a renormalizable model, at large $N$. A set of interactions of this type has been introduced in Ref.~\cite{Ferrari:2017jgw} with the name of \emph{melo-complete family}.
As we will explain further below, it turns out that besides the complete graph we need to include only the melonic bubbles (a straightforward generalization of the melonic bubbles of rank 3) and one new non-bipartite bubble:\footnote{Following the same logic for  rank-3, starting with the wheel interaction, we would have obtained the same action as in \eqref{eq:int-action-graph}. As in that case the set of interactions exhausts the sextic $U(N)^3$ invariants, we have chosen a different perspective in its presentation.}
\be
\begin{split} \label{eq:int-action-graph-rank5}
S_{\rm int}[\phi] = & \int d^d x  
\left( \f{\kappa_1}{6 N^{5}} \vcenter{\hbox{\tikzsetnextfilename{5simplex2}
\begin{tikzpicture}[line cap=round,line join=round,>=triangle 45,x=1.0cm,y=1.0cm,scale=0.25,node_style_l/.style={circle,draw=black,fill=black,scale=0.25}]
\draw [line width=1.pt,color=rouge] (-2.,3.9282032302755114)-- (2.,3.92820323027551);
\draw [line width=1.pt,color=vert] (2.,3.92820323027551)-- (4.,0.46410161513775483);
\draw [line width=1.pt,color=rouge] (4.,0.46410161513775483)-- (2.,-3.);
\draw [line width=1.pt,color=vert] (2.,-3.)-- (-2.,-3.);
\draw [line width=1.pt,color=rouge] (-2.,-3.)-- (-4.,0.46410161513775794);
\draw [line width=1.pt,color=vert] (-4.,0.46410161513775794)-- (-2.,3.9282032302755114);
\draw [line width=1.pt,color=bleu] (-2.,3.9282032302755114)-- (-2.,-3.);
\draw [line width=1.pt,color=bleu] (2.,3.92820323027551)-- (2.,-3.);
\draw [line width=1.pt,color=bleu] (-4.,0.46410161513775794)-- (4.,0.46410161513775483);
\draw [line width=1.pt,color=orange] (-2.,3.9282032302755114)-- (4.,0.46410161513775483);
\draw [line width=1.pt,color=orange] (-4.,0.46410161513775794)-- (2.,-3.);
\draw [line width=1.pt,color=orange] (-2.,-3.)-- (2.,3.92820323027551);
\draw [line width=1.pt,color=lavenderpurple] (2.,3.92820323027551)-- (-4.,0.46410161513775794);
\draw [line width=1.pt,color=lavenderpurple] (4.,0.46410161513775483)-- (-2.,-3.);
\draw [line width=1.pt,color=lavenderpurple] (-2.,3.9282032302755114)-- (2.,-3.);
\begin{scriptsize}
\draw [fill=black] (-2.,-3.) circle (3.75pt);
\draw [fill=black] (2.,-3.) circle (3.75pt);
\draw [fill=black] (4.,0.46410161513775483) circle (3.75pt);
\draw [fill=black] (2.,3.92820323027551) circle (3.75pt);
\draw [fill=black] (-2.,3.9282032302755114) circle (3.75pt);
\draw [fill=black] (-4.,0.46410161513775794) circle (3.75pt);
\end{scriptsize}
\end{tikzpicture}}}
+ \f{\kappa_2}{6 N^{8}} \vcenter{\hbox{\tikzsetnextfilename{long_pillow_rank52}
\begin{tikzpicture}[line cap=round,line join=round,>=triangle 45,x=1.0cm,y=1.0cm,scale=0.25]
\draw [line width=1.pt,color=vert] (2.,3.92820323027551)-- (4.,0.46410161513775483);
\draw [line width=1.pt,color=rouge] (4.,0.46410161513775483)-- (2.,-3.);
\draw [line width=1.pt,color=rouge] (-2.,-3.)-- (-4.,0.46410161513775794);
\draw [line width=1.pt,color=vert] (-4.,0.46410161513775794)-- (-2.,3.9282032302755114);
\draw [line width=1.pt,color=bleu] (-4.,0.46410161513775794)-- (4.,0.46410161513775483);
\draw [shift={(0.,0.7163553774951986)},line width=1.pt,color=rouge]  plot[domain=1.0138566155959863:2.1277360379938064,variable=\t]({1.*3.783644622504801*cos(\t r)+0.*3.783644622504801*sin(\t r)},{0.*3.783644622504801*cos(\t r)+1.*3.783644622504801*sin(\t r)});
\draw [shift={(0.,8.384780999581391)},line width=1.pt,color=bleu]  plot[domain=4.2905543607066985:5.134223600062681,variable=\t]({1.*4.884780999581392*cos(\t r)+0.*4.884780999581392*sin(\t r)},{0.*4.884780999581392*cos(\t r)+1.*4.884780999581392*sin(\t r)});
\draw [shift={(0.,0.21184785278031246)},line width=1.pt,color=vert]  plot[domain=4.1554492691857785:5.269328691583599,variable=\t]({1.*3.7836446225048*cos(\t r)+0.*3.7836446225048*sin(\t r)},{0.*3.7836446225048*cos(\t r)+1.*3.7836446225048*sin(\t r)});
\draw [shift={(0.,-7.456577769305882)},line width=1.pt,color=bleu]  plot[domain=1.1489617071169056:1.9926309464728869,variable=\t]({1.*4.884780999581395*cos(\t r)+0.*4.884780999581395*sin(\t r)},{0.*4.884780999581395*cos(\t r)+1.*4.884780999581395*sin(\t r)});
\draw [shift={(0.,2.5980762113533147)},line width=1.pt,color=lavenderpurple]  plot[domain=0.5868919041112808:2.5547007494785117,variable=\t]({1.*2.4019237886466853*cos(\t r)+0.*2.4019237886466853*sin(\t r)},{0.*2.4019237886466853*cos(\t r)+1.*2.4019237886466853*sin(\t r)});
\draw [shift={(0.,5.618802153517002)},line width=1.pt,color=orange]  plot[domain=3.8433514753343063:5.581426485435072,variable=\t]({1.*2.6188021535170023*cos(\t r)+0.*2.6188021535170023*sin(\t r)},{0.*2.6188021535170023*cos(\t r)+1.*2.6188021535170023*sin(\t r)});
\draw [shift={(0.,-7.070897253800446)},line width=1.pt,color=lavenderpurple]  plot[domain=1.0827696375788163:2.058823016010976,variable=\t]({1.*8.530897253800445*cos(\t r)+0.*8.530897253800445*sin(\t r)},{0.*8.530897253800445*cos(\t r)+1.*8.530897253800445*sin(\t r)});
\draw [shift={(0.,-1.6698729810778021)},line width=1.pt,color=orange]  plot[domain=3.7284845577010737:5.696293403068305,variable=\t]({1.*2.4019237886466853*cos(\t r)+0.*2.4019237886466853*sin(\t r)},{0.*2.4019237886466853*cos(\t r)+1.*2.4019237886466853*sin(\t r)});
\draw [shift={(0.,-4.690598923241491)},line width=1.pt,color=lavenderpurple]  plot[domain=0.701758821744514:2.4398338318452786,variable=\t]({1.*2.6188021535170036*cos(\t r)+0.*2.6188021535170036*sin(\t r)},{0.*2.6188021535170036*cos(\t r)+1.*2.6188021535170036*sin(\t r)});
\draw [shift={(0.,7.999100484075957)},line width=1.pt,color=orange]  plot[domain=4.224362291168609:5.200415669600769,variable=\t]({1.*8.530897253800443*cos(\t r)+0.*8.530897253800443*sin(\t r)},{0.*8.530897253800443*cos(\t r)+1.*8.530897253800443*sin(\t r)});
\begin{scriptsize}
\draw [fill=black] (-2.,-3.) circle (3.75pt);
\draw [fill=black] (2.,-3.) circle (3.75pt);
\draw [fill=black] (4.,0.46410161513775483) circle (3.75pt);
\draw [fill=black] (2.,3.92820323027551) circle (3.75pt);
\draw [fill=black] (-2.,3.9282032302755114) circle (3.75pt);
\draw [fill=black] (-4.,0.46410161513775794) circle (3.75pt);
\end{scriptsize}
\end{tikzpicture}}}
+ \f{\kappa_3}{6 N^{8}} \vcenter{\hbox{\tikzsetnextfilename{wheel_pillow2}
\begin{tikzpicture}[line cap=round,line join=round,>=triangle 45,x=1.0cm,y=1.0cm,scale=0.25,rotate=180,node_style_l/.style={circle,draw=black,fill=black,scale=0.25}]
\draw [line width=1.pt,color=rouge] (-2.,3.9282032302755114)-- (-4.,0.46410161513775794);
\draw [line width=1.pt,color=rouge] (-2.,-3.)-- (2.,-3.);
\draw [line width=1.pt,color=rouge] (2.,3.92820323027551)-- (4.,0.46410161513775483);
\draw [shift={(0.,0.7163553774951986)},line width=1.pt,color=vert]  plot[domain=1.0138566155959863:2.1277360379938064,variable=\t]({1.*3.783644622504801*cos(\t r)+0.*3.783644622504801*sin(\t r)},{0.*3.783644622504801*cos(\t r)+1.*3.783644622504801*sin(\t r)});
\draw [shift={(0.,2.5980762113533147)},line width=1.pt,color=bleu]  plot[domain=0.5868919041112808:2.5547007494785117,variable=\t]({1.*2.4019237886466853*cos(\t r)+0.*2.4019237886466853*sin(\t r)},{0.*2.4019237886466853*cos(\t r)+1.*2.4019237886466853*sin(\t r)});
\draw [shift={(0.,5.258330249197708)},line width=1.pt,color=orange]  plot[domain=3.728484557701074:5.696293403068305,variable=\t]({1.*2.4019237886466858*cos(\t r)+0.*2.4019237886466858*sin(\t r)},{0.*2.4019237886466858*cos(\t r)+1.*2.4019237886466858*sin(\t r)});
\draw [shift={(0.,7.1400510830558215)},line width=1.pt,color=lavenderpurple]  plot[domain=4.1554492691857785:5.269328691583599,variable=\t]({1.*3.7836446225048*cos(\t r)+0.*3.7836446225048*sin(\t r)},{0.*3.7836446225048*cos(\t r)+1.*3.7836446225048*sin(\t r)});
\draw [shift={(-1.8480762113533145,-0.6028856829700224)},line width=1.pt, color=bleu]  plot[domain=2.6812870065044763:4.649095851871707,variable=\t]({1.*2.4019237886466853*cos(\t r)+0.*2.4019237886466853*sin(\t r)},{0.*2.4019237886466853*cos(\t r)+1.*2.4019237886466853*sin(\t r)});
\draw [shift={(-0.2184581664017478,0.3379747339590355)},line width=1.pt,color=vert]  plot[domain=3.108251717989182:4.222131140387002,variable=\t]({1.*3.7836446225048017*cos(\t r)+0.*3.7836446225048017*sin(\t r)},{0.*3.7836446225048017*cos(\t r)+1.*3.7836446225048017*sin(\t r)});
\draw [shift={(-5.7815418335982525,-2.8738731188212747)},line width=1.pt,color=lavenderpurple]  plot[domain=-0.03334093560061113:1.0805384867972085,variable=\t]({1.*3.783644622504801*cos(\t r)+0.*3.783644622504801*sin(\t r)},{0.*3.783644622504801*cos(\t r)+1.*3.783644622504801*sin(\t r)});
\draw [shift={(-4.151923788646688,-1.9330127018922183)},line width=1.pt,color=orange]  plot[domain=-0.46030564708531685:1.5075031982819136,variable=\t]({1.*2.401923788646686*cos(\t r)+0.*2.401923788646686*sin(\t r)},{0.*2.401923788646686*cos(\t r)+1.*2.401923788646686*sin(\t r)});
\draw [shift={(1.8480762113533116,-0.6028856829700225)},line width=1.pt,color=bleu]  plot[domain=-1.5075031982819143:0.4603056470853159,variable=\t]({1.*2.401923788646686*cos(\t r)+0.*2.401923788646686*sin(\t r)},{0.*2.401923788646686*cos(\t r)+1.*2.401923788646686*sin(\t r)});
\draw [shift={(0.21845816640174584,0.3379747339590366)},line width=1.pt,color=vert]  plot[domain=-1.0805384867972094:0.03334093560061046,variable=\t]({1.*3.7836446225048026*cos(\t r)+0.*3.7836446225048026*sin(\t r)},{0.*3.7836446225048026*cos(\t r)+1.*3.7836446225048026*sin(\t r)});
\draw [shift={(5.7815418335982525,-2.87387311882128)},line width=1.pt,color=lavenderpurple]  plot[domain=2.0610541667925846:3.1749335891904034,variable=\t]({1.*3.7836446225048044*cos(\t r)+0.*3.7836446225048044*sin(\t r)},{0.*3.7836446225048044*cos(\t r)+1.*3.7836446225048044*sin(\t r)});
\draw [shift={(4.151923788646685,-1.93301270189222)},line width=1.pt,color=orange]  plot[domain=1.6340894553078786:3.601898300675109,variable=\t]({1.*2.4019237886466858*cos(\t r)+0.*2.4019237886466858*sin(\t r)},{0.*2.4019237886466858*cos(\t r)+1.*2.4019237886466858*sin(\t r)});
\draw[color=white] (-2,-4)--(2,-4);
\begin{scriptsize}
\draw [fill=black] (4.,0.46410161513775483) circle (3.75pt);
\draw [fill=black] (2.,3.92820323027551) circle (3.75pt);
\draw [fill=black] (-2.,3.9282032302755114) circle (3.75pt);
\draw [fill=black] (-4.,0.46410161513775794) circle (3.75pt);
\draw [fill=black] (2.,-3.) circle (3.75pt);
\draw [fill=black] (-2.,-3.) circle (3.75pt);
\end{scriptsize}
\end{tikzpicture}}}
 \right. \\
 & \left.
 + \f{\kappa_4}{6 N^{9}} \vcenter{\hbox{\tikzsetnextfilename{pillow_trace2}
\begin{tikzpicture}[line cap=round,line join=round,>=triangle 45,x=1.0cm,y=1.0cm,scale=0.25,node_style_l/.style={circle,draw=black,fill=black,scale=0.25}]
\draw [line width=1.pt,color=rouge] (-2.,1.5)-- (-2.,-1.5);
\draw [line width=1.pt,color=rouge] (2.,1.5)-- (2.,-1.5);
\draw [shift={(0.,-1.75-0.5)},line width=1.pt,color=vert]  plot[domain=1.0808390005411683:2.060753653048625,variable=\t]({1.*4.25*cos(\t r)+0.*4.25*sin(\t r)},{0.*4.25*cos(\t r)+1.*4.25*sin(\t r)});
\draw [shift={(0.,0.5-0.5)},line width=1.pt,color=bleu]  plot[domain=0.6435011087932844:2.498091544796509,variable=\t]({1.*2.5*cos(\t r)+0.*2.5*sin(\t r)},{0.*2.5*cos(\t r)+1.*2.5*sin(\t r)});
\draw [shift={(0.,3.5-0.5)},line width=1.pt,color=orange]  plot[domain=3.7850937623830774:5.639684198386302,variable=\t]({1.*2.5*cos(\t r)+0.*2.5*sin(\t r)},{0.*2.5*cos(\t r)+1.*2.5*sin(\t r)});
\draw [shift={(0.,5.75-0.5)},line width=1.pt,color=lavenderpurple]  plot[domain=4.222431654130961:5.202346306638418,variable=\t]({1.*4.25*cos(\t r)+0.*4.25*sin(\t r)},{0.*4.25*cos(\t r)+1.*4.25*sin(\t r)});
\draw [shift={(0.,-0.5+0.5)},line width=1.pt,color=bleu]  plot[domain=3.7850937623830774:5.639684198386302,variable=\t]({1.*2.5*cos(\t r)+0.*2.5*sin(\t r)},{0.*2.5*cos(\t r)+1.*2.5*sin(\t r)});
\draw [shift={(0.,1.75+0.5)},line width=1.pt,color=vert]  plot[domain=4.222431654130961:5.202346306638418,variable=\t]({1.*4.25*cos(\t r)+0.*4.25*sin(\t r)},{0.*4.25*cos(\t r)+1.*4.25*sin(\t r)});
\draw [shift={(0.,-5.75+0.5)},line width=1.pt,color=lavenderpurple]  plot[domain=1.0808390005411683:2.060753653048625,variable=\t]({1.*4.25*cos(\t r)+0.*4.25*sin(\t r)},{0.*4.25*cos(\t r)+1.*4.25*sin(\t r)});
\draw [shift={(0.,-3.5+0.5)},line width=1.pt,color=orange]  plot[domain=0.6435011087932844:2.498091544796509,variable=\t]({1.*2.5*cos(\t r)+0.*2.5*sin(\t r)},{0.*2.5*cos(\t r)+1.*2.5*sin(\t r)});
\draw [fill=black] (-2.,2.-0.5) circle (3.75pt);
\draw [fill=black] (-2.,-2.+0.5) circle (3.75pt);
\draw [fill=black] (2.,2.-0.5) circle (3.75pt);
\draw [fill=black] (2.,-2.+0.5) circle (3.75pt);

\draw [line width=1.pt,color=rouge] (-2.,0.+4.5)-- (2.,0.+4.5);
\draw [shift={(0.,-3.75+4.5)},line width=1.pt,color=orange]  plot[domain=1.0808390005411683:2.060753653048625,variable=\t]({1.*4.25*cos(\t r)+0.*4.25*sin(\t r)},{0.*4.25*cos(\t r)+1.*4.25*sin(\t r)});
\draw [shift={(0.,-1.5+4.5)},line width=1.pt,color=vert]  plot[domain=0.6435011087932844:2.498091544796509,variable=\t]({1.*2.5*cos(\t r)+0.*2.5*sin(\t r)},{0.*2.5*cos(\t r)+1.*2.5*sin(\t r)});
\draw [shift={(0.,1.5+4.5)},line width=1.pt,color=bleu]  plot[domain=3.7850937623830774:5.639684198386302,variable=\t]({1.*2.5*cos(\t r)+0.*2.5*sin(\t r)},{0.*2.5*cos(\t r)+1.*2.5*sin(\t r)});
\draw [shift={(0.,3.75+4.5)},line width=1.pt,color=lavenderpurple]  plot[domain=4.222431654130961:5.202346306638418,variable=\t]({1.*4.25*cos(\t r)+0.*4.25*sin(\t r)},{0.*4.25*cos(\t r)+1.*4.25*sin(\t r)});
\draw [fill=black] (-2.,0.+4.5) circle (3.75pt);
\draw [fill=black] (2.,0.+4.5) circle (3.75pt);
\end{tikzpicture}}}
 + \f{\kappa_5}{6 N^{10}} \vcenter{\hbox{\tikzsetnextfilename{triple_trace2}
\begin{tikzpicture}[line cap=round,line join=round,>=triangle 45,x=1.0cm,y=1.0cm,scale=0.25,node_style_l/.style={circle,draw=black,fill=black,scale=0.25}]
\draw [line width=1.pt,color=rouge] (-2.,0.)-- (2.,0.);
\draw [shift={(0.,-3.75)},line width=1.pt,color=orange]  plot[domain=1.0808390005411683:2.060753653048625,variable=\t]({1.*4.25*cos(\t r)+0.*4.25*sin(\t r)},{0.*4.25*cos(\t r)+1.*4.25*sin(\t r)});
\draw [shift={(0.,-1.5)},line width=1.pt,color=vert]  plot[domain=0.6435011087932844:2.498091544796509,variable=\t]({1.*2.5*cos(\t r)+0.*2.5*sin(\t r)},{0.*2.5*cos(\t r)+1.*2.5*sin(\t r)});
\draw [shift={(0.,1.5)},line width=1.pt,color=bleu]  plot[domain=3.7850937623830774:5.639684198386302,variable=\t]({1.*2.5*cos(\t r)+0.*2.5*sin(\t r)},{0.*2.5*cos(\t r)+1.*2.5*sin(\t r)});
\draw [shift={(0.,3.75)},line width=1.pt,color=lavenderpurple]  plot[domain=4.222431654130961:5.202346306638418,variable=\t]({1.*4.25*cos(\t r)+0.*4.25*sin(\t r)},{0.*4.25*cos(\t r)+1.*4.25*sin(\t r)});
\draw [fill=black] (-2.,0.) circle (3.75pt);
\draw [fill=black] (2.,0.) circle (3.75pt);

\draw [line width=1.pt,color=rouge] (-2.,0.+3)-- (2.,0.+3);
\draw [shift={(0.,-3.75+3)},line width=1.pt,color=orange]  plot[domain=1.0808390005411683:2.060753653048625,variable=\t]({1.*4.25*cos(\t r)+0.*4.25*sin(\t r)},{0.*4.25*cos(\t r)+1.*4.25*sin(\t r)});
\draw [shift={(0.,-1.5+3)},line width=1.pt,color=vert]  plot[domain=0.6435011087932844:2.498091544796509,variable=\t]({1.*2.5*cos(\t r)+0.*2.5*sin(\t r)},{0.*2.5*cos(\t r)+1.*2.5*sin(\t r)});
\draw [shift={(0.,1.5+3)},line width=1.pt,color=bleu]  plot[domain=3.7850937623830774:5.639684198386302,variable=\t]({1.*2.5*cos(\t r)+0.*2.5*sin(\t r)},{0.*2.5*cos(\t r)+1.*2.5*sin(\t r)});
\draw [shift={(0.,3.75+3)},line width=1.pt,color=lavenderpurple]  plot[domain=4.222431654130961:5.202346306638418,variable=\t]({1.*4.25*cos(\t r)+0.*4.25*sin(\t r)},{0.*4.25*cos(\t r)+1.*4.25*sin(\t r)});
\draw [fill=black] (-2.,0.+3) circle (3.75pt);
\draw [fill=black] (2.,0.+3) circle (3.75pt);

\draw [line width=1.pt,color=rouge] (-2.,0.-3)-- (2.,0.-3);
\draw [shift={(0.,-3.75-3)},line width=1.pt,color=orange]  plot[domain=1.0808390005411683:2.060753653048625,variable=\t]({1.*4.25*cos(\t r)+0.*4.25*sin(\t r)},{0.*4.25*cos(\t r)+1.*4.25*sin(\t r)});
\draw [shift={(0.,-1.5-3)},line width=1.pt,color=vert]  plot[domain=0.6435011087932844:2.498091544796509,variable=\t]({1.*2.5*cos(\t r)+0.*2.5*sin(\t r)},{0.*2.5*cos(\t r)+1.*2.5*sin(\t r)});
\draw [shift={(0.,1.5-3)},line width=1.pt,color=bleu]  plot[domain=3.7850937623830774:5.639684198386302,variable=\t]({1.*2.5*cos(\t r)+0.*2.5*sin(\t r)},{0.*2.5*cos(\t r)+1.*2.5*sin(\t r)});
\draw [shift={(0.,3.75-3)},line width=1.pt,color=lavenderpurple]  plot[domain=4.222431654130961:5.202346306638418,variable=\t]({1.*4.25*cos(\t r)+0.*4.25*sin(\t r)},{0.*4.25*cos(\t r)+1.*4.25*sin(\t r)});
\draw [fill=black] (-2.,0.-3) circle (3.75pt);
\draw [fill=black] (2.,0.-3) circle (3.75pt);
\end{tikzpicture}}}
  + \f{\kappa_6}{6 N^{7}} \vcenter{\hbox{\tikzsetnextfilename{triangle2}
\begin{tikzpicture}[line cap=round,line join=round,>=triangle 45,x=1.0cm,y=1.0cm,scale=0.25,node_style_l/.style={circle,draw=black,fill=black,scale=0.25}]

\node[node_style_l] (e) at (-2.,-3.){} ;
\node[node_style_l] (d) at (2,-3) {};
\node[node_style_l] (c) at (4.,0.46410161513775483) {};
\node[node_style_l] (a) at (-2.,3.9282032302755114) {};
\node[node_style_l] (b) at (2.,3.92820323027551) {};
\node[node_style_l] (f) at (-4.,0.46410161513775794) {};

\draw[line width=1.pt,color=rouge]  (a) edge   (b);
\draw[line width=1.pt,color=rouge] (c) edge (d);
\draw[line width=1.pt,color=rouge] (e) edge (f);
\draw[line width=1.pt,color=vert] (b) edge (c);
\draw[line width=1.pt,color=vert] (d) edge (e);
\draw[line width=1.pt,color=vert] (f) edge (a);
\draw[line width=1.pt,color=bleu] (a) edge (e);
\draw[line width=1.pt,color=bleu] (b) edge (d);
\draw[line width=1.pt,color=bleu] (f) edge (c);
\draw[line width=1.pt,color=orange] (a) edge [bend right] (b);
\draw[line width=1.pt,color=orange] (f) edge [bend right] (c);
\draw[line width=1.pt,color=orange] (e) edge [bend right] (d);
\draw[line width=1.pt,color=lavenderpurple] (a) edge [bend left] (b);
\draw[line width=1.pt,color=lavenderpurple] (f) edge [bend left] (c);
\draw[line width=1.pt,color=lavenderpurple] (e) edge [bend left] (d);
\draw[color=white] (-2,-4)--(2,-4);
\draw [fill=black] (e) circle (3.75pt);
\draw [fill=black] (2.,-3.) circle (3.75pt);
\draw [fill=black] (4.,0.46410161513775483) circle (3.75pt);
\draw [fill=black] (2.,3.92820323027551) circle (3.75pt);
\draw [fill=black] (-2.,3.9282032302755114) circle (3.75pt);
\draw [fill=black] (-4.,0.46410161513775794) circle (3.75pt);
\end{tikzpicture}}}
  \right)\,,
\end{split}
\ee
where a sum over color permutations should be understood. The conventions are detailed in App.~\ref{ap:conventions}.

\paragraph{Colored graphs and Feynman diagrams.}

\begin{figure}[htbp]
\centering
\captionsetup[subfigure]{labelformat=empty}
\vspace{-.5cm}
\begin{minipage}[c]{0.3\textwidth}
\subfloat[]{\tikzsetnextfilename{melon_freeenergy22}
\input{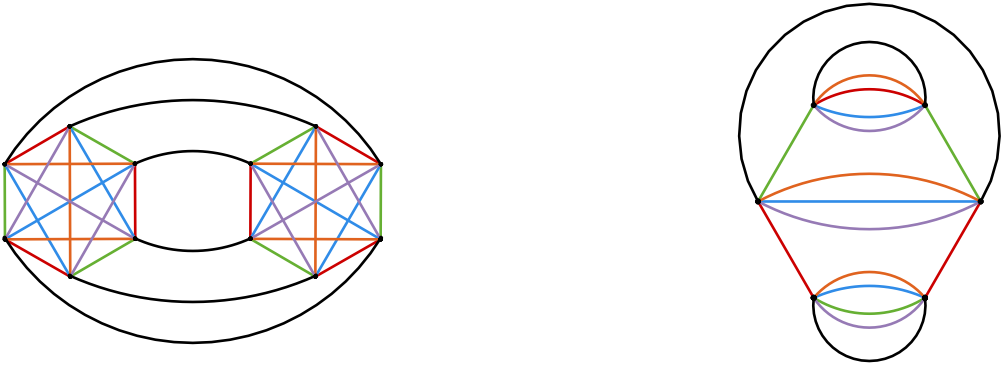}}
\end{minipage}
\begin{minipage}[c]{0.5\textwidth}
\subfloat[]{\tikzsetnextfilename{tadpole_long_pillow_rank52}
\begin{tikzpicture}[line cap=round,line join=round,>=triangle 45,x=1.0cm,y=1.0cm,scale=0.37]
\clip(-16.429953379333266,-9.231807183750373) rectangle (13.27043237065979,10.193681708463805);
\draw [line width=1.pt,color=vert] (2.,3.92820323027551)-- (4.,0.46410161513775483);
\draw [line width=1.pt,color=rouge] (4.,0.46410161513775483)-- (2.,-3.);
\draw [line width=1.pt,color=rouge] (-2.,-3.)-- (-4.,0.46410161513775794);
\draw [line width=1.pt,color=vert] (-4.,0.46410161513775794)-- (-2.,3.9282032302755114);
\draw [line width=1.pt,color=bleu] (-4.,0.46410161513775794)-- (4.,0.46410161513775483);
\draw [shift={(0.,0.7163553774951986)},line width=1.pt,color=rouge]  plot[domain=1.0138566155959863:2.1277360379938064,variable=\t]({1.*3.783644622504801*cos(\t r)+0.*3.783644622504801*sin(\t r)},{0.*3.783644622504801*cos(\t r)+1.*3.783644622504801*sin(\t r)});
\draw [shift={(0.,8.384780999581391)},line width=1.pt,color=bleu]  plot[domain=4.2905543607066985:5.134223600062681,variable=\t]({1.*4.884780999581392*cos(\t r)+0.*4.884780999581392*sin(\t r)},{0.*4.884780999581392*cos(\t r)+1.*4.884780999581392*sin(\t r)});
\draw [shift={(0.,0.21184785278031246)},line width=1.pt,color=vert]  plot[domain=4.1554492691857785:5.269328691583599,variable=\t]({1.*3.7836446225048*cos(\t r)+0.*3.7836446225048*sin(\t r)},{0.*3.7836446225048*cos(\t r)+1.*3.7836446225048*sin(\t r)});
\draw [shift={(0.,-7.456577769305882)},line width=1.pt,color=bleu]  plot[domain=1.1489617071169056:1.9926309464728869,variable=\t]({1.*4.884780999581395*cos(\t r)+0.*4.884780999581395*sin(\t r)},{0.*4.884780999581395*cos(\t r)+1.*4.884780999581395*sin(\t r)});
\draw [shift={(0.,2.5980762113533147)},line width=1.pt,color=orange]  plot[domain=0.5868919041112808:2.5547007494785117,variable=\t]({1.*2.4019237886466853*cos(\t r)+0.*2.4019237886466853*sin(\t r)},{0.*2.4019237886466853*cos(\t r)+1.*2.4019237886466853*sin(\t r)});
\draw [shift={(0.,5.618802153517002)},line width=1.pt,color=lavenderpurple]  plot[domain=3.8433514753343063:5.581426485435072,variable=\t]({1.*2.6188021535170023*cos(\t r)+0.*2.6188021535170023*sin(\t r)},{0.*2.6188021535170023*cos(\t r)+1.*2.6188021535170023*sin(\t r)});
\draw [shift={(0.,-7.070897253800446)},line width=1.pt,color=orange]  plot[domain=1.0827696375788163:2.058823016010976,variable=\t]({1.*8.530897253800445*cos(\t r)+0.*8.530897253800445*sin(\t r)},{0.*8.530897253800445*cos(\t r)+1.*8.530897253800445*sin(\t r)});
\draw [shift={(0.,-1.6698729810778021)},line width=1.pt,color=lavenderpurple]  plot[domain=3.7284845577010737:5.696293403068305,variable=\t]({1.*2.4019237886466853*cos(\t r)+0.*2.4019237886466853*sin(\t r)},{0.*2.4019237886466853*cos(\t r)+1.*2.4019237886466853*sin(\t r)});
\draw [shift={(0.,-4.690598923241491)},line width=1.pt,color=orange]  plot[domain=0.701758821744514:2.4398338318452786,variable=\t]({1.*2.6188021535170036*cos(\t r)+0.*2.6188021535170036*sin(\t r)},{0.*2.6188021535170036*cos(\t r)+1.*2.6188021535170036*sin(\t r)});
\draw [shift={(0.,7.999100484075957)},line width=1.pt,color=lavenderpurple]  plot[domain=4.224362291168609:5.200415669600769,variable=\t]({1.*8.530897253800443*cos(\t r)+0.*8.530897253800443*sin(\t r)},{0.*8.530897253800443*cos(\t r)+1.*8.530897253800443*sin(\t r)});
\draw [shift={(0.,4.183741176803939)},line width=1.pt]  plot[domain=-0.1270804326514412:3.268673086241234,variable=\t]({1.*2.016258823196061*cos(\t r)+0.*2.016258823196061*sin(\t r)},{0.*2.016258823196061*cos(\t r)+1.*2.016258823196061*sin(\t r)});
\draw [shift={(0.,-3.255537946528426)},line width=1.pt]  plot[domain=-3.2686730862412348:0.12708043265144067,variable=\t]({1.*2.0162588231960603*cos(\t r)+0.*2.0162588231960603*sin(\t r)},{0.*2.0162588231960603*cos(\t r)+1.*2.0162588231960603*sin(\t r)});
\draw [shift={(0.,2.8914255552766344)},line width=1.pt]  plot[domain=-0.5454271499964882:3.687019803586281,variable=\t]({1.*4.678878232052137*cos(\t r)+0.*4.678878232052137*sin(\t r)},{0.*4.678878232052137*cos(\t r)+1.*4.678878232052137*sin(\t r)});
\begin{scriptsize}
\draw [fill=black] (-2.,-3.) circle (2.5pt);
\draw [fill=black] (2.,-3.) circle (2.5pt);
\draw [fill=black] (4.,0.46410161513775483) circle (2.5pt);
\draw [fill=black] (2.,3.92820323027551) circle (2.5pt);
\draw [fill=black] (-2.,3.9282032302755114) circle (2.5pt);
\draw [fill=black] (-4.,0.46410161513775794) circle (2.5pt);
\draw [color=black] (-2.,-3.) circle (2.5pt);
\draw [fill=black] (2.,-3.) circle (2.5pt);
\draw [color=black] (-2.,-3.) circle (2.5pt);
\draw [fill=black] (2.,-3.) circle (2.5pt);
\draw [fill=black] (-2.,-3.) circle (2.5pt);
\draw [fill=black] (2.,-3.) circle (2.5pt);
\draw [fill=black] (-2.,-3.) circle (2.5pt);
\draw [fill=black] (2.,-3.) circle (2.5pt);
\draw [fill=black] (-4.,0.46410161513775794) circle (2.5pt);
\draw [fill=black] (4.,0.46410161513775483) circle (2.5pt);
\draw [fill=black] (-2.,-3.) circle (2.5pt);
\draw [fill=black] (2.,-3.) circle (2.5pt);
\end{scriptsize}
\end{tikzpicture}}
\end{minipage}
\vspace{-1.5cm}
\caption{Two examples of vacuum $6$-colored graphs.}
\label{fig:fund_vacuum_rank5}
\end{figure}

\begin{figure}[htbp]
\centering
\tikzsetnextfilename{rung2}
\begin{tikzpicture}[line cap=round,line join=round,>=triangle 45,x=1.0cm,y=1.0cm,scale=0.25]
\draw [line width=1.pt,color=rouge] (-2.,3.9282032302755114)-- (2.,3.92820323027551);
\draw [line width=1.pt,color=vert] (2.,3.92820323027551)-- (4.,0.46410161513775483);
\draw [line width=1.pt,color=rouge] (4.,0.46410161513775483)-- (2.,-3.);
\draw [line width=1.pt,color=vert] (2.,-3.)-- (-2.,-3.);
\draw [line width=1.pt,color=rouge] (-2.,-3.)-- (-4.,0.46410161513775794);
\draw [line width=1.pt,color=vert] (-4.,0.46410161513775794)-- (-2.,3.9282032302755114);
\draw [line width=1.pt,color=bleu] (-2.,3.9282032302755114)-- (-2.,-3.);
\draw [line width=1.pt,color=bleu] (2.,3.92820323027551)-- (2.,-3.);
\draw [line width=1.pt,color=bleu] (-4.,0.46410161513775794)-- (4.,0.46410161513775483);
\draw [line width=1.pt,color=orange] (-2.,3.9282032302755114)-- (4.,0.46410161513775483);
\draw [line width=1.pt,color=orange] (-4.,0.46410161513775794)-- (2.,-3.);
\draw [line width=1.pt,color=orange] (-2.,-3.)-- (2.,3.92820323027551);
\draw [line width=1.pt,color=lavenderpurple] (2.,3.92820323027551)-- (-4.,0.46410161513775794);
\draw [line width=1.pt,color=lavenderpurple] (4.,0.46410161513775483)-- (-2.,-3.);
\draw [line width=1.pt,color=lavenderpurple] (-2.,3.9282032302755114)-- (2.,-3.);
\draw [line width=1.pt,color=rouge] (16.,3.9282032302755123)-- (12.,3.9282032302755105);
\draw [line width=1.pt,color=vert] (12.,3.9282032302755105)-- (10.,0.4641016151377552);
\draw [line width=1.pt,color=rouge] (10.,0.4641016151377552)-- (12.,-3.);
\draw [line width=1.pt,color=vert] (12.,-3.)-- (16.,-3.);
\draw [line width=1.pt,color=rouge] (16.,-3.)-- (18.,0.46410161513775927);
\draw [line width=1.pt,color=vert] (18.,0.46410161513775927)-- (16.,3.9282032302755123);
\draw [line width=1.pt,color=bleu] (16.,3.9282032302755123)-- (16.,-3.);
\draw [line width=1.pt,color=bleu] (12.,3.9282032302755105)-- (12.,-3.);
\draw [line width=1.pt,color=bleu] (18.,0.46410161513775927)-- (10.,0.4641016151377552);
\draw [line width=1.pt,color=orange] (16.,3.9282032302755123)-- (10.,0.4641016151377552);
\draw [line width=1.pt,color=orange] (18.,0.46410161513775927)-- (12.,-3.);
\draw [line width=1.pt,color=orange] (16.,-3.)-- (12.,3.9282032302755105);
\draw [line width=1.pt,color=lavenderpurple] (12.,3.9282032302755105)-- (18.,0.46410161513775927);
\draw [line width=1.pt,color=lavenderpurple] (10.,0.4641016151377552)-- (16.,-3.);
\draw [line width=1.pt,color=lavenderpurple] (16.,3.9282032302755123)-- (12.,-3.);
\draw [line width=1.pt] (4.,0.46410161513775483)-- (10.,0.4641016151377552);
\draw [shift={(7.,-7.198557158514999)},line width=1.pt]  plot[domain=1.1484688490770039:1.9931238045127895,variable=\t]({1.*12.198557158515*cos(\t r)+0.*12.198557158515*sin(\t r)},{0.*12.198557158515*cos(\t r)+1.*12.198557158515*sin(\t r)});
\draw [shift={(7.,9.)},line width=1.pt]  plot[domain=4.317597860684928:5.1071801000844514,variable=\t]({1.*13.*cos(\t r)+0.*13.*sin(\t r)},{0.*13.*cos(\t r)+1.*13.*sin(\t r)});
\begin{scriptsize}
\draw [fill=black] (-2.,-3.) circle (3.75pt);
\draw [fill=black] (2.,-3.) circle (3.75pt);
\draw [fill=black] (4.,0.46410161513775483) circle (3.75pt);
\draw [fill=black] (2.,3.92820323027551) circle (3.75pt);
\draw [fill=black] (-2.,3.9282032302755114) circle (3.75pt);
\draw [fill=black] (-4.,0.46410161513775794) circle (3.75pt);
\draw [fill=black] (16.,3.9282032302755123) circle (3.75pt);
\draw [fill=black] (12.,3.9282032302755105) circle (3.75pt);
\draw [fill=black] (10.,0.4641016151377552) circle (3.75pt);
\draw [fill=black] (12.,-3.) circle (3.75pt);
\draw [fill=black] (16.,-3.) circle (3.75pt);
\draw [fill=black] (18.,0.46410161513775927) circle (3.75pt);
\end{scriptsize}
\end{tikzpicture}
\caption{$6$-colored graph corresponding to a two-loop correction to the six-point function, with external tensor contractions equivalent to $J_6$ (the prism).}
\label{fig:6colored}
\end{figure}
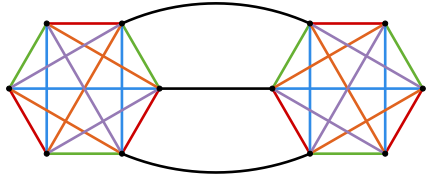

The expansion into Feynman diagrams is done similarly as for the previous model. Again, the propagators are represented by black edges. We give some examples of resulting $6$-colored graphs in Fig.~\ref{fig:fund_vacuum_rank5} and \ref{fig:6colored}.

\begin{figure}[htbp]
\centering
\tikzsetnextfilename{2rungs2}
\begin{tikzpicture}[line cap=round,line join=round,>=triangle 45,x=1.0cm,y=1.0cm,scale=0.25]
\draw [line width=1.pt,color=rouge] (-2.,3.9282032302755114)-- (2.,3.92820323027551);
\draw [line width=1.pt,color=vert] (2.,3.92820323027551)-- (4.,0.46410161513775483);
\draw [line width=1.pt,color=rouge] (4.,0.46410161513775483)-- (2.,-3.);
\draw [line width=1.pt,color=vert] (2.,-3.)-- (-2.,-3.);
\draw [line width=1.pt,color=rouge] (-2.,-3.)-- (-4.,0.46410161513775794);
\draw [line width=1.pt,color=vert] (-4.,0.46410161513775794)-- (-2.,3.9282032302755114);
\draw [line width=1.pt,color=bleu] (-2.,3.9282032302755114)-- (-2.,-3.);
\draw [line width=1.pt,color=bleu] (2.,3.92820323027551)-- (2.,-3.);
\draw [line width=1.pt,color=bleu] (-4.,0.46410161513775794)-- (4.,0.46410161513775483);
\draw [line width=1.pt,color=orange] (-2.,3.9282032302755114)-- (4.,0.46410161513775483);
\draw [line width=1.pt,color=orange] (-4.,0.46410161513775794)-- (2.,-3.);
\draw [line width=1.pt,color=orange] (-2.,-3.)-- (2.,3.92820323027551);
\draw [line width=1.pt,color=lavenderpurple] (2.,3.92820323027551)-- (-4.,0.46410161513775794);
\draw [line width=1.pt,color=lavenderpurple] (4.,0.46410161513775483)-- (-2.,-3.);
\draw [line width=1.pt,color=lavenderpurple] (-2.,3.9282032302755114)-- (2.,-3.);
\draw [line width=1.pt,color=rouge] (16.,3.9282032302755123)-- (12.,3.9282032302755105);
\draw [line width=1.pt,color=vert] (12.,3.9282032302755105)-- (10.,0.4641016151377552);
\draw [line width=1.pt,color=rouge] (10.,0.4641016151377552)-- (12.,-3.);
\draw [line width=1.pt,color=vert] (12.,-3.)-- (16.,-3.);
\draw [line width=1.pt,color=rouge] (16.,-3.)-- (18.,0.46410161513775927);
\draw [line width=1.pt,color=vert] (18.,0.46410161513775927)-- (16.,3.9282032302755123);
\draw [line width=1.pt,color=bleu] (16.,3.9282032302755123)-- (16.,-3.);
\draw [line width=1.pt,color=bleu] (12.,3.9282032302755105)-- (12.,-3.);
\draw [line width=1.pt,color=bleu] (18.,0.46410161513775927)-- (10.,0.4641016151377552);
\draw [line width=1.pt,color=orange] (16.,3.9282032302755123)-- (10.,0.4641016151377552);
\draw [line width=1.pt,color=orange] (18.,0.46410161513775927)-- (12.,-3.);
\draw [line width=1.pt,color=orange] (16.,-3.)-- (12.,3.9282032302755105);
\draw [line width=1.pt,color=lavenderpurple] (12.,3.9282032302755105)-- (18.,0.46410161513775927);
\draw [line width=1.pt,color=lavenderpurple] (10.,0.4641016151377552)-- (16.,-3.);
\draw [line width=1.pt,color=lavenderpurple] (16.,3.9282032302755123)-- (12.,-3.);
\draw [line width=1.pt] (4.,0.46410161513775483)-- (10.,0.4641016151377552);
\draw [shift={(7.,-7.198557158514999)},line width=1.pt]  plot[domain=1.1484688490770039:1.9931238045127895,variable=\t]({1.*12.198557158515*cos(\t r)+0.*12.198557158515*sin(\t r)},{0.*12.198557158515*cos(\t r)+1.*12.198557158515*sin(\t r)});
\draw [shift={(7.,9.)},line width=1.pt]  plot[domain=4.317597860684928:5.1071801000844514,variable=\t]({1.*13.*cos(\t r)+0.*13.*sin(\t r)},{0.*13.*cos(\t r)+1.*13.*sin(\t r)});
\draw [shift={(7.,-4.801442841485004)},line width=1.pt]  plot[domain=4.290061502666797:5.134716458102583,variable=\t]({1.*12.198557158514996*cos(\t r)+0.*12.198557158514996*sin(\t r)},{0.*12.198557158514996*cos(\t r)+1.*12.198557158514996*sin(\t r)});
\draw [shift={(7.,-21.)},line width=1.pt]  plot[domain=1.1760052070951352:1.965587446494658,variable=\t]({1.*13.*cos(\t r)+0.*13.*sin(\t r)},{0.*13.*cos(\t r)+1.*13.*sin(\t r)});
\draw [line width=1.pt,color=rouge] (-2.,-15.92820323027551)-- (2.,-15.92820323027551);
\draw [line width=1.pt,color=vert] (2.,-15.92820323027551)-- (4.,-12.464101615137755);
\draw [line width=1.pt,color=rouge] (4.,-12.464101615137755)-- (2.,-9.);
\draw [line width=1.pt,color=vert] (2.,-9.)-- (-2.,-9.);
\draw [line width=1.pt,color=rouge] (-2.,-9.)-- (-4.,-12.464101615137757);
\draw [line width=1.pt,color=vert] (-4.,-12.464101615137757)-- (-2.,-15.92820323027551);
\draw [line width=1.pt,color=bleu] (-2.,-15.92820323027551)-- (-2.,-9.);
\draw [line width=1.pt,color=bleu] (2.,-15.92820323027551)-- (2.,-9.);
\draw [line width=1.pt,color=bleu] (-4.,-12.464101615137757)-- (4.,-12.464101615137755);
\draw [line width=1.pt,color=orange] (-2.,-15.92820323027551)-- (4.,-12.464101615137755);
\draw [line width=1.pt,color=orange] (-4.,-12.464101615137757)-- (2.,-9.);
\draw [line width=1.pt,color=orange] (-2.,-9.)-- (2.,-15.92820323027551);
\draw [line width=1.pt,color=lavenderpurple] (2.,-15.92820323027551)-- (-4.,-12.464101615137757);
\draw [line width=1.pt,color=lavenderpurple] (4.,-12.464101615137755)-- (-2.,-9.);
\draw [line width=1.pt,color=lavenderpurple] (-2.,-15.92820323027551)-- (2.,-9.);
\draw [line width=1.pt,color=rouge] (16.,-15.928203230275512)-- (12.,-15.92820323027551);
\draw [line width=1.pt,color=vert] (12.,-15.92820323027551)-- (10.,-12.464101615137755);
\draw [line width=1.pt,color=rouge] (10.,-12.464101615137755)-- (12.,-9.);
\draw [line width=1.pt,color=vert] (12.,-9.)-- (16.,-9.);
\draw [line width=1.pt,color=rouge] (16.,-9.)-- (18.,-12.464101615137759);
\draw [line width=1.pt,color=vert] (18.,-12.464101615137759)-- (16.,-15.928203230275512);
\draw [line width=1.pt,color=bleu] (16.,-15.928203230275512)-- (16.,-9.);
\draw [line width=1.pt,color=bleu] (12.,-15.92820323027551)-- (12.,-9.);
\draw [line width=1.pt,color=bleu] (18.,-12.464101615137759)-- (10.,-12.464101615137755);
\draw [line width=1.pt,color=orange] (16.,-15.928203230275512)-- (10.,-12.464101615137755);
\draw [line width=1.pt,color=orange] (18.,-12.464101615137759)-- (12.,-9.);
\draw [line width=1.pt,color=orange] (16.,-9.)-- (12.,-15.92820323027551);
\draw [line width=1.pt,color=lavenderpurple] (12.,-15.92820323027551)-- (18.,-12.464101615137759);
\draw [line width=1.pt,color=lavenderpurple] (10.,-12.464101615137755)-- (16.,-9.);
\draw [line width=1.pt,color=lavenderpurple] (16.,-15.928203230275512)-- (12.,-9.);
\draw [line width=1.pt] (4.,-12.464101615137755)-- (10.,-12.464101615137755);
\draw [shift={(2.,-6.)},line width=1.pt]  plot[domain=2.498091544796509:3.7850937623830774,variable=\t]({1.*5.*cos(\t r)+0.*5.*sin(\t r)},{0.*5.*cos(\t r)+1.*5.*sin(\t r)});
\draw [shift={(11.57135869565218,-6.)},line width=1.pt]  plot[domain=-0.5954025490678632:0.5954025490678634,variable=\t]({1.*5.349099344990292*cos(\t r)+0.*5.349099344990292*sin(\t r)},{0.*5.349099344990292*cos(\t r)+1.*5.349099344990292*sin(\t r)});
\draw [shift={(7.,-10.533643493173694)},line width=1.pt]  plot[domain=1.0141213629921257:2.1274712905976676,variable=\t]({1.*17.033643493173695*cos(\t r)+0.*17.033643493173695*sin(\t r)},{0.*17.033643493173695*cos(\t r)+1.*17.033643493173695*sin(\t r)});
\begin{scriptsize}
\draw [fill=black] (-2.,-3.) circle (3.75pt);
\draw [fill=black] (2.,-3.) circle (3.75pt);
\draw [fill=black] (4.,0.46410161513775483) circle (3.75pt);
\draw [fill=black] (2.,3.92820323027551) circle (3.75pt);
\draw [fill=black] (-2.,3.9282032302755114) circle (3.75pt);
\draw [fill=black] (-4.,0.46410161513775794) circle (3.75pt);
\draw [fill=black] (16.,3.9282032302755123) circle (3.75pt);
\draw [fill=black] (12.,3.9282032302755105) circle (3.75pt);
\draw [fill=black] (10.,0.4641016151377552) circle (3.75pt);
\draw [fill=black] (12.,-3.) circle (3.75pt);
\draw [fill=black] (16.,-3.) circle (3.75pt);
\draw [fill=black] (18.,0.46410161513775927) circle (3.75pt);
\draw [fill=black] (2.,-15.92820323027551) circle (3.75pt);
\draw [fill=black] (12.,-15.92820323027551) circle (3.75pt);
\draw [fill=black] (2.,-9.) circle (3.75pt);
\draw [fill=black] (12.,-9.) circle (3.75pt);
\draw [fill=black] (-2.,-15.92820323027551) circle (3.75pt);
\draw [fill=black] (4.,-12.464101615137755) circle (3.75pt);
\draw [fill=black] (-2.,-9.) circle (3.75pt);
\draw [fill=black] (-4.,-12.464101615137757) circle (3.75pt);
\draw [fill=black] (16.,-15.928203230275512) circle (3.75pt);
\draw [fill=black] (10.,-12.464101615137755) circle (3.75pt);
\draw [fill=black] (16.,-9.) circle (3.75pt);
\draw [fill=black] (18.,-12.464101615137759) circle (3.75pt);
\end{scriptsize}
\end{tikzpicture}
\caption{Feynman diagram consisting in a ladder with two rungs and complete vertices. Its external tensor contractions are equivalent to $J_2$ (the long-pillow).}
\label{fig:2rungs}
\end{figure}
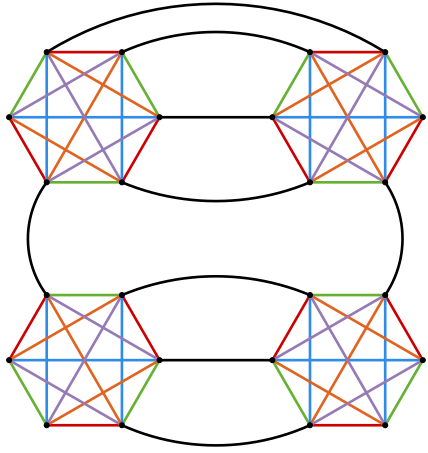

\paragraph{The large-$N$ expansion.} Like other tensor models, this model has also a $\frac{1}{N}$ expansion. First, we observe that every sextic interactions can be obtained as radiative corrections from the first interaction term $J_1$ (we call it the complete vertex, as its bubble is the complete graph on six vertices, also known as $K_6$). For example, the interaction $J_6$ (or the prism) is a rung with $3$ edges between two complete vertices (see Fig.~\ref{fig:6colored}), $J_2$ (or the long-pillow) and $J_4$ (or the pillow-dipole, our only double-trace interaction) are ladders made of two such rungs with different permutations of the colors between the rungs (see Fig.~\ref{fig:2rungs}). $J_5$ (or the triple-dipole, our only triple-trace interaction) is a ladder made of three rungs and $J_3$ a ladder made of four rungs. 

Then in any graph $\mathcal{G}$, we replace every interaction by their minimal representations in terms of complete vertices. This way, we obtain a new graph $\hat{\mathcal{G}}$ with only complete vertices. Since the rank of our model is a prime number, and the complete graph is the unique maximally single trace  (MST) invariant, we can use the result of Ref.~\cite{Ferrari:2017jgw} (see also \cite{Klebanov:2019jup}), where it has been proved that in this case, the leading order vacuum Feynman diagrams are the melons constructed with two mirrored complete MST interactions (see the diagram on the left in Fig.~\ref{fig:fund_vacuum_rank5}), and the usual iterative insertions of melonic two-point functions. Notice that unlike for the rank-3 wheel (which is MST, but  not a complete graph), the leading order diagrams include no tadpoles. This means that the leading order diagrams of our rank-5 model are melonic \textit{after} substituting every sextic interactions by their minimal representations in terms of the complete vertex. In terms of the original interactions, the leading order diagrams are again melon-tadpole diagrams (see Fig.~\ref{fig:melon_dtadpole}, with tadpoles now associated to  $J_b$ with $b \in [2,6]$), i.e.\ they are obtained by iterated insertions of melons and double tadpoles. The double tadpoles are based on the interactions $J_i$ vertices ($i \in [2,6]$) and the end vertices of melons are complete vertices. 
Therefore, the diagrammatics is somewhat similar to that of the quartic model \cite{Benedetti:2019eyl}, where the tetrahedron is a complete graph and it is associated to melonic diagrams, while the melonic graphs (pillow and double-trace) are associated to tadpoles.

Again, as explained for the previous model, we will ignore the effects of the tadpoles formed by $J_2$ to $J_6$, as tadpole corrections just renormalize the mass. We will also not include quartic interactions, assuming that they can be tuned to zero.

\begin{figure}[htbp]
\centering
\tikzsetnextfilename{i6non_melonic2}
\begin{tikzpicture}[line cap=round,line join=round,>=triangle 45,x=1.0cm,y=1.0cm,scale=0.25]
\draw [line width=1.pt,color=rouge] (-2.,3.9282032302755114)-- (2.,3.92820323027551);
\draw [line width=1.pt,color=vert] (2.,3.92820323027551)-- (4.,0.46410161513775483);
\draw [line width=1.pt,color=rouge] (4.,0.46410161513775483)-- (2.,-3.);
\draw [line width=1.pt,color=vert] (2.,-3.)-- (-2.,-3.);
\draw [line width=1.pt,color=rouge] (-2.,-3.)-- (-4.,0.46410161513775794);
\draw [line width=1.pt,color=vert] (-4.,0.46410161513775794)-- (-2.,3.9282032302755114);
\draw [line width=1.pt,color=orange] (-2.,3.9282032302755114)-- (-2.,-3.);
\draw [line width=1.pt,color=orange] (2.,3.92820323027551)-- (2.,-3.);
\draw [line width=1.pt,color=orange] (-4.,0.46410161513775794)-- (4.,0.46410161513775483);
\draw [line width=1.pt,color=lavenderpurple] (-2.,3.9282032302755114)-- (4.,0.46410161513775483);
\draw [line width=1.pt,color=lavenderpurple] (-4.,0.46410161513775794)-- (2.,-3.);
\draw [line width=1.pt,color=lavenderpurple] (-2.,-3.)-- (2.,3.92820323027551);
\draw [line width=1.pt,color=bleu] (2.,3.92820323027551)-- (-4.,0.46410161513775794);
\draw [line width=1.pt,color=bleu] (4.,0.46410161513775483)-- (-2.,-3.);
\draw [line width=1.pt,color=bleu] (-2.,3.9282032302755114)-- (2.,-3.);
\draw [line width=1.pt,color=rouge] (16.,3.9282032302755123)-- (12.,3.9282032302755105);
\draw [line width=1.pt,color=vert] (12.,3.9282032302755105)-- (10.,0.4641016151377552);
\draw [line width=1.pt,color=rouge] (10.,0.4641016151377552)-- (12.,-3.);
\draw [line width=1.pt,color=vert] (12.,-3.)-- (16.,-3.);
\draw [line width=1.pt,color=rouge] (16.,-3.)-- (18.,0.46410161513775927);
\draw [line width=1.pt,color=vert] (18.,0.46410161513775927)-- (16.,3.9282032302755123);
\draw [line width=1.pt,color=orange] (16.,3.9282032302755123)-- (16.,-3.);
\draw [line width=1.pt,color=orange] (12.,3.9282032302755105)-- (12.,-3.);
\draw [line width=1.pt,color=orange] (18.,0.46410161513775927)-- (10.,0.4641016151377552);
\draw [line width=1.pt,color=lavenderpurple] (16.,3.9282032302755123)-- (10.,0.4641016151377552);
\draw [line width=1.pt,color=lavenderpurple] (18.,0.46410161513775927)-- (12.,-3.);
\draw [line width=1.pt,color=lavenderpurple] (16.,-3.)-- (12.,3.9282032302755105);
\draw [line width=1.pt,color=bleu] (12.,3.9282032302755105)-- (18.,0.46410161513775927);
\draw [line width=1.pt,color=bleu] (10.,0.4641016151377552)-- (16.,-3.);
\draw [line width=1.pt,color=bleu] (16.,3.9282032302755123)-- (12.,-3.);
\draw [line width=1.pt] (4.,0.46410161513775483)-- (10.,0.4641016151377552);
\draw [shift={(7.,-7.198557158514999)},line width=1.pt]  plot[domain=1.1484688490770039:1.9931238045127895,variable=\t]({1.*12.198557158515*cos(\t r)+0.*12.198557158515*sin(\t r)},{0.*12.198557158515*cos(\t r)+1.*12.198557158515*sin(\t r)});
\draw [shift={(7.,-4.801442841485004)},line width=1.pt]  plot[domain=4.290061502666797:5.134716458102583,variable=\t]({1.*12.198557158514996*cos(\t r)+0.*12.198557158514996*sin(\t r)},{0.*12.198557158514996*cos(\t r)+1.*12.198557158514996*sin(\t r)});
\draw [line width=1.pt,color=rouge] (-2.,-15.92820323027551)-- (2.,-15.92820323027551);
\draw [line width=1.pt,color=vert] (2.,-15.92820323027551)-- (4.,-12.464101615137755);
\draw [line width=1.pt,color=rouge] (4.,-12.464101615137755)-- (2.,-9.);
\draw [line width=1.pt,color=vert] (2.,-9.)-- (-2.,-9.);
\draw [line width=1.pt,color=rouge] (-2.,-9.)-- (-4.,-12.464101615137757);
\draw [line width=1.pt,color=vert] (-4.,-12.464101615137757)-- (-2.,-15.92820323027551);
\draw [line width=1.pt,color=orange] (-2.,-15.92820323027551)-- (-2.,-9.);
\draw [line width=1.pt,color=orange] (2.,-15.92820323027551)-- (2.,-9.);
\draw [line width=1.pt,color=orange] (-4.,-12.464101615137757)-- (4.,-12.464101615137755);
\draw [line width=1.pt,color=bleu] (-2.,-15.92820323027551)-- (4.,-12.464101615137755);
\draw [line width=1.pt,color=bleu] (-4.,-12.464101615137757)-- (2.,-9.);
\draw [line width=1.pt,color=bleu] (-2.,-9.)-- (2.,-15.92820323027551);
\draw [line width=1.pt,color=lavenderpurple] (2.,-15.92820323027551)-- (-4.,-12.464101615137757);
\draw [line width=1.pt,color=lavenderpurple] (4.,-12.464101615137755)-- (-2.,-9.);
\draw [line width=1.pt,color=lavenderpurple] (-2.,-15.92820323027551)-- (2.,-9.);
\draw [line width=1.pt,color=rouge] (16.,-15.928203230275512)-- (12.,-15.92820323027551);
\draw [line width=1.pt,color=vert] (12.,-15.92820323027551)-- (10.,-12.464101615137755);
\draw [line width=1.pt,color=rouge] (10.,-12.464101615137755)-- (12.,-9.);
\draw [line width=1.pt,color=vert] (12.,-9.)-- (16.,-9.);
\draw [line width=1.pt,color=rouge] (16.,-9.)-- (18.,-12.464101615137759);
\draw [line width=1.pt,color=vert] (18.,-12.464101615137759)-- (16.,-15.928203230275512);
\draw [line width=1.pt,color=orange] (16.,-15.928203230275512)-- (16.,-9.);
\draw [line width=1.pt,color=orange] (12.,-15.92820323027551)-- (12.,-9.);
\draw [line width=1.pt,color=orange] (18.,-12.464101615137759)-- (10.,-12.464101615137755);
\draw [line width=1.pt,color=bleu] (16.,-15.928203230275512)-- (10.,-12.464101615137755);
\draw [line width=1.pt,color=bleu] (18.,-12.464101615137759)-- (12.,-9.);
\draw [line width=1.pt,color=bleu] (16.,-9.)-- (12.,-15.92820323027551);
\draw [line width=1.pt,color=lavenderpurple] (12.,-15.92820323027551)-- (18.,-12.464101615137759);
\draw [line width=1.pt,color=lavenderpurple] (10.,-12.464101615137755)-- (16.,-9.);
\draw [line width=1.pt,color=lavenderpurple] (16.,-15.928203230275512)-- (12.,-9.);
\draw [line width=1.pt] (4.,-12.464101615137755)-- (10.,-12.464101615137755);
\draw [shift={(2.,-6.)},line width=1.pt]  plot[domain=2.498091544796509:3.7850937623830774,variable=\t]({1.*5.*cos(\t r)+0.*5.*sin(\t r)},{0.*5.*cos(\t r)+1.*5.*sin(\t r)});
\draw [shift={(11.57135869565218,-6.)},line width=1.pt]  plot[domain=-0.5954025490678632:0.5954025490678634,variable=\t]({1.*5.349099344990292*cos(\t r)+0.*5.349099344990292*sin(\t r)},{0.*5.349099344990292*cos(\t r)+1.*5.349099344990292*sin(\t r)});
\draw [shift={(-2.,-6.)},line width=1.pt]  plot[domain=-0.6435011087932843:0.6435011087932843,variable=\t]({1.*5.*cos(\t r)+0.*5.*sin(\t r)},{0.*5.*cos(\t r)+1.*5.*sin(\t r)});
\draw [shift={(16.42864130434781,-6.)},line width=1.pt]  plot[domain=2.546190104521929:3.736995202657658,variable=\t]({1.*5.349099344990282*cos(\t r)+0.*5.349099344990282*sin(\t r)},{0.*5.349099344990282*cos(\t r)+1.*5.349099344990282*sin(\t r)});
\draw [shift={(7.,-1.4663565068263082)},line width=1.pt]  plot[domain=4.155714016581919:5.269063944187461,variable=\t]({1.*17.03364349317369*cos(\t r)+0.*17.03364349317369*sin(\t r)},{0.*17.03364349317369*cos(\t r)+1.*17.03364349317369*sin(\t r)});
\begin{scriptsize}
\draw [fill=black] (-2.,-3.) circle (3.75pt);
\draw [fill=black] (2.,-3.) circle (3.75pt);
\draw [fill=black] (4.,0.46410161513775483) circle (3.75pt);
\draw [fill=black] (2.,3.92820323027551) circle (3.75pt);
\draw [fill=black] (-2.,3.9282032302755114) circle (3.75pt);
\draw [fill=black] (-4.,0.46410161513775794) circle (3.75pt);
\draw [fill=black] (16.,3.9282032302755123) circle (3.75pt);
\draw [fill=black] (12.,3.9282032302755105) circle (3.75pt);
\draw [fill=black] (10.,0.4641016151377552) circle (3.75pt);
\draw [fill=black] (12.,-3.) circle (3.75pt);
\draw [fill=black] (16.,-3.) circle (3.75pt);
\draw [fill=black] (18.,0.46410161513775927) circle (3.75pt);
\draw [fill=black] (2.,-15.92820323027551) circle (3.75pt);
\draw [fill=black] (12.,-15.92820323027551) circle (3.75pt);
\draw [fill=black] (2.,-9.) circle (3.75pt);
\draw [fill=black] (12.,-9.) circle (3.75pt);
\draw [fill=black] (-2.,-15.92820323027551) circle (3.75pt);
\draw [fill=black] (4.,-12.464101615137755) circle (3.75pt);
\draw [fill=black] (-2.,-9.) circle (3.75pt);
\draw [fill=black] (-4.,-12.464101615137757) circle (3.75pt);
\draw [fill=black] (16.,-15.928203230275512) circle (3.75pt);
\draw [fill=black] (10.,-12.464101615137755) circle (3.75pt);
\draw [fill=black] (16.,-9.) circle (3.75pt);
\draw [fill=black] (18.,-12.464101615137759) circle (3.75pt);
\end{scriptsize}
\end{tikzpicture}
\caption{6-colored graph corresponding to a non-melonic Feynman diagram whose exterior tensor contractions are equivalent to $J_6$.}
\label{fig:non-melonicI6}
\end{figure}
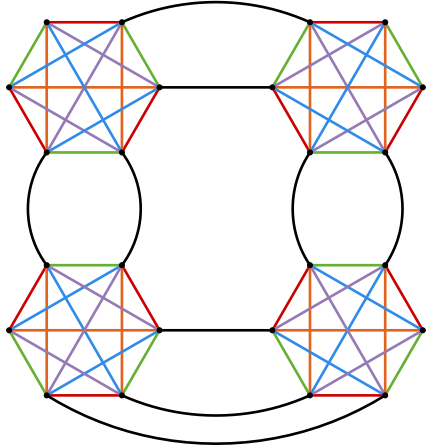

\paragraph{Radiative corrections to the prism interaction.} 
A comment is in order regarding the non-melonic interaction $J_6$. We presented in Figure~\ref{fig:6colored} a melonic contraction of two $J_1$ interactions that has $J_6$ as a boundary graph. It turns out that it is the only melonic diagram built with $J_1$ vertices that produces it. Indeed, we notice in $J_6$ the presence of two mirrored triangles (with edges red-green-blue in \eqref{eq:int-action-graph-rank5}) and each can result from 1, 2 or more complete graphs. The first case corresponds to Figure~\ref{fig:6colored}, but we see that the second case already requires non-melonic diagrams as in figure~\ref{fig:non-melonicI6}. In order to construct such a triangle from more that two $J_1$ vertices, we need at least two propagators (for the two colored edges that leave the nodes of the triangle) between each vertex, which in addition to at least two other propagators required to connect the mirror symmetric nodes of the two triangles, make the diagram non-melonic. 

\paragraph{Incomplete set of invariants.} As we said, the set of invariants we considered in the action is incomplete: there are more $O(N)^5$ invariants. However, it is closed. Indeed, we just showed that a $O(N)^5$ model with the complete interaction is dominated in the large-$N$ limit by melonic graphs. Therefore, it is enough to consider only the $O(N)^5$ invariants that can be generated from a melonic graph constructed with complete vertices. Those invariants are exactly $J_2$ to $J_6$ in the action of the model. The other $O(N)^5$ invariants will never be generated by a leading order six-point graph as they cannot be obtained from a melonic graph with complete vertices. Thus, at leading order in $N$, the set of invariants we consider is closed.

\paragraph{Rank-4 model.} 
Lastly, we notice that, as in rank 5, also in rank 4 there is a unique MST interaction \cite{Prakash:2019zia}. It turns out that the set of interactions it generates as radiative corrections are exactly of the same form as $J_2$ to $J_6$ in \eqref{eq:int-action-graph-rank5}, except that each multi-edge has one edge less than in rank 5 (for example, they can be obtained by removing the purple color in \eqref{eq:int-action-graph-rank5}).
Therefore, besides some different combinatorial factors, we do not expect important qualitative differences with respect to rank 5, and we chose to work with rank 5 as it contains a complete bubble, making the analogy to the quartic model  \cite{Benedetti:2019eyl} more evident.

\subsection{Renormalization: power counting}
\label{sec:powercount}

 We consider $\mathcal{G}$ a connected amputated Feynman diagram with $n(\mathcal{G})$ $6$-valent vertices, $E(\mathcal{G})$ edges and $r(\mathcal{G})$ external points. Computing the amplitude of the diagram $\mathcal{G}$ in momentum space, we get an independent integral $d^d p$ for every loop and a propagator $p^{-2\zeta}$ for every edge. Then, under a global rescaling of all the momenta by $t$, the amplitude is rescaled by:
\begin{equation}
t^{d(E(\mathcal{G})-n_6(\mathcal{G})-n_4(\mathcal{G})+1)-2\zeta E(\mathcal{G})}=t^{d(n_4(\mathcal{G})+2n_6(\mathcal{G})+1-\frac{r(\mathcal{G})}{2})-\zeta\left(4n_4(\mathcal{G})+6n_6(\mathcal{G})-r(\mathcal{G})\right)}=t^{d-\frac{r(\mathcal{G})}{2}(d-2\zeta)+n_6(\mathcal{G})(2d-6\zeta)+n_4(\mathcal{G})(d-4\zeta)}
\end{equation}
where we have used $2E(\mathcal{G})=6n(\mathcal{G})-r(\mathcal{G})$.

\paragraph{Short-range propagator.}

For $d=3$ and $\zeta=1$, the amplitude is rescaled as:
\begin{equation}
t^{3\left(1-\frac{r(\mathcal{G})}{6}\right)}\,.
\end{equation}

Thus, in $d=3$, the sextic interactions are marginal (the power counting does not depend on the number of internal vertices). The two-point and four-point diagrams are power divergent and the six-point diagrams are logarithmically divergent in the UV. Diagrams with more than eight external points are UV convergent. 

Therefore, in the following, we will use dimensional regularization, setting $d=3-\epsilon$. We will be interested in Wilson-Fisher type of fixed points, hence we will also consider $\epsilon$ finite, but small.

\paragraph{Long-range propagator.}

For $d<3$ and $\zeta=\frac{d}{3}$, the amplitude is rescaled as:

\begin{equation}
t^{d\left(1-\frac{r(\mathcal{G})}{6}\right)}\,.
\end{equation}

Again, the sextic interactions are marginal. The two-point and four-point diagrams are power divergent and the six-point diagrams are logarithmically divergent in the UV. Graphs with more than eight external point are UV convergent. 

We will again use dimensional regularization but in this case we will keep $d<3$ fixed and set $\zeta=\frac{d+\epsilon}{3}$.
In this case will be interested in fixed points that arise at $\epsilon=0$, as in Ref.~\cite{Benedetti:2019eyl}, by a different mechanism than in Wilson-Fisher.

\section{Two-point function}
\label{sec:SDeq}

\subsection{Rank $3$}

The standard Schwinger-Dyson equation (SDE) for the two-point function is, in momentum space:\footnote{We denote the momenta $p,q$ and so on. We define $\int_p \equiv \int \frac{d^d\; p}{(2\pi)^d}$.}
\be \label{eq:SDE}
G(p)^{-1}=C(p)^{-1}-\Sigma(p) \,,
\ee
where $G(p)$ is the Fourier transform of the full two-point function $N^{-3}\la\phib_{abc}(x)\phi_{abc}(0)\ra$, and $\Sigma(p)$ is the self-energy, i.e.\ the sum of non-trivial one-particle irreducible two-point diagrams. 

In a theory which is dominated by melon-tadpole diagrams, the self-energy at leading order in $1/N$ is obtained by summing up all the Feynman diagrams which can be obtained from those in Fig.~\ref{fig:SDE} by repeated insertions of either one of the two diagrams on internal lines. 
The resummation of all such diagrams can be represented by the same diagrams as in Fig.~\ref{fig:SDE}, but with the
edges decorated by the full two-point function. Therefore, it can be expressed in momentum space as:\footnote{See Footnote \ref{foot:trefoil} for the factor 3 in the double-tadpole contribution of the wheel.}
\be
\begin{split} \label{eq:Sigma-rank3}
\Sigma(p)&= 
\frac{\lambda_1^2}{4}\int_{q_1,q_2,q_3,q_4}G(q_1)G(q_2)G(q_3)G(q_4)G(p+q_1+q_2+q_3+q_4)\crcr
& \qquad -\frac{1}{2}(3\lambda_1+\lambda_2+\lambda_3+\lambda_4+\lambda_5)\left(\int_q G(q)\right)^2 \,.
\end{split}
\ee

\subsubsection{$\zeta=1$}

When $\zeta=1$, using a power counting argument, we see that the solution admits two regimes for  $d<3$. First, in the ultraviolet, there is a free scaling regime $G(p)^{-1}\sim p^2$: the free propagator dominates over the self energy. Second, in the infrared, there is an anomalous scaling regime $G(p)^{-1} \sim p^{2\Delta}$ with $\Delta=\frac{d}{3}$: the self energy dominates over the free propagator. Indeed, if we rescale the $q_i$ by $|p|$, the melon integral gives a global factor of $|p|^{4d-10\Delta}$ which must scale as $|p|^{2\Delta}$. This gives indeed $\Delta=\frac{d}{3}$.

We thus choose the following ansatz for the IR two-point function:\footnote{Notice that we choose a different convention for $\mathcal{Z}$ (and $Z$ below) than in Ref.~\cite{Benedetti:2019eyl}.}
\begin{equation}
G(p)=\frac{\mathcal{Z}}{p^{2d/3}}\,.
\label{eq:ansatz}
\end{equation}

Neglecting the free propagator in the IR, the SDE reduce to:
\begin{equation}
\frac{p^{2d/3}}{\mathcal{Z}}=-\frac{\lambda_1^2}{4}\mathcal{Z}^5 M_{d/3}(p)\,.
\end{equation}
The melon integral $M_{d/3}(p)$ is computed in App.~\ref{ap:melon}, giving
\begin{equation}
M_{d/3}(p)=-\frac{p^{2d/3}}{(4\pi)^{2d}}\frac{3}{d}\frac{\Gamma(1-\frac{d}{3})\Gamma(\frac{d}{6})^5}{\Gamma(\frac{d}{3})^5\Gamma(\frac{5d}{6})}~.
\end{equation}

We thus obtain:
\begin{equation}
\mathcal{Z}=\left(\frac{\lambda_1^2}{4(4\pi)^{2d}}\frac{3}{d}\frac{\Gamma(1-\frac{d}{3})\Gamma(\frac{d}{6})^5}{\Gamma(\frac{d}{3})^5\Gamma(\frac{5d}{6})}\right)^{-1/6} \,.
\end{equation}

\paragraph{Wave function renormalization.}

We introduce the wave function renormalization as $\phi=\phi_R\sqrt{Z}$ with $\phi$ the bare field and $\phi_R$ the renormalized field.
Notice that $Z$ is distinguished from $\mathcal{Z}$, as the latter is the full coefficient of the nonperturbative solution in the IR limit, while $Z$ is the usual perturbative wave function renormalization, to be fixed by a renormalization condition, as we will specify below.

After renormalization of the mass terms to zero, we have for the expansion of the renormalized two-point function at lowest order:
\begin{equation}
\Gamma^{(2)}_R(p)\equiv G_R(p)^{-1} =Zp^2-\frac{\lambda_1^2}{4} Z^{-5} M_1(p) \,.
\label{eq:gamma2}
\end{equation}
The integral $M_1(p)$ is computed in App.~\ref{ap:melon}, Eq.~\eqref{eq:M-1}. At leading order in $\epsilon$, we have:
\begin{equation}
M_1(p)=-p^{2-4\epsilon}\frac{2\pi^2}{3\epsilon(4\pi)^{6}} + \mathcal{O}(1)\,.
\end{equation}
At last, we fix $Z$ such that
\be
\lim_{\epsilon\to 0}\frac{d\Gamma^{(2)}_R(p)}{dp^2}|_{p^2=\mu^2}=1\,,
\ee
with $\mu$ the renormalization scale. 
At quadratic order in $\lambda_1$, we obtain:
\begin{equation}
Z=1+\frac{\lambda_1^2}{4}\tilde{M}_1(\mu) =1-\mu^{-4\epsilon}\frac{\lambda_1^2\pi^2}{6\epsilon(4\pi)^{6}}\,,
\label{eq:wavef3}
\end{equation}
with $\tilde{M}_1(\mu)=\frac{d}{dp^2}M_1(p)|_{p^2=\mu^2}$.

\subsubsection{$\zeta=\frac{d}{3}$}

The value of $\zeta$ in this case is chosen to match the infrared scaling of the two-point function. We now have only one regime and the full SDE is solved by the ansatz:
\begin{equation}
G(p)=\frac{\mathcal{Z}}{p^{2d/3}} \,.
\label{eq:ansatz2}
\end{equation}
For the vertex renormalization in Sec.~\ref{sec:betas} we will use analytic regularization, keeping $d<3$ fixed and setting $\zeta=\frac{d+\epsilon}{3}$, but since the two-point function is finite,  as we will now see, we can here set $\epsilon=0$.

The computations are the same as in the IR limit of the previous section, but we do not neglect the free propagator. Thus, we obtain:
\begin{equation}
\frac{1}{\mathcal{Z}^6}-\frac{1}{\mathcal{Z}^5}=\frac{\lambda_1^2}{4(4\pi)^{2d}}\frac{3\Gamma(1-\frac{d}{3})\Gamma(\frac{d}{6})^5}{d\Gamma(\frac{d}{3})^5\Gamma(\frac{5d}{6})} \,.
\label{Z-norm-d/3}
\end{equation}
At the first non-trivial order in the coupling constant, this gives:
\begin{equation}  \label{eq:wavef3-LR}
\mathcal{Z}=1-\frac{\lambda_1^2}{4(4\pi)^{2d}}\frac{3\Gamma(1-\frac{d}{3})\Gamma(\frac{d}{6})^5}{d\Gamma(\frac{d}{3})^5\Gamma(\frac{5d}{6})}+\mathcal{O}(\lambda_1^4).
\end{equation}
This expression is finite for $d<3$. Moreover, as we did not neglect the free propagator, $\mathcal{Z}$ has an expansion in $\lambda_1$, as the perturbative wave function renormalization, with which it can be identified in our non-minimal subtraction scheme. Therefore, in the case $\zeta=d/3$, the wave function renormalization is finite.

\subsection{Rank $5$}

For the $O(N)^5$ model, the Schwinger-Dyson equation in the large-$N$ limit is:
\be
G(p)^{-1}=C(p)^{-1}-\Sigma(p) \,,
\ee
with
\be
\begin{split} \label{eq:Sigma-rank5}
\Sigma(p)&= 
\frac{\kappa_1^2}{6}\int_{q_1,q_2,q_3,q_4}G(q_1)G(q_2)G(q_3)G(q_4)G(p+q_1+q_2+q_3+q_4)\\
& \qquad -(\kappa_2+\kappa_3+\kappa_4+\kappa_5+\kappa_6)\left(\int_q G(q)\right)^2  \,.
\end{split}
\ee
The only differences with the rank-3 model are the combinatorial factors in front of the melon and tadpole integrals. Thus, we can use the results of the previous section.

\subsubsection{$\zeta=1$}

In the IR limit, the SDE is solved again by $G(p)=\frac{\mathcal{Z}}{p^{2d/3}}$ with:
\begin{equation}
\mathcal{Z}=\left(\frac{\kappa_1^2}{6(4\pi)^{2d}}\frac{3}{d}\frac{\Gamma(1-\frac{d}{3})\Gamma(\frac{d}{6})^5}{\Gamma(\frac{d}{3})^5\Gamma(\frac{5d}{6})}\right)^{-1/6} \,.
\end{equation}
The wave function renormalization is given by:
\begin{equation} \label{eq:wavef5}
Z=1+\frac{\kappa_1^2}{6}\tilde{M}_1(\mu) =1-\mu^{-4\epsilon}\frac{\lambda_1^2\pi^2}{9\epsilon (4\pi)^{6}}\,.
\end{equation}

\subsubsection{$\zeta=\frac{d}{3}$}

The full SDE is solved again by $G(p)=\frac{\mathcal{Z}}{p^{2\zeta}}$ with:
\begin{equation} \label{eq:wavef5-LR}
\mathcal{Z}=1-\frac{\kappa_1^2}{2(4\pi)^{2d}}\frac{\Gamma(1-\frac{d}{3})\Gamma(\frac{d}{6})^5}{d\Gamma(\frac{d}{3})^5\Gamma(\frac{5d}{6})}+ \mathcal{O}(\kappa_1^3) \,.
\end{equation}
This is directly the wave function renormalization which is thus finite.

\section{2PI effective action and four-point kernels}
\label{sec:kernels}

In this section, we compute the four-point kernels of both models using the 2PI formalism (see \cite{Benedetti:2018goh}).
We will make use of them first for showing that indeed there is no need of counterterms with quartic interactions, and then, in the next section, for the discussion of the all-orders beta functions for the sextic couplings.

\subsection{Rank 3}

In rank 3 and at leading order in $1/N$, the 2PI effective action is given by: 
\begin{equation}
\label{eq:2PIrank3}
\begin{split}
-\G^{2PI}[\mbG] = &-\f{1}{6}\left(\f{3 \l_1}{N^3}\d^{(1)}_{\mba\mbd\mbb\mbc\mbe\mbf}+\sum_{i=2}^5 \f{\l_i}{N^{3+\rho(I_i)}}\d^{(i)}_{\mba\mbb;\mbc\mbd;\mbe\mbf}\right) \int\dd x ~\mbG_{(\mba,x)(\mbb,x)}\mbG_{(\mbc,x)(\mbd,x)}\mbG_{(\mbe,x)(\mbf,x)} \\
&+\f{1}{2}\left(\f{\l_1}{6N^3}\right)^2 3\, \d^{(1)}_{\mba\mbb\mbc\mbd\mbe\mbf} \d^{(1)}_{\mbg\mbh\mbj\mbk\mbm\mbn} \times\\
&\quad \int\dd x\dd y ~\mbG_{(\mba,x)(\mbg,y)}\mbG_{(\mbb,x)(\mbh,y)} \mbG_{(\mbc,x)(\mbj,y)}\mbG_{(\mbd,x)(\mbk,y)}\mbG_{(\mbe,x)(\mbm,y)}\mbG_{(\mbf,x)(\mbn,y)} \,.
\end{split}
\end{equation}
This is obtained by summing the contributions of the leading-order vacuum diagrams which are also two-particle irreducible (2PI), i.e.\ cannot be disconnected by cutting two lines, and with arbitrary propagator $\mbG$ on each line. As we already know, all the leading-order vacuum diagrams are obtained from the diagrams in Fig.~\ref{fig:fund_vacuum} by repeated insertions of the two-point diagrams in Fig.~\ref{fig:SDE}, but since all such insertions lead to two-particle reducible diagrams, we are left with just the two fundamental diagrams of Fig.~\ref{fig:fund_vacuum}, whose evaluation leads to Eq.~\eqref{eq:2PIrank3}.

One recovers the self-energy from (using the further condensed notation $A=(\mba,x)$):
\begin{equation} \label{eq:Sigma2PI-complex}
\Sigma[\mbG]_{AB} = -\f{\d\G^{2PI}[\mbG]}{\d\mbG_{AB}} \,,
\end{equation}
which can be seen to reproduce Eq.~\eqref{eq:Sigma-rank3} in momentum space.

\begin{figure}[htbp]
\captionsetup[subfigure]{labelformat=empty}
\vspace{-.5cm}
\begin{minipage}[c]{0.4\textwidth}
\subfloat[]{\tikzsetnextfilename{kernel_wheel2}
\begin{tikzpicture}[line cap=round,line join=round,>=triangle 45,x=1.0cm,y=1.0cm, scale=0.25]
\clip(-12.508068585376279,-9.815087403149695) rectangle (23.50912630459411,12.492092186958157);
\draw [line width=1.pt,color=rouge] (-3.444101615137754,1.9841016151377573)-- (0.02,3.984101615137755);
\draw [line width=1.pt,color=vert] (0.02,3.984101615137755)-- (3.4841016151377553,1.9841016151377542);
\draw [line width=1.pt,color=rouge] (3.4841016151377553,1.9841016151377542)-- (3.4841016151377553,-2.015898384862246);
\draw [line width=1.pt,color=vert] (3.4841016151377553,-2.015898384862246)-- (0.02,-4.015898384862245);
\draw [line width=1.pt,color=rouge] (0.02,-4.015898384862245)-- (-3.4441016151377557,-2.0158983848622425);
\draw [line width=1.pt,color=vert] (-3.4441016151377557,-2.0158983848622425)-- (-3.444101615137754,1.9841016151377573);
\draw [line width=1.pt,color=bleu] (-3.4441016151377557,-2.0158983848622425)-- (3.4841016151377553,1.9841016151377542);
\draw [line width=1.pt,color=bleu] (0.02,-4.015898384862245)-- (0.02,3.984101615137755);
\draw [line width=1.pt,color=bleu] (-3.444101615137754,1.9841016151377573)-- (3.4841016151377553,-2.015898384862246);
\draw [line width=1.pt,color=rouge] (19.444101615137754,1.984101615137756)-- (15.98,3.984101615137754);
\draw [line width=1.pt,color=vert] (15.98,3.984101615137754)-- (12.515898384862245,1.9841016151377537);
\draw [line width=1.pt,color=rouge] (12.515898384862245,1.9841016151377537)-- (12.515898384862245,-2.0158983848622465);
\draw [line width=1.pt,color=vert] (12.515898384862245,-2.0158983848622465)-- (15.98,-4.015898384862246);
\draw [line width=1.pt,color=rouge] (15.98,-4.015898384862246)-- (19.444101615137754,-2.015898384862244);
\draw [line width=1.pt,color=vert] (19.444101615137754,-2.015898384862244)-- (19.444101615137754,1.984101615137756);
\draw [line width=1.pt,color=bleu] (19.444101615137754,-2.015898384862244)-- (12.515898384862245,1.9841016151377537);
\draw [line width=1.pt,color=bleu] (15.98,-4.015898384862246)-- (15.98,3.984101615137754);
\draw [line width=1.pt,color=bleu] (19.444101615137754,1.984101615137756)-- (12.515898384862245,-2.0158983848622465);
\draw [shift={(8.,-5.1778543393248)},line width=1.pt]  plot[domain=1.008223238915605:2.1333694146741884,variable=\t]({1.*8.46681470897191*cos(\t r)+0.*8.46681470897191*sin(\t r)},{0.*8.46681470897191*cos(\t r)+1.*8.46681470897191*sin(\t r)});
\draw [shift={(8.,-10.802495306012025)},line width=1.pt]  plot[domain=1.0759124552393733:2.06568019835042,variable=\t]({1.*16.802495306012023*cos(\t r)+0.*16.802495306012023*sin(\t r)},{0.*16.802495306012023*cos(\t r)+1.*16.802495306012023*sin(\t r)});
\draw [shift={(8.,5.1778543393248)},line width=1.pt]  plot[domain=4.149815892505398:5.2749620682639815,variable=\t]({1.*8.46681470897191*cos(\t r)+0.*8.46681470897191*sin(\t r)},{0.*8.46681470897191*cos(\t r)+1.*8.46681470897191*sin(\t r)});
\draw [shift={(8.,10.802495306012025)},line width=1.pt]  plot[domain=4.217505108829166:5.207272851940213,variable=\t]({1.*16.802495306012023*cos(\t r)+0.*16.802495306012023*sin(\t r)},{0.*16.802495306012023*cos(\t r)+1.*16.802495306012023*sin(\t r)});
\draw [line width=1.pt] (-3.444101615137754,1.9841016151377573)-- (-3.418555679390061,6.2051790936510205);
\draw [line width=1.pt] (-3.444101615137754,-1.9841016151377573)-- (-3.418555679390061,-6.2051790936510205);
\draw [line width=1.pt] (19.444101615137754,1.984101615137756)-- (19.41855567939006,6.205179093651019);
\draw [line width=1.pt] (19.444101615137754,-1.9841016151377586)-- (19.41855567939006,-6.205179093651022);
\draw [line width=1.pt] (42.330628271232506,2.261130007650736)-- (42.25399420649868,6.481589093409948);
\begin{scriptsize}
\draw [color=black, fill=white] (-3.444101615137754,1.9841016151377573) circle (5pt);
\draw [color=black, fill=white] (0.02,3.984101615137755) circle (5pt);
\draw [color=black] (3.4841016151377553,1.9841016151377542) circle (5pt);
\draw [fill=black] (3.4841016151377553,-2.015898384862246) circle (5pt);
\draw [fill=black] (-3.4441016151377557,-2.0158983848622425) circle (5pt);
\draw [fill=black] (-3.4441016151377557,-2.0158983848622425) circle (5pt);
\draw [color=black, fill=white] (-3.444101615137754,1.9841016151377573) circle (5pt);
\draw [fill=black] (-3.4441016151377557,-2.0158983848622425) circle (5pt);
\draw [color=black, fill=white] (3.4841016151377553,1.9841016151377542) circle (5pt);
\draw [fill=black] (0.02,3.984101615137755) circle (5pt);
\draw [fill=black] (19.444101615137754,1.984101615137756) circle (5pt);
\draw [fill=black] (12.515898384862245,1.9841016151377537) circle (5pt);
\draw [color=black] (12.515898384862245,1.9841016151377537) circle (5pt);
\draw [color=black, fill=white] (15.98,-4.015898384862246) circle (5pt);
\draw [color=black, fill=white] (19.444101615137754,-2.015898384862244) circle (5pt);
\draw [color=black] (19.444101615137754,1.984101615137756) circle (5pt);
\draw [color=black] (12.515898384862245,1.9841016151377537) circle (5pt);
\draw [color=black, fill=white] (15.98,-4.015898384862246) circle (5pt);
\draw [color=black, fill=white] (15.98,3.984101615137754) circle (5pt);
\draw [color=black] (3.4841016151377553,-1.9841016151377542) circle (5pt);
\draw [color=black, fill=white] (12.515898384862245,-1.9841016151377537) circle (5pt);
\draw [color=black, fill=white] (0.02,-3.984101615137755) circle (5pt);
\draw [fill=black] (15.98,-3.984101615137754) circle (5pt);
\draw [fill=black] (-3.418555679390061,6.2051790936510205) circle (5pt);
\draw[color=black] (-3.1534448862987965,6.9058290468207915) node {$a$};
\draw [color=black] (-3.444101615137754,-1.9841016151377573) circle (5pt);
\draw [fill=black, fill=white] (-3.418555679390061,-6.2051790936510205) circle (5pt);
\draw[color=black] (-3.1534448862987965,-6.9058290468207915) node {$d$};
\draw [fill=black, fill=white] (19.41855567939006,6.205179093651019) circle (5pt);
\draw[color=black] (19.683956289991578,6.9058290468207915) node {$b$};
\draw [color=black] (19.444101615137754,-1.9841016151377586) circle (5pt);
\draw [fill=black] (19.41855567939006,-6.205179093651022) circle (5pt);
\draw[color=black] (19.683956289991578,-6.9058290468207915) node {$c$};
\draw [color=black] (42.330628271232506,2.261130007650736) circle (5pt);
\draw [fill=black] (42.25399420649868,6.481589093409948) circle (5pt);
\draw[color=black] (-12.186148336622601,12.908694861815858) node {$Z'_2$};
\draw[color=black] (43.18413444901011,4.803879187311478) node {$b'_2$};
\end{scriptsize}
\end{tikzpicture}}
\end{minipage}
\hspace{.5cm}
\begin{minipage}[c]{0.4\textwidth}
\subfloat[]{\tikzsetnextfilename{kernel_wheel_dt2}
\begin{tikzpicture}[line cap=round,line join=round,>=triangle 45,x=1.0cm,y=1.0cm,scale=0.25]
\clip(-9.97057956578846,-10.496800871098664) rectangle (26.04661532418193,11.810378719009192);
\draw [line width=1.pt,color=rouge] (-3.444101615137754,1.9841016151377573)-- (0.02,3.984101615137755);
\draw [line width=1.pt,color=vert] (0.02,3.984101615137755)-- (3.4841016151377553,1.9841016151377542);
\draw [line width=1.pt,color=rouge] (3.4841016151377553,1.9841016151377542)-- (3.4841016151377553,-2.015898384862246);
\draw [line width=1.pt,color=vert] (3.4841016151377553,-2.015898384862246)-- (0.02,-4.015898384862245);
\draw [line width=1.pt,color=rouge] (0.02,-4.015898384862245)-- (-3.4441016151377557,-2.0158983848622425);
\draw [line width=1.pt,color=vert] (-3.4441016151377557,-2.0158983848622425)-- (-3.444101615137754,1.9841016151377573);
\draw [line width=1.pt,color=bleu] (-3.4441016151377557,-2.0158983848622425)-- (3.4841016151377553,1.9841016151377542);
\draw [line width=1.pt,color=bleu] (0.02,-4.015898384862245)-- (0.02,3.984101615137755);
\draw [line width=1.pt,color=bleu] (-3.444101615137754,1.9841016151377573)-- (3.4841016151377553,-2.015898384862246);
\draw [line width=1.pt,color=rouge] (19.444101615137754,1.984101615137756)-- (15.98,3.984101615137754);
\draw [line width=1.pt,color=vert] (15.98,3.984101615137754)-- (12.515898384862245,1.9841016151377537);
\draw [line width=1.pt,color=rouge] (12.515898384862245,1.9841016151377537)-- (12.515898384862245,-2.0158983848622465);
\draw [line width=1.pt,color=vert] (12.515898384862245,-2.0158983848622465)-- (15.98,-4.015898384862246);
\draw [line width=1.pt,color=rouge] (15.98,-4.015898384862246)-- (19.444101615137754,-2.015898384862244);
\draw [line width=1.pt,color=vert] (19.444101615137754,-2.015898384862244)-- (19.444101615137754,1.984101615137756);
\draw [line width=1.pt,color=bleu] (19.444101615137754,-2.015898384862244)-- (12.515898384862245,1.9841016151377537);
\draw [line width=1.pt,color=bleu] (15.98,-4.015898384862246)-- (15.98,3.984101615137754);
\draw [line width=1.pt,color=bleu] (19.444101615137754,1.984101615137756)-- (12.515898384862245,-2.0158983848622465);
\draw [shift={(8.,-5.1778543393248)},line width=1.pt]  plot[domain=1.008223238915605:2.1333694146741884,variable=\t]({1.*8.46681470897191*cos(\t r)+0.*8.46681470897191*sin(\t r)},{0.*8.46681470897191*cos(\t r)+1.*8.46681470897191*sin(\t r)});
\draw [shift={(8.,-10.802495306012025)},line width=1.pt]  plot[domain=1.0759124552393733:2.06568019835042,variable=\t]({1.*16.802495306012023*cos(\t r)+0.*16.802495306012023*sin(\t r)},{0.*16.802495306012023*cos(\t r)+1.*16.802495306012023*sin(\t r)});
\draw [shift={(8.,5.1778543393248)},line width=1.pt]  plot[domain=4.149815892505398:5.2749620682639815,variable=\t]({1.*8.46681470897191*cos(\t r)+0.*8.46681470897191*sin(\t r)},{0.*8.46681470897191*cos(\t r)+1.*8.46681470897191*sin(\t r)});
\draw [line width=1.pt] (-3.444101615137754,1.9841016151377573)-- (-3.418555679390061,6.2051790936510205);
\draw [line width=1.pt] (19.444101615137754,1.984101615137756)-- (19.41855567939006,6.205179093651019);
\draw [line width=1.pt] (42.330628271232506,2.261130007650736)-- (42.25399420649868,6.481589093409948);
\draw [shift={(8.,4.480539494266125)},line width=1.pt]  plot[domain=3.6578984641268115:5.766879496642567,variable=\t]({1.*13.159451618322569*cos(\t r)+0.*13.159451618322569*sin(\t r)},{0.*13.159451618322569*cos(\t r)+1.*13.159451618322569*sin(\t r)});
\draw [line width=1.pt] (0.02,-3.984101615137755)-- (4.00454652716535,-5.383949861481413);
\draw [line width=1.pt] (15.98,-3.9841016151377557)-- (11.99545347283465,-5.383949861481414);
\begin{scriptsize}
\draw [color=black] (-3.444101615137754,1.9841016151377573) circle (5pt);
\draw [color=black] (0.02,3.984101615137755) circle (5pt);
\draw [fill=black] (0.02,3.984101615137755) circle (5pt);
\draw [color=black, fill=white] (3.4841016151377553,1.9841016151377542) circle (5pt);
\draw [fill=black] (3.4841016151377553,-2.015898384862246) circle (5pt);
\draw [fill=black] (-3.4441016151377557,-2.0158983848622425) circle (5pt);
\draw [color=black] (-3.444101615137754,1.9841016151377573) circle (5pt);
\draw [fill=black] (-3.4441016151377557,-2.0158983848622425) circle (5pt);
\draw [color=black, fill=white] (-3.444101615137754,1.9841016151377573) circle (5pt);
\draw [fill=black] (19.444101615137754,1.984101615137756) circle (5pt);
\draw [fill=black] (12.515898384862245,1.9841016151377537) circle (5pt);
\draw [color=black] (12.515898384862245,1.9841016151377537) circle (5pt);
\draw [color=black] (15.98,-4.015898384862246) circle (5pt);
\draw [color=black, fill=white] (19.444101615137754,-2.015898384862244) circle (5pt);
\draw [color=black] (19.444101615137754,1.984101615137756) circle (5pt);
\draw [color=black] (12.515898384862245,1.9841016151377537) circle (5pt);
\draw [color=black] (15.98,-4.015898384862246) circle (5pt);
\draw [color=black, fill=white] (15.98,3.984101615137754) circle (5pt);
\draw [color=black] (19.444101615137754,1.984101615137756) circle (5pt);
\draw [color=black] (3.4841016151377553,-1.9841016151377542) circle (5pt);
\draw [color=black, fill=white] (12.515898384862245,-1.9841016151377537) circle (5pt);
\draw [color=black, fill=white] (0.02,-3.984101615137755) circle (5pt);
\draw [fill=black] (15.98,-3.984101615137754) circle (5pt);
\draw [fill=black] (-3.418555679390061,6.2051790936510205) circle (5pt);
\draw[color=black] (-3.1534448862987965,6.905829046820792) node {$a$};
\draw [color=black] (-3.444101615137754,-1.9841016151377573) circle (5pt);
\draw [color=black] (19.444101615137754,1.984101615137756) circle (5pt);
\draw [fill=black, fill=white] (19.41855567939006,6.205179093651019) circle (5pt);
\draw[color=black] (19.683956289991578,6.905829046820792) node {$b$};
\draw [color=black] (19.444101615137754,-1.9841016151377586) circle (5pt);
\draw [color=black] (42.330628271232506,2.261130007650736) circle (5pt);
\draw [fill=black] (42.25399420649868,6.481589093409948) circle (5pt);
\draw[color=black] (-9.648659317034781,12.226981393866893) node {$Z'_2$};
\draw[color=black] (43.18413444901011,4.803879187311479) node {$b'_2$};
\draw [fill=black] (4.00454652716535,-5.383949861481413) circle (5pt);
\draw[color=black] (4.269657320256615,-4.683299908311642) node {$c$};
\draw [color=black] (15.98,-3.9841016151377557) circle (5pt);
\draw [fill=black, fill=white] (11.99545347283465,-5.383949861481414) circle (5pt);
\draw[color=black] (12.260854083436165,-4.683299908311642) node {$d$};
\end{scriptsize}
\end{tikzpicture}}
\end{minipage}
\vspace{-.5cm}
\caption{The contractions corresponding to a 4-point rung in the ladder making the 4-point function (a pillow on the left and a double trace on the right).}
\label{fig:kernel_wheel}
\end{figure}
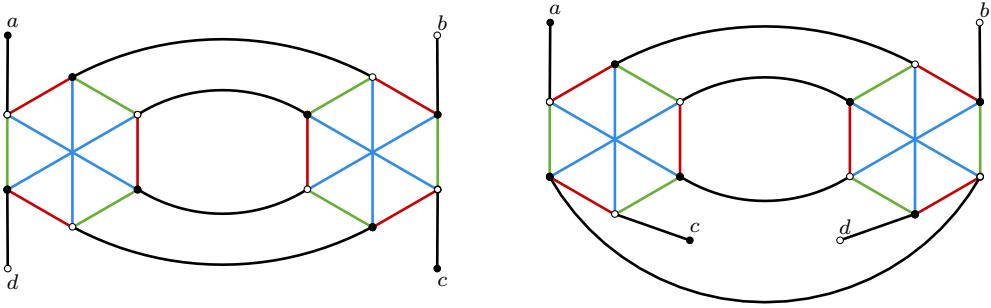

The right-amputated four-point kernel on-shell is obtained by taking two derivatives of $\G^{2PI}[\mbG]$ with respect to $\mbG$  and then multiplying by two full propagators on the left:
\be \label{eq:defK}
K[\mbG]_{AB,CD} = \mbG_{AA'}\mbG_{BB'}\f{\d\Sigma[\mbG]_{CD}}{\d\mbG_{A'B'}}\,.
\ee
Applying such definition to Eq.~\eqref{eq:2PIrank3} we obtain:
\be
\begin{split}
K_{(\mba,x)(\mbb,y)(\mbc,z)(\mbd,w)}=&\int dx'dy' \, G_{xx'}G_{yy'}\left[ 
 -\f{1}{3}(9\l_1+ 2\l_2 + 3\l_3 + \l_4)\delta_{x'y'}\delta_{x'z}\delta_{x'w}G_{x'x'}\hat{\delta}^p_{\mba \mbb;\mbc\mbd}\right.\crcr
&\quad -\f{1}{3}(\l_2 + 2 \l_4+3\l_5)\delta_{x'y'}\delta_{x'z}\delta_{x'w}G_{x'x'}\hat{\delta}^d_{\mba \mbb;\mbc\mbd} \crcr
&\quad \left.+\frac{\l_1^2}{4}G_{x'y'}^4(3\hat{\delta}^p_{\mba \mbb;\mbc\mbd} \delta_{x'w}\delta_{y'z}+ 2\hat{\delta}^d_{\mba \mbb;\mbc\mbd}\delta_{x'z}\delta_{y'w})\right] \,,
\end{split}
\ee
where we defined the rescaled pillow and double-trace contraction operators $\hat{\delta}^p_{\mba \mbb;\mbc\mbd}=\frac{1}{N^2}\delta^p_{\mba \mbb;\mbc\mbd}$ and $\hat{\delta}^d_{\mba \mbb;\mbc\mbd}=\frac{1}{N^3}\delta^d_{\mba \mbb;\mbc\mbd}$ (see App.~\ref{ap:conventions}).
The colored graphs corresponding to the last line are depicted in Fig.\ref{fig:kernel_wheel}.

In momentum space this four-point kernel becomes:
\be
\begin{split}
K_{(\mba,p_1)(\mbb,p_2)(\mbc,p_3)(\mbd,p_4)}=&(2\pi)^d\delta(p_1+p_2+p_3+p_4)G(p_1)G(p_2)\left[ -\f{1}{3}(9\l_1+ 2\l_2 + 3\l_3 + \l_4)\int_q G(q)\hat{\delta}^p_{\mba \mbb;\mbc\mbd} \right.\crcr
&\quad -\f{1}{3}(\l_2 + 2 \l_4+3\l_5)\int_q G(q)\hat{\delta}^d_{\mba \mbb;\mbc\mbd} \crcr
&\quad +\frac{\l_1^2}{4}\left(3\hat{\delta}^p_{\mba \mbb;\mbc\mbd} \int_{q_1,q_2,q_3}G(q_1)G(q_2)G(q_3)G(-p_1-p_4-q_1-q_2-q_3)\right.\\
&\qquad \left.\left.  + 2\hat{\delta}^d_{\mba \mbb;\mbc\mbd}\int_{q_1,q_2,q_3}G(q_1)G(q_2)G(q_3)G(-p_1-p_3-q_1-q_2-q_3)\right) \right] \,.
\end{split}
\ee
For convenience, we introduce also a reduced kernel, with the tadpoles set to zero, i.e.:
\begin{equation}
\hat{K}_{(\mba,x)(\mbb,y)(\mbc,z)(\mbd,w)}=\frac{\l_1^2}{4}G_{zw}^4(3\hat{\delta}^p_{\mba \mbb;\mbc\mbd} G_{xw}G_{yz} + 2\hat{\delta}^d_{\mba \mbb;\mbc\mbd} G_{xz}G_{yw}) \,.
\label{kernel-rank3}
\end{equation}
In fact, since the propagator is massless, the tadpoles are zero in dimensional regularization, hence the reduced kernel is all we need.

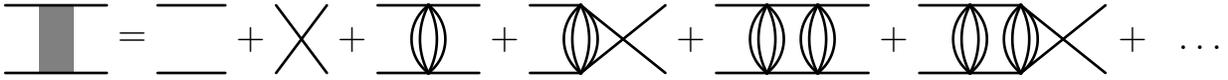
\begin{figure}[htbp]
\centering
\vspace{.4cm}
\tikzsetnextfilename{ladders2}
\begin{tikzpicture}[line cap=round,line join=round,>=triangle 45,x=1.0cm,y=1.0cm,scale=0.225]
\fill[line width=2.pt,color=black,fill=gray] (-4.,2.) -- (-2.,2.) -- (-2.,-2.) -- (-4.,-2.) -- cycle;
\draw [line width=1.pt] (-6.,2.)-- (0.,2.);
\draw [line width=1.pt] (-6.,-2.)-- (0.,-2.);
\draw [line width=1.pt] (3.,2.)-- (7.,2.);
\draw [line width=1.pt] (3.,-2.)-- (7.,-2.);
\draw [line width=1.pt] (10.,2.)-- (13.,-2.);
\draw [line width=1.pt] (10.,-2.)-- (13.,2.);
\draw [line width=1.pt] (16.,2.)-- (22.,2.);
\draw [line width=1.pt] (16.,-2.)-- (22.,-2.);
\draw [shift={(22.75,0.)},line width=1.pt]  plot[domain=2.651635327336065:3.631549979843521,variable=\t]({1.*4.25*cos(\t r)+0.*4.25*sin(\t r)},{0.*4.25*cos(\t r)+1.*4.25*sin(\t r)});
\draw [shift={(15.25,0.)},line width=1.pt]  plot[domain=-0.48995732625372845:0.4899573262537283,variable=\t]({1.*4.25*cos(\t r)+0.*4.25*sin(\t r)},{0.*4.25*cos(\t r)+1.*4.25*sin(\t r)});
\draw [shift={(20.5,0.)},line width=1.pt]  plot[domain=2.214297435588181:4.068887871591405,variable=\t]({1.*2.5*cos(\t r)+0.*2.5*sin(\t r)},{0.*2.5*cos(\t r)+1.*2.5*sin(\t r)});
\draw [shift={(17.5,0.)},line width=1.pt]  plot[domain=-0.9272952180016123:0.9272952180016122,variable=\t]({1.*2.5*cos(\t r)+0.*2.5*sin(\t r)},{0.*2.5*cos(\t r)+1.*2.5*sin(\t r)});
\draw [shift={(29.5,0.)},line width=1.pt]  plot[domain=2.214297435588181:4.068887871591405,variable=\t]({1.*2.5*cos(\t r)+0.*2.5*sin(\t r)},{0.*2.5*cos(\t r)+1.*2.5*sin(\t r)});
\draw [shift={(31.75,0.)},line width=1.pt]  plot[domain=2.651635327336065:3.631549979843521,variable=\t]({1.*4.25*cos(\t r)+0.*4.25*sin(\t r)},{0.*4.25*cos(\t r)+1.*4.25*sin(\t r)});
\draw [shift={(24.25,0.)},line width=1.pt]  plot[domain=-0.48995732625372845:0.4899573262537283,variable=\t]({1.*4.25*cos(\t r)+0.*4.25*sin(\t r)},{0.*4.25*cos(\t r)+1.*4.25*sin(\t r)});
\draw [shift={(26.5,0.)},line width=1.pt]  plot[domain=-0.9272952180016123:0.9272952180016122,variable=\t]({1.*2.5*cos(\t r)+0.*2.5*sin(\t r)},{0.*2.5*cos(\t r)+1.*2.5*sin(\t r)});
\draw [line width=1.pt] (25.,2.)-- (28.,2.);
\draw [line width=1.pt] (25.,-2.)-- (28.,-2.);
\draw [line width=1.pt] (28.,2.)-- (33.,-2.);
\draw [line width=1.pt] (28.,-2.)-- (33.,2.);
\draw [line width=1.pt] (36.,2.)-- (45.,2.);
\draw [line width=1.pt] (36.,-2.)-- (45.,-2.);
\draw [shift={(42.75,0.)},line width=1.pt]  plot[domain=2.651635327336065:3.631549979843521,variable=\t]({1.*4.25*cos(\t r)+0.*4.25*sin(\t r)},{0.*4.25*cos(\t r)+1.*4.25*sin(\t r)});
\draw [shift={(35.25,0.)},line width=1.pt]  plot[domain=-0.48995732625372845:0.4899573262537283,variable=\t]({1.*4.25*cos(\t r)+0.*4.25*sin(\t r)},{0.*4.25*cos(\t r)+1.*4.25*sin(\t r)});
\draw [shift={(40.5,0.)},line width=1.pt]  plot[domain=2.214297435588181:4.068887871591405,variable=\t]({1.*2.5*cos(\t r)+0.*2.5*sin(\t r)},{0.*2.5*cos(\t r)+1.*2.5*sin(\t r)});
\draw [shift={(37.5,0.)},line width=1.pt]  plot[domain=-0.9272952180016123:0.9272952180016122,variable=\t]({1.*2.5*cos(\t r)+0.*2.5*sin(\t r)},{0.*2.5*cos(\t r)+1.*2.5*sin(\t r)});
\draw [shift={(43.5,0.)},line width=1.pt]  plot[domain=2.214297435588181:4.068887871591405,variable=\t]({1.*2.5*cos(\t r)+0.*2.5*sin(\t r)},{0.*2.5*cos(\t r)+1.*2.5*sin(\t r)});
\draw [shift={(45.75,0.)},line width=1.pt]  plot[domain=2.651635327336065:3.631549979843521,variable=\t]({1.*4.25*cos(\t r)+0.*4.25*sin(\t r)},{0.*4.25*cos(\t r)+1.*4.25*sin(\t r)});
\draw [shift={(38.25,0.)},line width=1.pt]  plot[domain=-0.48995732625372845:0.4899573262537283,variable=\t]({1.*4.25*cos(\t r)+0.*4.25*sin(\t r)},{0.*4.25*cos(\t r)+1.*4.25*sin(\t r)});
\draw [shift={(40.5,0.)},line width=1.pt]  plot[domain=-0.9272952180016123:0.9272952180016122,variable=\t]({1.*2.5*cos(\t r)+0.*2.5*sin(\t r)},{0.*2.5*cos(\t r)+1.*2.5*sin(\t r)});
\draw [shift={(52.5,0.)},line width=1.pt]  plot[domain=2.214297435588181:4.068887871591405,variable=\t]({1.*2.5*cos(\t r)+0.*2.5*sin(\t r)},{0.*2.5*cos(\t r)+1.*2.5*sin(\t r)});
\draw [shift={(54.75,0.)},line width=1.pt]  plot[domain=2.651635327336065:3.631549979843521,variable=\t]({1.*4.25*cos(\t r)+0.*4.25*sin(\t r)},{0.*4.25*cos(\t r)+1.*4.25*sin(\t r)});
\draw [shift={(47.25,0.)},line width=1.pt]  plot[domain=-0.48995732625372845:0.4899573262537283,variable=\t]({1.*4.25*cos(\t r)+0.*4.25*sin(\t r)},{0.*4.25*cos(\t r)+1.*4.25*sin(\t r)});
\draw [shift={(49.5,0.)},line width=1.pt]  plot[domain=-0.9272952180016123:0.9272952180016122,variable=\t]({1.*2.5*cos(\t r)+0.*2.5*sin(\t r)},{0.*2.5*cos(\t r)+1.*2.5*sin(\t r)});
\draw [shift={(55.5,0.)},line width=1.pt]  plot[domain=2.214297435588181:4.068887871591405,variable=\t]({1.*2.5*cos(\t r)+0.*2.5*sin(\t r)},{0.*2.5*cos(\t r)+1.*2.5*sin(\t r)});
\draw [shift={(57.75,0.)},line width=1.pt]  plot[domain=2.651635327336065:3.631549979843521,variable=\t]({1.*4.25*cos(\t r)+0.*4.25*sin(\t r)},{0.*4.25*cos(\t r)+1.*4.25*sin(\t r)});
\draw [shift={(50.25,0.)},line width=1.pt]  plot[domain=-0.48995732625372845:0.4899573262537283,variable=\t]({1.*4.25*cos(\t r)+0.*4.25*sin(\t r)},{0.*4.25*cos(\t r)+1.*4.25*sin(\t r)});
\draw [shift={(52.5,0.)},line width=1.pt]  plot[domain=-0.9272952180016123:0.9272952180016122,variable=\t]({1.*2.5*cos(\t r)+0.*2.5*sin(\t r)},{0.*2.5*cos(\t r)+1.*2.5*sin(\t r)});
\draw [line width=1.pt] (48.,2.)-- (54.,2.);
\draw [line width=1.pt] (48.,-2.)-- (54.,-2.);
\draw [line width=1.pt] (54.,2.)-- (59.,-2.);
\draw [line width=1.pt] (54.,-2.)-- (59.,2.);
\begin{scriptsize}
\draw [fill=black] (19.,2.) circle (2.5pt);
\draw [fill=black] (19.,-2.) circle (2.5pt);
\draw [fill=black] (28.,2.) circle (2.5pt);
\draw [fill=black] (28.,-2.) circle (2.5pt);
\draw [fill=black] (39.,2.) circle (2.5pt);
\draw [fill=black] (39.,-2.) circle (2.5pt);
\draw [fill=black] (42.,2.) circle (2.5pt);
\draw [fill=black] (42.,-2.) circle (2.5pt);
\draw [fill=black] (51.,2.) circle (2.5pt);
\draw [fill=black] (51.,-2.) circle (2.5pt);
\draw [fill=black] (54.,2.) circle (2.5pt);
\draw [fill=black] (54.,-2.) circle (2.5pt);
\end{scriptsize}
\node at (1.5,0) {\Large $=$};
\node at (8.5,0) {\Large $+$};
\node at (14.5,0) {\Large $+$};
\node at (23.5,0) {\Large $+$};
\node at (34.5,0) {\Large $+$};
\node at (46.5,0) {\Large $+$};
\node at (63,0) {\Large $+  ~~\dots$};
\end{tikzpicture}
\caption{The full forward four-point function as a series of ladders. The crossings should be omitted for our rank-3 model, because it is built on complex fields, with bipartite graphs. However, they contribute for the rank-5 model, which has real fields.
}
\label{fig:ladders}
\end{figure}

The full four-point function  at leading order in $1/N$ is obtained by summing ladders of arbitrary lenghts with the (reduced) four-point kernel acting as rung (see \cite{Benedetti:2019eyl,Gurau:2019qag}). 
More precisely, defining the forward four-point function as
\be \label{eq:fw4pt}
\cF_{(\mba,x)(\mbb,y)(\mbc,z)(\mbd,w)} \equiv  \langle{\phi_{\mba}(x) \bar{\phi}_{\mbb}(y) \phi_{\mbc}(z) \bar{\phi}_{\mbd}(w)} \rangle - G(x-y) G(z-w) \d_{\mba\mbb} \d_{\mbc\mbd} \,,
\ee
one finds that at leading order in $1/N$ it is given by a geometric series on the (reduced) kernel:
\be \label{eq:fw4pt-ladders}
\cF_{(\mba,x)(\mbb,y)(\mbc,z)(\mbd,w)} = \int dz'dw' \,  (\mathbf{1} - \hat{K})^{-1}_{(\mba,x)(\mbb,y)(\mbc,z')(\mbd,w')} 
 \, G_{w'w}G_{z'z}\,.
\ee
We represent the series of ladder diagrams in Fig.~\ref{fig:ladders}, where the crossings do not contribute here because we consider a bipartite model with $U(N)^3$ symmetry.

For dimensional reasons, the propagators being massless and by the use of dimensional regularization, we do not expect the four-point function to require a renormalization of the quartic couplings, which are dimensionful. We verify this explicitly at lowest order in perturbation theory, that is by considering the fully amputated four-point kernel, with $G(q)$ replaced by the bare propagator $C(q)$.
Therefore, the reduced kernel writes:
\begin{equation}
\frac{\lambda_1^2}{4}\mathcal{Z}^4 (3\hat{\delta}^p_{\mba \mbb;\mbc\mbd}U_{\zeta}(p_1+p_4) + 2\hat{\delta}^d_{\mba \mbb;\mbc\mbd}U_{\zeta}(p_1+p_3))\,,
\end{equation}
with 
\begin{equation}
U_{\zeta}(p_1+p_4)=\int_{q_1,q_2,q_3}\frac{1}{q_1^{2\zeta}q_2^{2\zeta}q_3^{2\zeta}(p_1+p_4+q_1+q_2+q_3)^{2\zeta}}\,.
\end{equation}

Using Eq.~\eqref{eq:intG}, we find:
\begin{equation}
U_{\zeta}(p_1+p_4)=\frac{|p_1+p_4|^{3d-8\zeta}}{(4\pi)^{3d/2}}\frac{\Gamma(d/2-\zeta)^4\Gamma(4\zeta-3d/2)}{\Gamma(\zeta)^4\Gamma(2d-4\zeta)}\,.
\end{equation}

\paragraph{Standard propagator.} For $\zeta=1$ and $d=3-\epsilon$, this is finite (no poles in $\epsilon$):
\begin{equation}
U_1(p_1+p_4)=-\frac{|p_1+p_4| \pi}{4} \,.
\end{equation}

\paragraph{Long-range propagator.} For $\zeta=d/3$ and $d<3$, this is also finite:
\begin{equation}
U_{d/3}(p_1+p_4)=\frac{|p_1+p_4|^{d/3}}{(4\pi)^{3d/2}}\frac{\Gamma(d/6)^4\Gamma(-d/6)}{\Gamma(d/3)^4\Gamma(2d/3)} \,.
\end{equation}

In both cases, there are no divergences in the four-point kernel. We thus do not need to renormalize the four-point couplings and we can take them to be zero from the beginning.

\subsection{Rank 5}

In rank $5$ and at leading order in $1/N$, the 2PI effective action is given by: 
\begin{equation}
\label{eq:2PIrank5}
\begin{split}
-\G^{2PI}[\mbG] = &-\f{1}{6}\left(\sum_{i=2}^6 \f{\kappa_i}{N^{5+\rho(J_i)}}\d^{(i)}_{\mba\mbb;\mbc\mbd;\mbe\mbf}\right) \int\dd x ~\mbG_{(\mba,x)(\mbb,x)}\mbG_{(\mbc,x)(\mbd,x)}\mbG_{(\mbe,x)(\mbf,x)} \\
&+\f{1}{2}\left(\f{\kappa_1}{6N^5}\right)^2 \d^{(1)}_{\mba\mbb\mbc\mbd\mbe\mbf} \d^{(1)}_{\mbg\mbh\mbj\mbk\mbm\mbn}\times\\
&\qquad \int\dd x\dd y ~\mbG_{(\mba,x)(\mbg,y)}\mbG_{(\mbb,x)(\mbh,y)} \mbG_{(\mbc,x)(\mbj,y)}\mbG_{(\mbd,x)(\mbk,y)}\mbG_{(\mbe,x)(\mbm,y)}\mbG_{(\mbf,x)(\mbn,y)} \,.
\end{split}
\end{equation}

One recovers the self-energy from:
\begin{equation}
\Sigma[\mbG] = -2\f{\d\G^{2PI}[\mbG]}{\d\mbG} \,,
\end{equation}
where the extra factor  2 with respect to Eq.~\eqref{eq:Sigma2PI-complex} is due to the difference between real and complex fields.
The amputated four-point kernel is still obtained by derivating the self energy with respect to $\mbG$.

The right-amputated four-point kernel on-shell is then:
\be
\begin{split}
K_{(\mba,x)(\mbb,y)(\mbc,z)(\mbd,w)}=&G_{xx'}G_{yy'}\left[ -2\left(\kappa_{6}+\kappa_{3}+\frac{2\kappa_{2}}{3}+\frac{\kappa_{4}}{3}\right)\delta_{x'y'}\delta_{x'z}\delta_{x'w}G_{x'x'}\hat{\delta}^p_{\mba\mbb;\mbc\mbd} \right.\crcr
&-2\left(\frac{\kappa_{2}}{3}+\frac{2\kappa_{4}}{3}+\kappa_{5}\right)\delta_{x'y'}\delta_{x'z}\delta_{x'w}G_{x'x'}\hat{\delta}^d_{\mba\mbb;\mbc\mbd} \\
&\left.+\frac{5\kappa_1^2}{6}\delta_{x'w}\delta_{y'z}G_{x'y'}^4\hat{\delta}^p_{\mba\mbb;\mbc\mbd}\right] \,. 
\end{split}
\ee

The structure is the same as for the rank-$3$ model: the only difference comes from the combinatorial factors. Then, the Feynman amplitudes are the same as before and there are still no divergences. We can thus again take the four-point couplings to be zero from the beginning. Eliminating also the tadpoles, the four-point kernel is reduced to:
\begin{equation}
\hat{K}_{(\mba,x)(\mbb,y)(\mbc,z)(\mbd,w)}=G_{xw}G_{yz}\frac{5\kappa_1^2}{6}G_{zw}^4\hat{\delta}^p_{\mba\mbb;\mbc\mbd} \,.
\end{equation}

\section{Beta functions}
\label{sec:betas}

We have seen in Sec.~\ref{sec:SDeq} that the Schwinger-Dyson equations for the two-point functions admit a conformal IR limit for $\zeta=1$, and a conformal solution valid at all scales for $\zeta=d/3$. The argument is by now quite standard in theories with a melonic large-$N$ limit, and in one dimension, for the SYK model or its tensor generalizations, it is sufficient for concluding that the theory is conformal (in the IR limit or at all scales). However, for field theories in higher dimensions we should also consider the renormalization of the couplings. In particular, it is not possible to restrict the model to having only one interaction (the one leading to melonic diagrams), as we have seen that other interactions are generated by radiative corrections, and these lead to a renormalization group flow of the other couplings, which hence cannot be set to zero. In fact, in order to claim that we found a non-trivial conformal field theory, we should identify an interacting fixed point of the renormalization group.\footnote{In principle  a fixed point provides us only with a scale invariant theory, full conformal invariance having to be proved case by case or on the basis of the available theorems in dimensions two and four. See for example \cite{Nakayama:2013is} for a review.}
Therefore, in this section we will study the beta functions for the full actions \eqref{eq:int-action-graph} and \eqref{eq:int-action-graph-rank5}, and their relative fixed points.

We will explain the general structure of the beta functions in the rank-3 case. As we will see, the rank-5 case is very similar, except for the presence of an additional type of interaction, $J_6$ (the prism), a difference which however turns out to be crucial.

\subsection{Rank $3$}

The amputated connected six-point function can be decomposed into the different interaction terms:
\begin{equation}
\Gamma^{(6)}(p_1,\ldots , p_6)=\sum_{i=1}^5\Gamma^{(6,i)}(p_1,\ldots , p_6)\hat{\delta}^{i}\,.
\end{equation}

The renormalized sextic couplings $g_i$ are defined in terms of the  bare expansion of the six-point functions by the renormalization condition:
\begin{equation}
\label{eq:renorcouplings}
g_i=\mu^{-2\epsilon}Z^3\Gamma^{(6,i)}(p_1,\ldots , p_6)
\end{equation}
where the power of the renormalization scale $\m$ is  fixed by demanding that $g_i$ are dimensionless, and it is the same both for $\zeta=1$ in $d=3-\epsilon$ dimensions  and for $\zeta=\frac{d+\epsilon}{3}$ in general $d<3$.
For the external momenta we choose a symmetric subtraction point (generalizing the quartic case, see \cite{Brezin:1974eb,Kleinert:2001ax}):
\be
p_i \cdot p_j =\frac{\mu^2}{9}\left(6\delta_{ij}-1\right) \,.
\ee
This choice of external momenta satisfies the momentum conservation, $\sum_{i=1}^6 p_i=0$, and it is non-exceptional, in the sense that $\sum_{i\in I} p_i\neq 0$ for any subset $I$ of the set of indices $\{1,\ldots,6\}$, therefore avoiding IR divergences in all diagrams.

The beta functions are defined by $\b_i = \m\p_\m g_i$. We will obtain them by differentiating the bare expansion with respect to $\mu$ (using the fact that the bare couplings are independent of the renormalization scale $\mu$) and then replacing the bare couplings by their expressions in terms of the renormalized one. 

At leading order in $1/N$ the wheel vertex does not receive any radiative corrections, i.e.:
\be
g_1=\mu^{-2\epsilon}Z^3\l_1\,.
\ee
Since $Z=1+\cO(\l_1^2)$, the inverse $\l_1(g_1)$ is guaranteed to exist in the perturbative expansion, at least.\footnote{For the long-range model, it is actually easier to write the inverse relation, because at $\epsilon=0$ we can solve the exact equation \eqref{Z-norm-d/3} by multiplying it by $\mathcal{Z}^6$ and using $\mathcal{Z}^6\l_1^2=g_1^2$:
\begin{equation*}
\mathcal{Z} = 1- \f{g_1^2}{g_c^2}\,, \;\;\;\; g_c^{-2} = \frac{1}{4(4\pi)^{2d}}\frac{3\Gamma(1-\frac{d}{3})\Gamma(\frac{d}{6})^5}{d\Gamma(\frac{d}{3})^5\Gamma(\frac{5d}{6})} \,.
\end{equation*}
Therefore, $\l_1=g_1/\mathcal{Z}^3$ exists for $g_1<g_c$.
\label{foot:g_c}
}
Its beta function will then be
\be
\beta_1 \equiv \m \partial_\m g_1 = (-2\epsilon+3 \eta)  g_1\,,
\ee
where we defined the anomalous dimension $\eta= (\m\partial_\m \ln Z)|_{\l_1(g_1)}$. Clearly, if $\epsilon=0$ and $Z$ is finite, as in the long-range case $\z=d/3$ with $d<3$, then the beta function is identically zero, and we have a chance of finding a one-parameter family of fixed points, as in Ref.~\cite{Benedetti:2019eyl}. On the other hand, for $\epsilon>0$, in order to find a non-trivial fixed point we have to rely on a Wilson-Fisher type of cancellation between the tree level term and the quantum corrections, hence we need $\eta\neq 0$, that is, we need a short-range propagator.

\begin{figure}[htbp]
\centering
\captionsetup[subfigure]{labelformat=empty}
\subfloat[]{\tikzsetnextfilename{bare_vertex2}
\begin{tikzpicture}[line cap=round,line join=round,>=triangle 45,x=1.0cm,y=1.0cm,scale=0.5]
\draw [line width=1.pt,dash pattern=on 5pt off 5pt] (0.,0.)-- (0.,1.);
\draw [line width=1.pt,dash pattern=on 5pt off 5pt] (0.,0.)-- (0.7071067811865475,0.7071067811865475);
\draw [line width=1.pt,dash pattern=on 5pt off 5pt] (0.,0.)-- (-0.7071067811865474,0.7071067811865476);
\draw [line width=1.pt,dash pattern=on 5pt off 5pt] (0.,0.)-- (0.7071067811865475,-0.7071067811865475);
\draw [line width=1.pt,dash pattern=on 5pt off 5pt] (0.,0.)-- (-0.7071067811865474,-0.7071067811865476);
\draw [line width=1.pt,dash pattern=on 5pt off 5pt] (0.,0.)-- (0.,-1.);
\draw [color=black,fill=white] (0.,0.) circle (3.5pt);
\draw[white] (-1,-3)--(1,-3);
\draw[white] (-1,3)--(1,3);
\end{tikzpicture}}
\hspace{1cm}
\subfloat[]{\tikzsetnextfilename{rung32}
\begin{tikzpicture}[line cap=round,line join=round,>=triangle 45,x=1.0cm,y=1.0cm,scale=0.5]
\draw [shift={(1.5,0.)},line width=1.pt]  plot[domain=2.214297435588181:4.068887871591405,variable=\t]({1.*2.5*cos(\t r)+0.*2.5*sin(\t r)},{0.*2.5*cos(\t r)+1.*2.5*sin(\t r)});
\draw [shift={(-1.5,0.)},line width=1.pt]  plot[domain=-0.9272952180016123:0.9272952180016122,variable=\t]({1.*2.5*cos(\t r)+0.*2.5*sin(\t r)},{0.*2.5*cos(\t r)+1.*2.5*sin(\t r)});
\draw [line width=1.pt] (0.,2.)-- (0.,-2.);
\draw [line width=1.pt,dash pattern=on 5pt off 5pt] (0.,2.)-- (0.,3.);
\draw [line width=1.pt,dash pattern=on 5pt off 5pt] (0.,-2.)-- (0.,-3.);
\draw [line width=1.pt,dash pattern=on 5pt off 5pt] (0.,2.)-- (0.7071067811865475,2.7071067811865475);
\draw [line width=1.pt,dash pattern=on 5pt off 5pt] (0.,2.)-- (-0.7071067811865471,2.7071067811865475);
\draw [line width=1.pt,dash pattern=on 5pt off 5pt] (0.,-2.)-- (0.7071067811865475,-2.7071067811865475);
\draw [line width=1.pt,dash pattern=on 5pt off 5pt] (0.,-2.)-- (-0.7071067811865471,-2.7071067811865475);
\begin{scriptsize}
\draw [fill=black] (0.,2.) circle (2.5pt);
\draw [fill=black] (0.,-2.) circle (2.5pt);
\end{scriptsize}
\end{tikzpicture}}
\hspace{1cm}
\subfloat[]{\tikzsetnextfilename{order32}
\begin{tikzpicture}[line cap=round,line join=round,>=triangle 45,x=1.0cm,y=1.0cm,scale=0.5]
\draw [shift={(3.75,0.)},line width=1.pt]  plot[domain=2.651635327336065:3.631549979843521,variable=\t]({1.*4.25*cos(\t r)+0.*4.25*sin(\t r)},{0.*4.25*cos(\t r)+1.*4.25*sin(\t r)});
\draw [shift={(1.5,0.)},line width=1.pt]  plot[domain=2.214297435588181:4.068887871591405,variable=\t]({1.*2.5*cos(\t r)+0.*2.5*sin(\t r)},{0.*2.5*cos(\t r)+1.*2.5*sin(\t r)});
\draw [shift={(-1.5,0.)},line width=1.pt]  plot[domain=-0.9272952180016123:0.9272952180016123,variable=\t]({1.*2.5*cos(\t r)+0.*2.5*sin(\t r)},{0.*2.5*cos(\t r)+1.*2.5*sin(\t r)});
\draw [shift={(-3.75,0.)},line width=1.pt]  plot[domain=-0.48995732625372845:0.4899573262537281,variable=\t]({1.*4.25*cos(\t r)+0.*4.25*sin(\t r)},{0.*4.25*cos(\t r)+1.*4.25*sin(\t r)});
\draw [line width=1.pt] (0.,2.)-- (3.,0.);
\draw [line width=1.pt] (0.,-2.)-- (3.,0.);
\draw [line width=1.pt,dash pattern=on 5pt off 5pt] (0.,2.)-- (-1.,2.);
\draw [line width=1.pt,dash pattern=on 5pt off 5pt] (0.,-2.)-- (-1.,-2.);
\draw [line width=1.pt,dash pattern=on 5pt off 5pt] (3.,0.)-- (3.866025403784439,0.5);
\draw [line width=1.pt,dash pattern=on 5pt off 5pt] (3.,0.)-- (3.5,0.8660254037844386);
\draw [line width=1.pt,dash pattern=on 5pt off 5pt] (3.,0.)-- (3.866025403784439,-0.5);
\draw [line width=1.pt,dash pattern=on 5pt off 5pt] (3.,0.)-- (3.5,-0.8660254037844386);
\draw [fill=black] (0.,2.) circle (2.5pt);
\draw [fill=black] (0.,-2.) circle (2.5pt);
\draw [color=black,fill=white] (3.,0.) circle (3.5pt);
\draw[white] (-1,-3)--(1,-3);
\draw[white] (-1,3)--(1,3);
\end{tikzpicture}}
\caption{The three diagrams that contribute to the bare expansion of the six-point couplings up to order three (bare coupling, $D$ and $S$, see Sec.~\ref{sec:betas1}). The black circles represent wheel vertices and the white circles can be any of the $I_i$ interactions (including the wheel itself).}
\label{fig:bare3}
\end{figure}
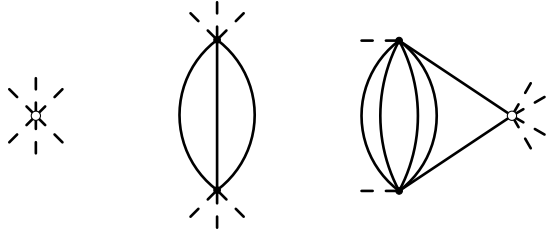

We now compute the bare expansion of the other couplings. The expansion starts of course at tree level, with a bare vertex with any $I_i$ interaction.
At order two in the coupling constants, there is only one diagram which contributes: two wheel vertices connected by three internal edges (we call this Feynman diagram the \textit{candy}). At order three, we have one more diagram: two wheel vertices connected to each other by four internal edges and each of them connected by another internal edge to a vertex with any $I_i$ interaction (including the wheel itself). These diagrams are the only tadpole-free six-point diagrams that can be obtained by cutting edges of vacuum melon-tadpole diagrams, at this order in the couplings, and they are pictured in Fig.~\ref{fig:bare3}.

We can actually construct the leading order $6$-point graphs at all orders using the forward four-point function introduced in Eq.~\eqref{eq:fw4pt-ladders}. 
Indeed, the amputated connected six-point functions can be obtained by deleting  three different lines in the vacuum diagrams, without disconnecting the diagrams. On the other hand, vacuum diagrams are given in Fig.~\ref{fig:fund_vacuum}, with lines decorated by melonic and tadpole insertions, but we should not leave any closed tadpoles otherwise the diagram will evaluate to zero in dimensional regularization. Therefore, we can have at most one tadpole vertex; this fact does not limit the number of wheel vertices, as they can appear in melonic insertions as well, but it has the important consequence that the couplings $\l_2$ to $\l_5$ appear at most linearly in $\Gamma^{(6)}$.
Equivalently, we can just consider the two diagrams in Fig.~\ref{fig:fund_vacuum} with only melonic insertions. Furthermore, for the trefoil on the right of Fig.~\ref{fig:fund_vacuum}, we should cut an internal line on each of the three (decorated) leaves. At last, we should notice that each time we delete a line in a melonic two-point function, we generate a ladder diagram. In fact, starting from the SDE equation $G=(C^{-1}-\Sigma[G])^{-1}$, and using \eqref{eq:defK}, we obtain
\be
\f{\d G_{AB}}{\d C_{EF}} = (1-K)^{-1}_{AB,A'B'} G_{A'E'} C^{-1}_{E'E} G_{B'F'} C^{-1}_{F'F} + (E\leftrightarrow F)\,.
\ee
When using this formula on vacuum diagrams, we should then strip off the factors $G\cdot C^{-1}$ in order to obtain amputated $n$-point functions. We thus obtain the right-amputated version of Eq.~\eqref{eq:fw4pt-ladders}.

\begin{figure}[htbp]
\centering
\captionsetup[subfigure]{labelformat=empty}
\subfloat[]{\tikzsetnextfilename{bare_vertex_kernel2}
\begin{tikzpicture}[line cap=round,line join=round,>=triangle 45,x=1.0cm,y=1.0cm,scale=0.25]
\fill[line width=1.pt,color=black,fill=gray] (-2.5,4.330127018922194) -- (2.5,4.330127018922194) -- (1.5,2.598076211353316) -- (-1.5,2.598076211353316) -- cycle;
\fill[line width=1.pt,color=black,fill=gray] (-3.,0.) -- (-1.5,-2.5980762113533156) -- (-2.5,-4.330127018922193) -- (-5.,0.) -- cycle;
\fill[line width=1.pt,color=black,fill=gray] (3.,0.) -- (5.,0.) -- (2.5,-4.3301270189221945) -- (1.5,-2.598076211353317) -- cycle;
\draw [line width=1.pt] (-3.,5.196152422706632)-- (0.,0.);
\draw [line width=1.pt] (0.,0.)-- (3.,5.196152422706632);
\draw [line width=1.pt] (0.,0.)-- (-6.,0.);
\draw [line width=1.pt] (0.,0.)-- (-3.,-5.196152422706631);
\draw [line width=1.pt] (0.,0.)-- (3.,-5.196152422706634);
\draw [line width=1.pt] (0.,0.)-- (6.,0.);
\begin{scriptsize}
\draw [color=black,fill=white] (0.,0.) circle (4.5pt);\end{scriptsize}
\end{tikzpicture}}
\hspace{1cm}
\subfloat[]{\tikzsetnextfilename{candy_kernel2}
\begin{tikzpicture}[line cap=round,line join=round,>=triangle 45,x=1.0cm,y=1.0cm,scale=0.25]
\draw [shift={(3.75,0.)},line width=1.pt]  plot[domain=2.651635327336065:3.631549979843521,variable=\t]({1.*4.25*cos(\t r)+0.*4.25*sin(\t r)},{0.*4.25*cos(\t r)+1.*4.25*sin(\t r)});
\draw [shift={(-3.75,0.)},line width=1.pt]  plot[domain=-0.48995732625372845:0.4899573262537283,variable=\t]({1.*4.25*cos(\t r)+0.*4.25*sin(\t r)},{0.*4.25*cos(\t r)+1.*4.25*sin(\t r)});
\draw [line width=1.pt] (0.,2.)-- (0.,-2.);
\draw [line width=1.pt] (0.,-2.)-- (-6.928203230275509,-6.);
\draw [line width=1.pt] (0.,-2.)-- (6.92820323027551,-6.);
\draw [line width=1.pt] (0.,2.)-- (-5.65685424949238,7.656854249492381);
\draw [line width=1.pt] (0.,2.)-- (-6.928203230275509,-2.);
\draw [line width=1.pt] (0.,2.)-- (6.928203230275509,-2.);
\draw [line width=1.pt] (0.,-2.)-- (-5.65685424949238,3.656854249492381);
\fill[line width=1.pt,color=black,fill=gray] (-3.031088913245535,0.25) -- (-3.031088913245535,-3.75) -- (-4.763139720814412,-4.75) -- (-4.763139720814412,-0.75) -- cycle;
\fill[line width=1.pt,color=black,fill=gray] (-3.535533905932737,5.535533905932738) -- (-2.121320343559642,4.121320343559643) -- (-2.121320343559642,0.12132034355964276) -- (-3.535533905932737,1.535533905932738) -- cycle;
\fill[line width=1.pt,color=black,fill=gray] (3.031088913245535,0.25) -- (3.0310889132455348,-3.75) -- (4.7631397208144115,-4.75) -- (4.763139720814412,-0.75) -- cycle;
\begin{scriptsize}
\draw [fill=black] (0.,2.) circle (2.5pt);
\draw [fill=black] (0.,-2.) circle (2.5pt);
\end{scriptsize}
\end{tikzpicture}}
\hspace{1cm}
\subfloat[]{\tikzsetnextfilename{3kernels2}
\begin{tikzpicture}[line cap=round,line join=round,>=triangle 45,x=1.0cm,y=1.0cm,scale=0.25]
\fill[line width=1.pt,color=black,fill=gray] (-2.,6.) -- (2.,6.) -- (2.,4.) -- (-2.,4.) -- cycle;
\fill[line width=1.pt,color=black,fill=gray] (-4.196152422706632,-4.732050807568876) -- (-6.196152422706632,-1.2679491924311213) -- (-4.464101615137754,-0.2679491924311217) -- (-2.4641016151377553,-3.7320508075688767) -- cycle;
\fill[line width=1.pt,color=black,fill=gray] (6.196152422706632,-1.2679491924311213) -- (4.196152422706632,-4.732050807568876) -- (2.4641016151377553,-3.7320508075688767) -- (4.464101615137754,-0.2679491924311217) -- cycle;
\draw [line width=1.pt] (-2.,2.)-- (-2.,8.);
\draw [line width=1.pt] (2.,2.)-- (2.,8.);
\draw [line width=1.pt] (-0.7320508075688779,-2.7320508075688767)-- (-5.9282032302755105,-5.732050807568876);
\draw [line width=1.pt] (-2.7320508075688767,0.7320508075688779)-- (-7.928203230275509,-2.2679491924311206);
\draw [line width=1.pt] (0.7320508075688759,-2.732050807568877)-- (5.928203230275505,-5.7320508075688785);
\draw [line width=1.pt] (2.732050807568877,0.7320508075688759)-- (7.928203230275507,-2.267949192431126);
\draw [shift={(-3.3727438137666783,1.947254548785837)},line width=1.pt]  plot[domain=-1.0856020347111377:0.03840448351454007,variable=\t]({1.*1.3737567691765782*cos(\t r)+0.*1.3737567691765782*sin(\t r)},{0.*1.3737567691765782*cos(\t r)+1.*1.3737567691765782*sin(\t r)});
\draw [shift={(3.3727438137666774,1.9472545487858353)},line width=1.pt]  plot[domain=3.103188170075252:4.227194688300931,variable=\t]({1.*1.3737567691765769*cos(\t r)+0.*1.3737567691765769*sin(\t r)},{0.*1.3737567691765769*cos(\t r)+1.*1.3737567691765769*sin(\t r)});
\draw [shift={(0.,-3.894509097571673)},line width=1.pt]  plot[domain=1.0087930676820573:2.132799585907735,variable=\t]({1.*1.3737567691765782*cos(\t r)+0.*1.3737567691765782*sin(\t r)},{0.*1.3737567691765782*cos(\t r)+1.*1.3737567691765782*sin(\t r)});
\end{tikzpicture}}
\caption{The three types of diagrams contributing to the bare expansion of the six-point couplings in the large-$N$ limit at all order in the coupling constants. The black circles represent wheel vertices and the white circles can be any of the $I_i$ interactions (including the wheel). The grey squares represent the full forward four-point function.}
\label{fig:allorders}
\end{figure}
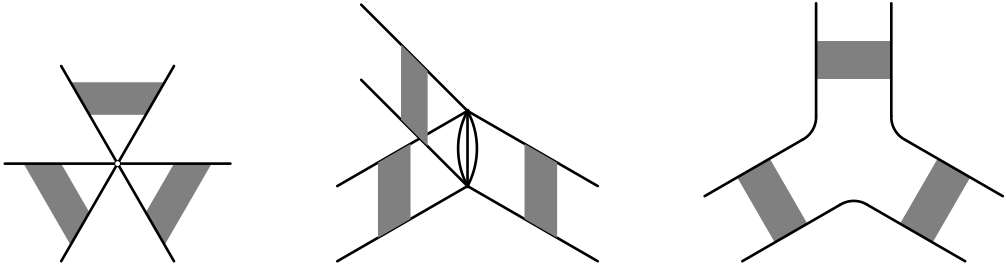

In conclusion, we then have three different types of leading-order $6$-point graphs. First, we can connect three full forward four-point functions on every pairs of external legs of the bare vertex of Fig.~\ref{fig:bare3}, thus obtaining the graph on the left of Fig.~\ref{fig:allorders}. We can also do the same with the candy and obtain the graph in the middle of Fig.~\ref{fig:allorders}.  Finally, we can also connect three full forward four-point functions and obtain the graph on the right of Fig.~\ref{fig:allorders}. The last two have been encountered for example in  \cite{Gross:2017hcz,Gross:2017aos}, where they have been called \emph{contact} and \emph{planar} diagrams, respectively.

This implies that renormalized couplings $g_i$, with $i>1$, have a bare expansion of the form:
\be
g_i = \m^{-2\epsilon} Z^3 \left( \l_i +  A_i(\l_1^2 )+ \sum_{j=1\ldots 5}  B_{ij}(\l_1^2) \l_j \right)\,,
\ee
with $A_i(x)$ and $B_{ij}(x)$ starting at linear order in $x$. The term $ \l_i + \sum_{j=1\ldots 5}  B_{ij}(\l_1^2) \l_j$ is a resummation of contribution from the graphs on the left of Fig.~\ref{fig:allorders}, while $A_i(\l_1^2 )$ resums the other two. Although we could at least write the relative Feynman integral expressions in terms of forward four-point functions and six-point kernels, as we will not need them, and the combinatorics is different for different bubbles, we will not be more precise than that.

For $i>1$ the relation between bare and renormalized couplings is linear and thus it can be easily inverted:
\be
\l_i = (\mathbf{1}+B)^{-1}_{ij} \left( \f{\m^{2\epsilon} g_j}{Z^3} - A_j \right)\big|_{\l_1=\l_1(g_1)}\,,
\ee
where $\mathbf{1}_{ij}=\delta_{ij}$.

Using the fact that the flow of $g_1$ is independent of the others, then one arrives at the conclusion that the beta functions of the other couplings are linear combinations of the couplings, with coefficients that are functions of $g_1^2$:
\be
\beta_i = (-2\epsilon+3 \eta)  g_i + \tilde{A}_i(g_1^2) + \sum_j  \tilde{B}_{ij}(g_1^2)  g_j \,,
\ee
where 
\begin{align}
\tilde{A}_i(g_1^2)  &=  \m^{-2\epsilon} Z^3 \left(  \m\p_\m A_i - \sum_{j,k}  (\m\p_\m B_{ij}) (\mathbf{1}+B)^{-1}_{jk} A_k \right)\big|_{\l_1=\l_1(g_1)} \,,\\
\tilde{B}_{ij}(g_1^2) &=  \sum_{j}  (\m\p_\m B_{ij}) (\mathbf{1}+B)^{-1}_{jk} \,.
\end{align}

As we saw above, the combination $-2\epsilon+3 \eta$ is either identically zero, or it is zero at the fixed point of $g_1$.
In order to find the fixed points we are left with a linear problem.

In the following we will compute explicitly the beta functions at lowest order in the perturbative expansion, i.e.\ we will only include the contribution of the diagrams in Fig.~\ref{fig:bare3}.

\subsubsection{The $\zeta=1$ case}
\label{sec:betas1}

We will now compute the beta function only up to order $3$ in the coupling constants for $\zeta=1$. 

Expanding the six-point functions of Eq. \eqref{eq:renorcouplings} to order three, we have the bare expansions:
\begin{align}
g_{2}&=\mu^{-2\epsilon}Z^3\left(\l_{2}-\frac{9}{2}D_1(\mu)\l_1^2+S_1(\mu)\l_1^2\left(\frac{9}{2}\l_1 + \frac{1}{2}\l_2\right)\right)\,,\crcr
g_{3}&=\mu^{-2\epsilon}Z^3\left(\l_{3}+ S_1(\mu)\frac{3}{4}\l_1^2\l_{3}\right)\,,\crcr
g_{4}&=\mu^{-2\epsilon}Z^3\left(\l_{4}+ S_1(\mu)\l_1^2\left(\f{27}{4}\l_1 +\frac{5}{2} \l_2+3\l_3+\f{7}{4}\l_4\right)\right)\,,\crcr
g_{5}&=\mu^{-2\epsilon}Z^3\left(\l_{5}-D_1(\mu)\f{1}{2}\l_1^2+ S_1(\mu)\l_1^2\left(\f{3}{4}\l_2+2\l_4+\f{15}{4}\l_5\right)\right)\,,
\end{align}
where $Z$ is given in Eq.~\eqref{eq:wavef3}, $D_1(\mu)$ is the candy integral, and $S_1(\mu)$ the integral corresponding to the Feynman diagram on the right of Fig.~\ref{fig:bare3}. The integrals are both computed in App.~\ref{ap:betafun4}.
It is convenient to rescale the couplings as $\tilde{g}_i=g_i/(4\pi)^d$ (and we immediately forget the $\sim$). To compute the beta functions we need:
\begin{gather}
\mu\partial_{\mu} D_1=-\frac{4\pi}{(4\pi)^3}\mu^{-2\epsilon}+\mathcal{O}(\epsilon)\,,\\
\mu\partial_{\mu} S_1 =\frac{8\pi^{2}}{(4\pi)^{6}}\mu^{-4\epsilon}+\mathcal{O}(\epsilon)\,, \\
\mu\partial_{\mu} Z=\mu^{-4\epsilon}\frac{2g_1^2\pi^2}{3}\,.
\end{gather}
Then, using $\mu\partial_{\mu} \l_i=0$, the beta functions $\b_i\equiv \mu\partial_{\mu} g_i$ come as:
\begin{align}
\b_1 &= -2 g_1\left(\eps-g_1^2\pi^2\right)\,, \\
\b_2 &= -2 g_2\left(\eps-g_1^2\pi^2\right) + 4g_1^2\pi^2\left(\f{9}{2\pi}+9g_1+g_2\right)\,,\\
\b_3 &= -2 g_3\left(\eps-4\, g_1^2\pi^2\right)\,,\\
\b_4 &= -2 g_4\left(\eps-g_1^2\pi^2\right) + g_1^2\pi^2\left(54g_1 +20g_2 + 24g_3 + 14g_4\right)\,,\\
\b_5 &= -2 g_5\left(\eps-g_1^2\pi^2\right)+g_1^2\pi^2\left(\f{2}{\pi} +6g_2+16g_4+30g_5\right)\,.
\end{align}
First, we notice that if $g_1=0$, then all the other coupling have a trivial running, dictated by their canonical dimension ($2\epsilon$).
In such case, for $\epsilon>0$ we have only the trivial fixed point,  $g^*_i=0~\forall i$.
For $\epsilon=0$ instead, we have a 4-dimensional manifold of fixed points. This is a generalization of the vector model case, where the $(\phi_i \phi_i)^3$ interaction is exactly marginal at large $N$, and which in fact corresponds to the case in which we retain only the triple-trace term $I_5$ in our action. In that case it is known that at some critical coupling non-perturbative effects lead to vacuum instability with a consequent breaking of conformal invariance  \cite{Bardeen:1983rv,Amit:1984ri}. It would be interesting to study the vacuum stability of our model with $g_1=0$ in order to explore the possibility of a similar phenomenon, but we leave this for future work.

We are here interested in melonic fixed points, with $g_1\neq 0$, for which we need  $\epsilon>0$.
In that case, we obtain two interacting fixed points:
\begin{gather}
g^*_1=\pm\f{\sqrt{\eps}}{\pi}; \qquad g^*_2=\f{9}{2\pi}\left(-1\mp 2\sqrt{\eps}\right); \qquad g^*_3=0;
\\
g^*_4=\f{9}{7\pi}\left(5 \pm 7\sqrt{\eps}\right);\qquad g^*_5 = \f{-109\mp 126\sqrt{\eps}}{42\pi}.
\end{gather}
The standard linear stability analysis of the system of beta functions consists in diagonalizing the stability matrix $\cB_{ij} \equiv \p\b_i/\p g_j|_{g*}$,
thus identifying the scaling operators and their scaling dimensions, from its right-eigenvectors and eigenvalues, respectively. 
In the present case, we find the slightly unusual situation of having a non-diagonalizable stability matrix.
In fact, we find that both melonic fixed points have the same eigenvalues (critical exponents):
\begin{equation}
(4 \eps;\; 4 \eps;\; 6 \eps;\; 14 \eps;\; 30 \eps)\,,
\end{equation}
with the $4 \eps$ eigenvalue having algebraic multiplicty two, but geometric multiplicity one; hence the stability matrix is not diagonalizable.
In terms of the couplings $\{g_1,g_2,g_3,g_4,g_5\}$, the associated eigendirections are, respectively:
\begin{gather}
 \{0,1,0,-2,1\};\quad \{\mp\f{1}{18\sqrt{\eps}}, \f{2(- 632 \pm 4095\sqrt{\eps})}{12285\eps}, 0, \f{1124\mp 4095\sqrt{\eps}}{24570 \eps} ,0 \}; \\
 \{0,0,\frac{1}{2},-\frac{3}{2},1\} \quad
 \{0,0,0,-1,1\};\quad \{0,0,0,0,1\}\,,
\end{gather} 
with the first two forming a Jordan chain of length two.
Each (generalized) eigendirection, by its scalar product with the vector of renormalized operators arranged in the same order as the corresponding couplings, defines a scaling operator $\cO_i$ of dimension $\Delta_i=d+\theta_i$, with the $\theta_i$ being the critical exponent associated to that eigendirection.
As our critical exponents are all positive, all the scaling operators are irrelevant at the fixed points, and therefore the latter are infrared stable.
The fact that the stability matrix is not diagonalizable implies that the fixed point theory is a logarithmic conformal field theory (see for example \cite{Hogervorst:2016itc}). Therefore, although we find real exponents, as opposed to the complex ones of the quartic model \cite{Giombi:2017dtl}, the fixed-point theory is still non-unitary.

\subsubsection{The $\zeta=\frac{d}{3}$ case}
Using the results of App.~\ref{ap:betafun4}, along with the fact that there is no diverging wave-function renormalization in this case, the bare expansion gives:
\begin{align}
g_{1}&=\mu^{-2\epsilon}\cZ^3\l_{1}\,,\crcr
g_{2}&=\mu^{-2\epsilon}\cZ^3\left(\l_{2}-\frac{9}{2}\cZ^3D_{d/3}(\mu)\l_1^2+\cZ^6S_{d/3}(\mu)\l_1^2\left(\frac{9}{2}\l_1 + \frac{1}{2}\l_2\right)\right)\,,\crcr
g_{3}&=\mu^{-2\epsilon}\cZ^3\left(\l_{3}+ \cZ^6S_{d/3}(\mu)\frac{3}{4}\l_1^2\l_{3}\right)\,,\crcr
g_{4}&=\mu^{-2\epsilon}\cZ^3\left(\l_{4}+\cZ^6S_{d/3}(\mu)\l_1^2\left(\f{27}{4}\l_1 + \frac{5}{2}\l_2+3\l_3+\f{7}{4}\l_4\right)\right)\,,\crcr
g_{5}&=\mu^{-2\epsilon}\cZ^3\left(\l_{5}-\cZ^3D_{d/3}(\mu)\f{1}{2}\l_1^2+ \cZ^6S_{d/3}(\mu)\l_1^2\left(\f{3}{4}\l_2+2\l_4+\f{15}{4}\l_5\right)\right)\,,
\end{align}
with $\cZ$ given in Eq.~\eqref{eq:wavef3-LR}.
After rescaling of the coupling constants by $(4\pi)^d$, the beta functions  at $\epsilon=0$ read:
\begin{align}
\b_1 &= 0\,, \\
\b_2 &= g_1^2\f{\Gamma(d/6)^3}{\Gamma(d/3)^3\Gamma(d/2)}\left(-\f{\Gamma(-d/6)\Gamma(d/6)}{\Gamma(d/3)\Gamma(2d/3)}\left(9g_1+g_2\right)+9\right)\,,\\
\b_3 &= - 3g_1^2 g_3\frac{\Gamma(-d/6) \Gamma(d/6)^4}{2 \Gamma(d/3)^4 \Gamma(d/2)\Gamma(2d/3)}\,,\\
\b_4 &= -g_1^2 \frac{\Gamma(-d/6) \Gamma(d/6)^4}{\Gamma(d/3)^4 \Gamma(d/2)\Gamma(2d/3)}\left(\frac{27}{2}g_1+5g_2+6g_3+\frac{7}{2}g_4\right)\,,\\
\b_5 &= g_1^2\f{\Gamma(d/6)^3}{\Gamma(d/3)^3\Gamma(d/2)}\left(-\frac{2\Gamma(-d/6)\Gamma(d/6)}{\Gamma(d/3)\Gamma(2d/3)}\left(\frac{3}{4} g_2+2g_4+\frac{15}{4}g_5\right)+1\right)\,.
\end{align}
This time, in addition to a 4-dimensional manifold of fixed points (set by $g_1^*=0$, and thus analogue to what we discussed for the case $\zeta=1$ at  $\epsilon=0$), we also find a line of fixed points parametrized by the exactly marginal coupling $g_1$: 
\begin{gather}
g_2^*=-9g_1+\f{9\Gamma(d/3)\Gamma(2d/3)}{\Gamma(-d/6)\Gamma(d/6)};\qquad g_3^*=0;\\
g_4^*=9g_1 - \frac{90}{7}\f{2\Gamma(d/3)\Gamma(2d/3)}{\Gamma(-d/6)\Gamma(d/6)};\\
g_5^*=-3g_1 + \f{109\Gamma(d/3)\Gamma(2d/3)}{21\Gamma(-d/6)\Gamma(d/6)}.
\end{gather}
The critical exponents are: 
\begin{equation}
 \left( \f{15 g_1^2\a}{ 2};\;
   \f{7 g_1^2\a}{ 2 };\; 
   \f{3 g_1^2\a}{2 };\;
    g_1^2\a\right)\,,
\end{equation}
with
\be
\a= - \f{\Gamma(-d/6) \Gamma(d/6)^4}{
  \Gamma(d/3)^4 \Gamma(d/2) \Gamma(2 d/3)} >0\,, \;\;\; \text{for } d<3\,.
\ee
The respective eigendirections in terms of $\{g_2,g_3,g_4,g_5\}$ are: 
\begin{equation}
\{0,0,0,1\};\quad \{0,0,-1,1\};\quad\{0,1,-3,2\};\quad\{1,0,-2,1\}.
\end{equation}
Since the critical exponents are again positive, the eigendirections correspond again to irrelevant perturbations.
In this case, the stability matrix is diagonalizable, with real exponents, hence we have so far no signal of non-unitarity.

\subsection{Rank $5$}

The diagrams contributing to the six-point function at large $N$ are again the ones of Fig.~\ref{fig:bare3} (or Fig.~\ref{fig:allorders} at all orders). However, now the black vertices represent the complete interaction and the white vertices represent only the other interactions $J_i$ for $i>1$. This will slightly change the bare expansion of the couplings and their beta functions. 

\subsubsection{The $\zeta=1$ case}

There is no radiative corrections for the coupling of the complete interaction, the renormalized coupling is just rescaled by the wave function renormalization of Eq.~\eqref{eq:wavef5}:
\begin{equation}
g_1=\mu^{-2\epsilon}Z^3\kappa_1 \,.
\end{equation}

Then, we obtain the following bare expansions up to order three in the coupling constants:
\begin{align}
g_{2}&=\mu^{-2\epsilon}Z^3\left(\kappa_{2}+\kappa_1^2S_1(\mu)\left(2\kappa_{6}+\frac{2}{3}\kappa_{2}\right)\right)\,,\crcr
g_{3}&=\mu^{-2\epsilon}Z^3\left(\kappa_{3}+\kappa_1^2\kappa_{3}S_1(\mu)\right)\,,\crcr
g_{4}&=\mu^{-2\epsilon}Z^3\left(\kappa_{4}+\kappa_1^2S_1(\mu)\left(3\kappa_{6}+4\kappa_{3}+\frac{10}{3}\kappa_{2}+\frac{7}{3}\kappa_{4}\right)\right)\,,\crcr
g_{5}&=\mu^{-2\epsilon}Z^3\left(\kappa_{5}+\kappa_1^2S_1(\mu)\left(\kappa_{2}+\frac{8}{3}\kappa_{4}+5\kappa_{5}\right)\right)\,,\crcr
g_{6}&=\mu^{-2\epsilon}Z^3\left(\kappa_{6}-\frac{10}{3}\kappa_1^2 D_1(\mu)\right) \,,
\end{align}
with $D_{1}(\mu)$ and $S_{1}(\mu)$ defined in the previous section.

Let us rescale all the coupling constants as $\tilde{\kappa_1}=\frac{\kappa_1}{(4\pi)^d}$ and forget about the tilde.
Then the beta functions are:
\begin{align}
\beta_{g_1}&=-2\epsilon g_1+\frac{4}{3}\pi^2g_1^3 \,,\crcr  
\beta_{g_{2}}&=-2\epsilon g_{2}+\frac{4}{3}\pi^2g_1^2g_{2}+\frac{8\pi^2}{3}g_1^2\left(6g_{6}+2g_{2}\right)\,, \crcr
\beta_{g_{3}}&=-2\epsilon g_{3}+\frac{4}{3}\pi^2g_1^2g_{3}+8\pi^2g_1^2g_{3} \,, \crcr
\beta_{g_{4}}&=-2\epsilon g_{4}+\frac{4}{3}\pi^2g_1^2g_{4}+\frac{8\pi^2}{3}g_1^2\left(9g_{6}+12g_{3}+10g_{2}+7g_{4}\right)\,,\crcr
\beta_{g_{5}}&=-2\epsilon g_{5}+\frac{4}{3}\pi^2g_1^2g_{5}+\frac{8\pi^2}{3}g_1^2\left(3g_{2}+8g_{4}+15g_{5}\right)\,, \crcr
\beta_{g_{6}}&=-2\epsilon g_{6}+\frac{4}{3}\pi^2g_1^2g_{6}+\frac{40\pi}{3}g_1^2 \,.
\end{align}

The only fixed point when $\epsilon \neq 0$ is the trivial one: $g_i^*=0,~\forall i$. We do not find any Wilson-Fisher like fixed point. 
This is due to the beta function of the prism. The non-zero fixed point of $\beta_{g_1}$ is $g_1^*=\frac{\sqrt{\epsilon}}{2\pi}$. If we put it in the beta function of the prism, we obtain an expression independent of $g_6$ and proportional to $g_1^2$. This would imply $g_1=0$ which is incompatible with $g_1^*=\frac{\sqrt{\epsilon}}{2\pi}$ when $\epsilon \neq 0$. This solution is not a fixed point of the whole system.  

\subsubsection{The $\zeta=\frac{d}{3}$ case}

When $\zeta=d/3$, the wave function renormalization is finite and equal to $\mathcal{Z}$, given in Eq.~\eqref{eq:wavef5-LR}. In this case the bare expansion is:
\begin{align}
g_1&=\mu^{-2\epsilon}\mathcal{Z}^3\kappa_1  \,,\crcr
g_{2}&=\mu^{-2\epsilon}\left(\mathcal{Z}^3\kappa_{2}+\kappa_1^2\mathcal{Z}^9S_{d/3}(\mu)\left(2\kappa_{6}+\frac{2}{3}\kappa_{2}\right)\right)\,,\crcr
g_{3}&=\mu^{-2\epsilon}\left(\mathcal{Z}^3\kappa_{3}+\kappa_1^2\kappa_{3}\mathcal{Z}^9S_{d/3}(\mu)\right)\,,\crcr
g_{4}&=\mu^{-2\epsilon}\left(\mathcal{Z}^3\kappa_{4}+\kappa_1^2\mathcal{Z}^9S_{d/3}(\mu)\left(3\kappa_{6}+4\kappa_{3}+\frac{10}{3}\kappa_{2}+\frac{7}{3}\kappa_{4}\right)\right)\,,\crcr
g_{5}&=\mu^{-2\epsilon}\left(\mathcal{Z}^3\kappa_{5}+\kappa_1^2\mathcal{Z}^9S_{d/3}(\mu)\left(\kappa_{2}+\frac{8}{3}\kappa_{4}+5\kappa_{5}\right)\right)\,,\crcr
g_{6}&=\mu^{-2\epsilon}\left(\mathcal{Z}^3\kappa_{6}-\frac{10}{3}\mathcal{Z}^6\kappa_1^2 D_{d/3}(\mu)\right)\,.
\end{align}

Then, the beta function of the complete interaction is again exactly zero. The other beta functions are, after rescaling of the coupling constants by $(4\pi)^d$:
\begin{align}
\beta_{g_2}&=-2g_1^2\frac{\Gamma(d/6)^4 \Gamma(-d/6)}{\Gamma(d/3)^4\Gamma(d/2)\Gamma(2d/3)}\left(2g_6+\frac{2}{3}g_2\right) \,,\crcr
\beta_{g_3}&=-2g_1^2g_3\frac{\Gamma(d/6)^4 \Gamma(-d/6)}{\Gamma(d/3)^4\Gamma(d/2)\Gamma(2d/3)} \,,\crcr
\beta_{g_4}&=-2g_1^2\frac{\Gamma(d/6)^4 \Gamma(-d/6)}{\Gamma(d/3)^4\Gamma(d/2)\Gamma(2d/3)}\left(3g_6+4g_3+\frac{10}{3}g_2+\frac{7}{3}g_4 \right)\,,\crcr
\beta_{g_5}&=-2g_1^2\frac{\Gamma(d/6)^4 \Gamma(-d/6)}{\Gamma(d/3)^4\Gamma(d/2)\Gamma(2d/3)}\left(g_2+\frac{8}{3}g_4+5g_5 \right) \,,\crcr
\beta_{g_6}&=\frac{20}{3}\frac{\Gamma(d/6)^3}{\Gamma(d/3)^3\Gamma(d/2)}g_1^2 \,.
\end{align}

The beta function for $g_6$ admits a unique fixed point with $g_1^*=0$. The other beta functions are then exactly zero. 
Starting from nonzero couplings, we find that the flow is driven by $g_6$ flowing to minus infinity in the IR, and the other couplings flow towards: 
\begin{equation}
g_2^*=-3g_6; \quad  g_3^* = 0; \quad g_4^*=3g_6; \quad g_5^* = -3g_6.
\end{equation}

\section{Spectrum of operators}
\label{sec:BSeq}

For the rank-3 case we found IR fixed points with non-zero wheel coupling, both in the short-range and long-range versions of the model.
In order to better understand the conformal field theory at such IR fixed points,\footnote{We assume here that our fixed points correspond to conformal field theories.} we wish to compute the spectrum of operators that appear in the operator-product expansion (OPE) of $\phi_{abc}(x)\bar{\phi}_{abc}(0)$. Schematically, these are expected to be the bilinear operators $\phi_{abc}(\partial^2)^n\bar{\phi}^{abc}$, and their spectrum can  be obtained using the conformal Bethe-Salpeter (BS) equation \cite{Klebanov:2016xxf,Giombi:2017dtl}, or equivalently, the spectral decomposition of the four-point function \cite{Liu:2018jhs,Gurau:2019qag,Benedetti:2019ikb}.

The four-point function of our CFT can be written in a standard representation-theoretic form as \cite{Simmons-Duffin:2017nub,Liu:2018jhs,Gurau:2019qag}:
\begin{equation} \label{eq:4pt}
\begin{split}
  \frac{1}{N^6} \langle{\phi_{abc}(x_1) \bar{\phi}_{abc}(x_2) \phi_{a'b'c'}(x_3) \bar{\phi}_{a'b'c'}(x_4)} \rangle
 = & G(x_1-x_2) G(x_3-x_4)
 + \crcr
    +  \frac{1}{N^3}\sum_J 
  \int_{\frac{d}{2}-\imath \infty}^{\frac{d}{2}+\imath\infty} \frac{dh}{2\pi \imath} 
  &
  \;\frac{1}{1-k_{\zeta}(h,J)} \; \mu_{\Delta_{\phi}}^d(h,J)
     \cG^{\Delta_{\phi}}_{h,J}(x_i) + (\text{non-norm.})\,,
\end{split}
\end{equation}
with $\cG^{\Delta_{\phi}}_{h,J}(x_i) $ the conformal block, $\mu_{\Delta_{\phi}}^d(h,J)$ the measure, and $k_{\zeta}(h,J)$ the eigenvalues of the two particle irreducible four-point kernel. The non-normalizable contributions are due to operators with dimension $h<d/2$, and they should be treated separately \cite{Simmons-Duffin:2017nub}.\footnote{As we will see below, we will actually encounter an operator with dimension $h_0<d/2$. We will be cavalier in its treatment.} The subleading term is  the most interesting part, and it is related to the forward  four-point function that we introduced in Eq.\eqref{eq:fw4pt}. The appearance of $k_{\zeta}(h,J)$ should be clear from Eq.\eqref{eq:fw4pt-ladders}.
Closing the contour to the right, we pick poles at $k_{\zeta}(h,J)=1$ (other poles are spurious and they cancel out \cite{Simmons-Duffin:2017nub}), and we recover an operator-product expansion in the $t$-channel ($12\to34$):
\begin{equation} 
  \frac{1}{N^6} \langle{\phi_{abc}(x_1) \bar{\phi}_{abc}(x_2) \phi_{a'b'c'}(x_3) \bar{\phi}_{a'b'c'}(x_4)} \rangle
 =  G(x_1-x_2) G(x_3-x_4)
 +  \frac{1}{N^3} \sum_{m,J} c_{m,J}^2  \; \cG^{\Delta_{\phi}}_{h_{m,J},J}(x_i) \,,
\end{equation}
where $h_{m,J}$ are the poles of $(1-k_{\zeta}(h,J))^{-1}$, and the squares of the OPE coefficients are the residues at the poles \cite{Liu:2018jhs,Gurau:2019qag,Benedetti:2019ikb}.
We will limit ourselves to just studying the location of the poles, i.e.\ the spectrum of operators in the OPE.

Eigenfunctions of the kernel are known to take the form of three-point functions of two fundamental scalars with an operator. For example, in the case of spin zero we have:
\begin{equation}
    v_0(x_0,x_1,x_2) = \la\cO_h(x_0)\phi_{abc}(x_1)\bar{\phi}_{abc}(x_2)\ra = \f{C_{\cO\phi\bar{\phi}}}{(x_{01}^2 x_{02}^2)^{h/2}(x_{12}^2)^{\f12(\f{d}3-h)}} \,.
    \label{3-pt}
\end{equation}
Therefore, we need to find the eigenvalues $k(h,J)$ of the kernel from the equation:
\begin{equation}
    k_{\zeta}(h,J) v_J(x_0,x_1,x_2) = \int\,d^dx_3\,d^dx_4\, K(x_1,x_2;x_3,x_4)v_J(x_0,x_3,x_4),
    \label{BS}
\end{equation}
where the form of the kernel is obtained from \eqref{kernel-rank3} to be
\begin{equation}
    K(x_1,x_2;x_3,x_4) = \f{\l_1^2}{4} \left[3G(x_{14})G(x_{23})+2G(x_{13})G(x_{24})\right]G(x_{34})^4\,,
    \label{kernel}
\end{equation}
and since we integrate over $x_3$ and $x_4$, both terms can be combined into one.

\subsection{$\z=1$}

Since the corresponding integrals are simpler to solve in position space, we wish to set up the eigenvalue equation in position space. For that, we need the two-point function in position space, which for the case $\zeta=1$ is as follows:
\begin{align}
 G(x) &= \int\,\f{d^d p }{(2\pi)^d} e^{-i p\cdot x}\,G(p)=\cZ\int\,\f{d^d p }{(2\pi)^d} \f{e^{-i p\cdot x}}{p^{2d/3}}\nn\\
 &= \cZ\f{2^{d/3}}{(4\pi)^{d/2}}\f{\Gamma(\f{d}6)}{\Gamma(\f{d}3)}\f1{(x^2)^{d/6}}= F_1 \f1{(x^2)^{d/6}} \,,
 \label{2-pt-x}
\end{align}
where $F_1 =\cZ\f{2^{d/3}}{(4\pi)^{d/2}}\f{\Gamma(\f{d}6)}{\Gamma(\f{d}3)} $.
To perform the integrals at $J=0$, we shall use the following identity \cite{Giombi:2017dtl},
\begin{equation}
    \int\,d^d x_0 \f1{(x_{01}^2)^{\a_1}(x_{02}^2)^{\a_2}(x_{03}^2)^{\a_3}} = \f{L_d(\a_1,\a_2)}{(x_{12}^2)^{d/2-\a_3}(x_{13}^2)^{d/2-\a_2}(x_{23}^2)^{d/2-\a_1}},
    \label{identity}
\end{equation}
with $\a_1 + \a_2 +\a_3 =d$, and $L_d(\a_1,\a_2)=\pi^{d/2}\f{\Gamma(\f{d}2-\a_1)\Gamma(\f{d}2-\a_2)\Gamma(\f{d}2-\a_3)}{\Gamma(\a_1)\Gamma(\a_2)\Gamma(\a_3)}$.

To solve for the eigenvalues, let us first perform the integral over $x_3$ using \eqref{identity},
\begin{equation}
     \int\,d^d x_3 \f1{(x_{03}^2)^{h/2}(x_{23}^2)^{d/6}(x_{34}^2)^{5d/6-h/2}}=\f{L_d\big(\f{h}2,\f{d}6\big)}{(x_{02}^2)^{-d/3+h/2}(x_{04}^2)^{d/3}(x_{24}^2)^{d/2-h/2}}
     \label{int-1}
\end{equation}
with,
\begin{equation}
   L_d\bigg(\f{h}2,\f{d}6\bigg) =\pi^{d/2}\f{\Gamma(\f{d}3)\Gamma(-\f{d}3+\f{h}2)\Gamma(\f{d}2-\f{h}2)}{\Gamma(\f{d}6)\Gamma(\f{5d}6-\f{h}2)\Gamma(\f{h}2)}\,.
   \label{part-1}
\end{equation}
Now, performing the remaining integral over $x_4$, we get
\begin{equation}
     \int\,d^d x_4 \f1{(x_{04}^2)^{d/3+h/2}(x_{24}^2)^{d/2-h/2}(x_{14}^2)^{d/6}}=\f{L_d(\f{d}3+\f{h}2,\f{d}2-\f{h}2)}{(x_{02}^2)^{d/3}(x_{01}^2)^{h/2}(x_{12}^2)^{d/6-h/2}},
     \label{int-2}
\end{equation}
with
\begin{equation}
     L_d\bigg(\f{d}3+\f{h}2,\f{d}2-\f{h}2\bigg) =\pi^{d/2}\f{\Gamma(\f{d}3)\Gamma(\f{h}2)\Gamma(\f{d}6-\f{h}2)}{\Gamma(\f{d}3+\f{h}2)\Gamma(\f{d}2-\f{h}2)\Gamma(\f{d}6)}.
   \label{part-2}
\end{equation}
Collecting the terms from the first and second integrals, and combining their coefficients from Eq.~\eqref{2-pt-x}, Eq.~\eqref{part-1} and Eq.~\eqref{part-2}, we get the $J=0$ eigenvalues of the kernel to be
\begin{align}
    k_1(h,0) &= \f54\, \l_1^2\,F_1^6\, \pi^d\,\f{\Gamma(\f{d}3)^2\Gamma(-\f{d}3+\f{h}2)\Gamma(\f{d}6-\f{h}2)}{\Gamma(\f{d}6)^2\Gamma(\f{5d}6-\f{h}2)\Gamma(\f{d}3 + \f{h}2)}\nn\\
    &=\f54 \l_1^2\bigg(\f1{\pi^{d}}\f{4}{\lambda_1^2}\f{d}{3}\f{\Gamma(\frac{d}{6})\Gamma(\frac{5d}{6})}{\Gamma(1-\frac{d}{3})\Gamma(\frac{d}{3})}\bigg)\bigg(\pi^d\f{\Gamma(\f{d}3)^2\Gamma(-\f{d}3+\f{h}2)\Gamma(\f{d}6-\f{h}2)}{\Gamma(\f{d}6)^2\Gamma(\f{5d}6-\f{h}2)\Gamma(\f{d}3 + \f{h}2)}\bigg)
    \nn\\
    &= -5\times \f{\Gamma(\f{5d}6)\Gamma(\f{d}3)\Gamma(-\f{d}3+\f{h}2)\Gamma(\f{d}6-\f{h}2)}{\Gamma(-\f{d}3)\Gamma(\f{d}6)\Gamma(\f{5d}6-\f{h}2)\Gamma(\f{d}3 + \f{h}2)}.
    \label{eq:eigenvalue}
\end{align}
To find the spectrum of the bilinears, we must solve the above equation for $k_1(h,0)=1$, with $d=3-\epsilon$. We use the method of Ref.~\cite{Benedetti:2019ikb},  setting $h=1+2n+2z$, and treating $z$ as a perturbation of the classical dimension, which is justified for small $\epsilon$. 

For $n=0$ and $n=1$, we find the following solutions:
\begin{align}
h_{0}&=1+\frac{29}{3}\epsilon + \mathcal{O}(\epsilon^2), \crcr
h_1&=3+3\epsilon + \mathcal{O}(\epsilon^2).
\end{align}
We also find a solution corresponding to the mixing with a quartic operator (as we deduce from the dimension at $\epsilon=0$):
\begin{equation}
h_{q}=2-\frac{32}{3}\epsilon + \mathcal{O}(\epsilon^2).
\end{equation}
Lastly, for $n>1$ we find:
\begin{equation}
h_n=1+2n-\frac{\epsilon}{3}+\frac{20}{3n(n-1)(4 n^2-1)}\epsilon^2+\mathcal{O}(\epsilon^3)~~, ~~ n>1.
\end{equation}
The solutions we just found are exactly the ones found in Ref.~\cite{Giombi:2017dtl}, which is not surprising, as their equation 4.6 for $q=6$, giving the eigenvalues of the kernel, is the same as Eq.~\eqref{eq:eigenvalue}. However, it was assumed to hold for rank $5$, but as we have seen, it turns out that in that case there is no Wilson-Fisher fixed point, hence no interacting CFT to which these equations might apply. On the other hand, we have shown here that we still recover the same spectrum for the model in rank $3$, which admits a melonic Wilson-Fisher fixed point.

\begin{figure}[htbp]
\centering
\captionsetup[subfigure]{labelformat=empty}
\subfloat[]{\includegraphics[scale=0.45]{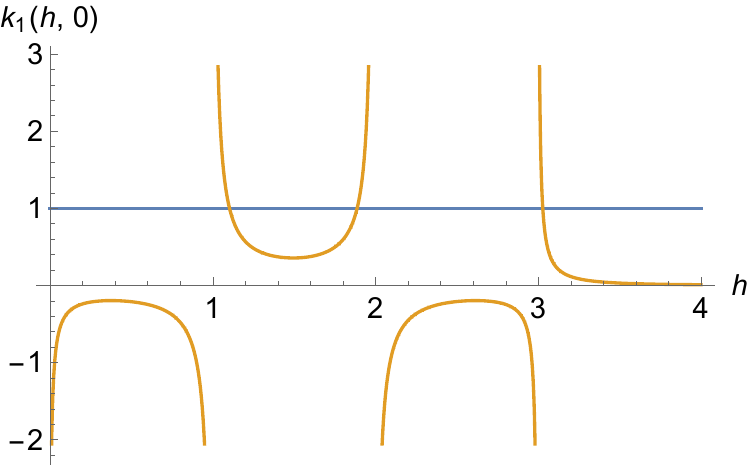}}
\subfloat[]{\includegraphics[scale=0.45]{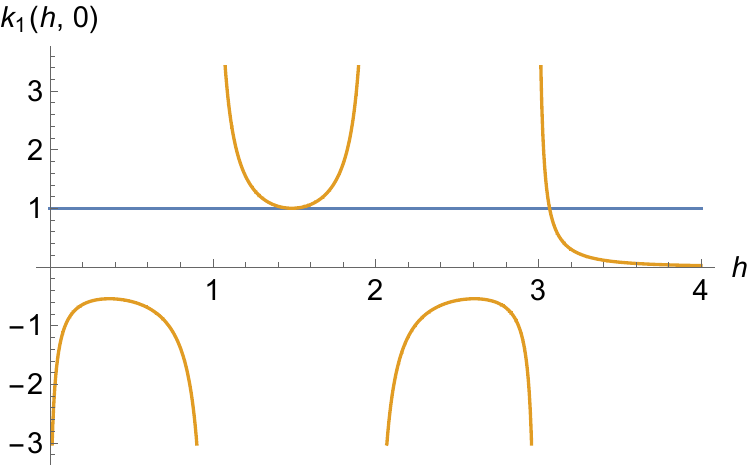}}
\subfloat[]{\includegraphics[scale=0.45]{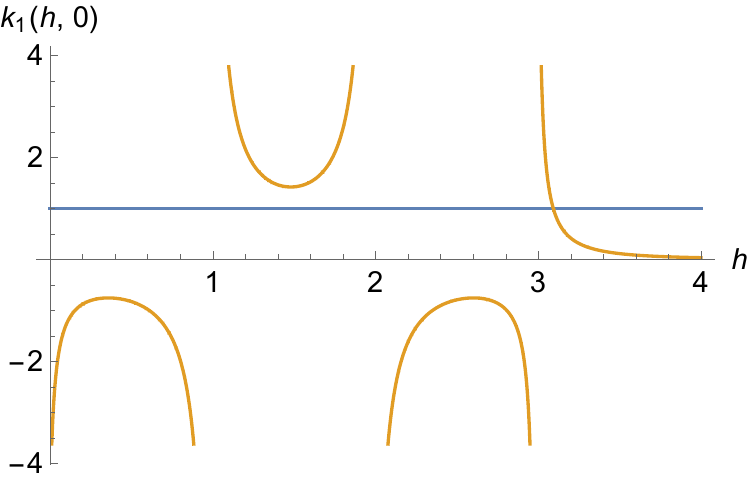}}
\caption{Plots of $k_1(h,0)$ at $d=3-\epsilon$ for, from left to right, $\epsilon=0.01$, $\epsilon=0.02819$, and $\epsilon=0.04$. On the left panel, the intersections with the blue line  correspond to $h_0$, $h_q$ and $h_1$. On the middle panel, $h_0$ and $h_q$ have merged, and on the right panel only $h_1$ remains.}
\label{fig:complexsol}
\end{figure}

As $\epsilon>0$, all the solutions we found are real. If we send $\epsilon$ to zero, we recover the classical dimensions $h^{classical}_n=1+2n$ of the bilinear operators $\phi_{abc}(\partial^2)^n\phi^{abc}$, except for $h_{q}$ corresponding to a quartic operator. 
However, this is only true for $\epsilon$ small enough. As $\epsilon$ increases, the two solutions $h_{0}$ and $h_q$ merge and become complex, see Fig.~\ref{fig:complexsol}. This happens around $\epsilon=0.02819$. Again, the same phenomenon appeared in Ref.~\cite{Giombi:2017dtl}.

\paragraph{Higher spins.}
We can also compute the spectrum of bilinears at higher spin. Using \cite{Gurau:2019qag}, $k_1$ becomes:
\begin{equation} \label{eq:f1J-zeta1}
k_1(h,J)=-5\times \f{\Gamma(\f{5d}6)\Gamma(\f{d}3)\Gamma(-\f{d}3+\f{h}2+J/2)\Gamma(\f{d}6-\f{h}2+J/2)}{\Gamma(-\f{d}3)\Gamma(\f{d}6)\Gamma(\f{5d}6-\f{h}2+J/2)\Gamma(\f{d}3 + \f{h}2+J/2)}.
\end{equation}
We find the following solutions for $k_1(h,J)=1$:
\begin{align}
\label{eq:h0J-zeta1}
h_{0,J}&=1+J-\frac{4J^2+29}{3(4J^2-1)}\epsilon + \mathcal{O}(\epsilon^2), \\
h_{1,J}&=3+J+\frac{-4J^2-8J+27}{3(2J+3)(2J+1)}\epsilon + \mathcal{O}(\epsilon^2), \\
h_{n,J}&=1+2n+J-\frac{\epsilon}{3}+\frac{5\epsilon^2}{3n(n-1)(n+1/2+J)(n-1/2+J)} + \mathcal{O}(\epsilon^3)~~, \quad n>1.
\end{align}
Notice that these can all be written in the form $h_{n,J}=d-2+2n+J+2z_{n,J}$, with $d=3-\epsilon$.

For $J=0$, we recover the solutions we found in the beginning of this section, except for $h_q$. This is due to the fact that the factor $\Gamma(-\f{d}3+\f{h}2+J/2)$ in Eq.~\eqref{eq:f1J-zeta1} only leads to a singularity for $h>0$ if $J=0$. Therefore, for $J> 0$, we only have dimensions corresponding to bilinear operators and no longer have a dimension corresponding to a quartic operator. 

One can check at leading order in $\epsilon$ from Eq.~\eqref{eq:h0J-zeta1}, or to all orders directly from Eq.~\eqref{eq:f1J-zeta1}, that the spin-2 operator with $n=0$ has the classical dimension $h_{0,2}=3-\epsilon=d$, as expected from a conserved energy-momentum tensor.

\subsection{$\z=\frac{d}{3}$}

The computation of the spectrum of bilinears of the long range model with the modified propagator goes exactly along the same lines as the one with the normal propagator. 
The only difference lies in the structure of the two-point function. The position space expression for the renormalized propagator (or two-point function) is:
\begin{equation}
	G(x) = \f{F_{d/3}}{(x^2)^{d/6}},\hspace{.2cm}F_{d/3} =\cZ\f{2^{d/3}}{(4\pi)^{d/2}}\f{\Gamma(\f{d}6)}{\Gamma(\f{d}3)},
\end{equation}
where $\cZ$ is the solution of \eqref{Z-norm-d/3}. \\
Once again we solve the same eigenvalue Eq.~\eqref{BS} using the same kernel \eqref{kernel}. The resulting eigenvalue, for $J=0$, is:
\be
\begin{split}
	k_{d/3}(h,0) &= \f54\, \l_1^2\,F_{d/3}^6\, \pi^d\,\f{\Gamma(\f{d}3)^2\Gamma(-\f{d}3+\f{h}2)\Gamma(\f{d}6-\f{h}2)}{\Gamma(\f{d}6)^2\Gamma(\f{5d}6-\f{h}2)\Gamma(\f{d}3 + \f{h}2)}\\
	&=\f54\, \l_1^2\,\cZ^6\f1{(4\pi)^{2d}}\bigg(\f{\Gamma(\f{d}6)}{\Gamma(\f{d}3)}\bigg)^4 \,\f{\Gamma(-\f{d}3+\f{h}2)\Gamma(\f{d}6-\f{h}2)}{\Gamma(\f{5d}6-\f{h}2)\Gamma(\f{d}3 + \f{h}2)}\\
	&= \f54 g_1^2\bigg(\f{\Gamma(\f{d}6)}{\Gamma(\f{d}3)}\bigg)^4 \,\f{\Gamma(-\f{d}3+\f{h}2)\Gamma(\f{d}6-\f{h}2)}{\Gamma(\f{5d}6-\f{h}2)\Gamma(\f{d}3 + \f{h}2)} \,,
	\label{eigenvalue-2}
\end{split}
\ee
where in the last line we used the renormalized coupling defined in \ref{sec:betas1}, namely  $g_1 = \f1{(4\pi)^d}\l_1\,\cZ^3$.

In order to find the OPE spectrum we have to solve for $k_{d/3}(h,0)=1$. The main difference with respect to the previous case is that the spectrum will now depend on the value of the exactly marginal coupling, which will replace $\epsilon$ in the role of small parameter. 

Again we use the method of Ref.~\cite{Benedetti:2019ikb} to solve $k_{d/3}(h,0)=1$, and we find the following solutions:
\begin{align}
h_0&=\frac{d}{3}+\frac{15\Gamma(1-d/6)}{d\Gamma(2d/3)\Gamma(d/2)}\left(\frac{\Gamma(d/6)}{\Gamma(d/3)}\right)^4g_1^2 + \mathcal{O}(g_1^4)\,, \crcr
h_n&=\frac{d}{3}+2n+\frac{(-1)^{n+1}}{n!}\frac{5\Gamma(n-d/6)}{2\Gamma(2d/3-n)\Gamma(d/2+n)}\left(\frac{\Gamma(d/6)}{\Gamma(d/3)}\right)^4g_1^2+ \mathcal{O}(g_1^4)\,.
\end{align}
Notice that at $g_1=0$, we recover the classical dimensions $h^{classical}_n=d/3+2n$. At $g_1 \neq 0$, all dimensions are real, and they are greater than $d/3$ for $g_1^2>0$ and small. 

As before, there is also a solution corresponding to a quartic operator:
\begin{equation}
h_{q}=\frac{2d}{3}-\frac{15\Gamma(1-d/6)}{d\Gamma(2d/3)\Gamma(d/2)}\left(\frac{\Gamma(d/6)}{\Gamma(d/3)}\right)^4g_1^2 + \mathcal{O}(g_1^4)\,.
\end{equation}

\begin{figure}[htbp]
\centering
\includegraphics[scale = 0.5]{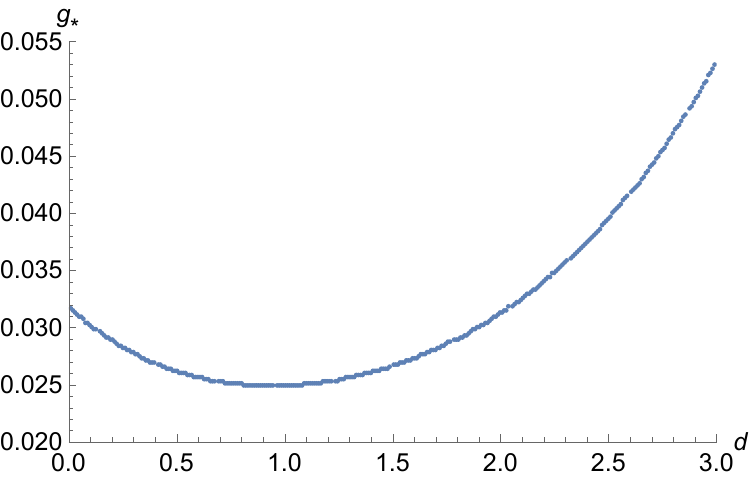}
\hspace{1cm}
\includegraphics[scale = 0.5]{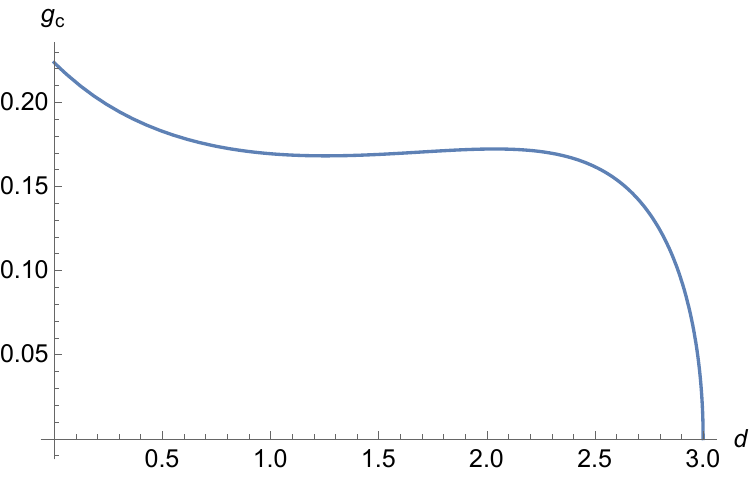}
\caption{Plots of $g_\star$ and $g_c$ as functions of $d$. The two curves cross at $d\simeq 2.97$.}
\label{fig:g_star}
\end{figure}

The plots of $k_{d/3}(h,0)$ are qualitatively similar to those in Fig.~\ref{fig:complexsol}, and we find the appearance of a pair of complex solutions for $g_1> g_\star >0$. For $d=2$, we have $g_\star \simeq 0.0313$, which is smaller than the value $g_c$ defined in Footnote \ref{foot:g_c}, at which the relation between bare $\l_1$ and renormalized $g_1$ becomes non-invertible, and which for $d=2$ is $g_c\simeq 0.1722$. A similar situation is found for any $d\lesssim 2.97$, while for $d\gtrsim 2.97$ we find $g_c<g_\star$. Comparative plots of $g_\star$ and $g_c$ as functions of $d$ are shown in Fig.~\ref{fig:g_star}. Therefore, for $d\lesssim 2.97$, the scenario differs from the one encountered in Ref.~\cite{Benedetti:2019eyl}, as in the present case the complex transition lies within the regime of validity of the fixed point solution. Furthermore, at the transition where the first two solutions of $k_{d/3}(h,0)=1$ merge (and then become complex) their value is $d/2$, within numerical precision. Such transition thus seems to be compatible with the scenario advanced in Ref.~\cite{Kim:2019upg}, where the appearance of complex dimensions of the form $d/2+\im f$ for a given operator has been conjectured to be a signal that such operator acquires a non-zero vacuum expectation value.

\paragraph{Higher spins.}

Again we can compute the spectrum of bilinears for spin $J>0$. The eigenvalue becomes:
\begin{equation}
k_{d/3}(h,J) = \f54 g_1^2\bigg(\f{\Gamma(\f{d}6)}{\Gamma(\f{d}3)}\bigg)^4 \,\f{\Gamma(-\f{d}3+\f{h}2+J/2)\Gamma(\f{d}6-\f{h}2+J/2)}{\Gamma(\f{5d}6-\f{h}2+J/2)\Gamma(\f{d}3 + \f{h}2+J/2)}\,.
\end{equation}
We find the following solutions for $k_{d/3}(h,J)=1$:
\begin{align}
h_{0,J}&=\frac{d}{3}+J-\frac{5}{2}\frac{\Gamma(-d/6+J)}{\Gamma(2d/3)\Gamma(d/2+J)}\left(\frac{\Gamma(d/6)}{\Gamma(d/3)}\right)^4g_1^2 + \mathcal{O}(g_1^4)\,, \crcr
h_{n,J}&=\frac{d}{3}+2n+J+\frac{(-1)^{n+1}}{n!}\frac{5\Gamma(n-d/6+J)}{2\Gamma(2d/3-n)\Gamma(d/2+n+J)}\left(\frac{\Gamma(d/6)}{\Gamma(d/3)}\right)^4g_1^2+ \mathcal{O}(g_1^4)\,.
\end{align}

Again, when $J=0$ we recover the dimensions we computed in the beginning of this section, except for the one corresponding to a quartic operator.

However, differently from the $\zeta=1$ case, and as in  \cite{Benedetti:2019ikb}, we find no spin-two operator of dimension $d$. This is due to the fact that the energy momentum tensor is not a local operator.

\section{Conclusions}
\label{sec:concl}

In this paper, we presented an analysis of the melonic large-$N$ limit in various versions of bosonic tensor models with sextic interactions.
We considered explicitly tensors of rank 3 and 5, but we expect rank 4 to behave similarly to rank 5. And we chose as free propagator either the standard short-range propagator, or a critical long-range propagator. We discussed in detail some standard properties of melonic theories, as the closed Schwinger-Dyson equation for the two-point function, and the Bethe-Salpeter equation for the spectrum of bilinear operators. However, as we emphasized, the conformal solution of these equations are only justified if the quantum field theory actually admits a fixed point of the renormalization group. In this respect, we found a striking difference between the rank-3 and rank-5 models, as only the former (both in the short-range and long-range versions) admits a non-trivial (and real) fixed point for $d<3$, with an interaction leading to melonic dominance. 
The rank-5 model instead has only one trivial (i.e.\ non-interacting) fixed point. It would be interesting to check whether such conclusion would remain valid after including in the action \eqref{eq:int-action-graph-rank5} the other possible sextic interactions that we have omitted by restricting to a melo-complete family.

Comparing our findings for the short range model with those of the sextic model in Ref.~\cite{Giombi:2017dtl}, we observe similar results for two-point function and spectrum  of operators. However, we do so for the rank-3 model, where such analysis is justified by the existence of a melonic fixed point, whereas their analysis was formally based on a rank-5 model, which we showed is inconsistent. The fact that we find the same result is not a coincidence: our kernel eigenvalue \eqref{eq:eigenvalue} coincides with the $q=6$ case of the eigenvalue computed in Ref.~\cite{Giombi:2017dtl} for a general $q$-valent melonic theory. Such eigenvalue depends only on the assumption that a $q$-valent interaction leads to melonic dominance. The latter can for example be obtained with a rank-$(q-1)$ model with a complete interaction, as assumed in Ref.~\cite{Giombi:2017dtl}. However, as argued in Ref.~\cite{Prakash:2019zia}, and as we saw also here, rank $q-1$ is not necessary: a $q$-valent interaction can lead to a melonic limit even in a tensor model of rank $r<q-1$ (in which case the model was called \emph{subchromatic} in Ref.~\cite{Prakash:2019zia}); this is the case of our rank-3 model with wheel interaction.

Comparing instead our long-range model to the quartic long-range model of Ref.~\cite{Benedetti:2019eyl}, we see some similarity but also an important difference: on one hand, both models admit a line of fixed points, parametrized by the interaction that leads to melonic dominance; on the other, in the quartic case, the fixed point and conformal dimensions are real only for purely imaginary tetrahedral coupling \cite{Benedetti:2019eyl}, while in our sextic model, we have a real fixed point and real spectrum for a real wheel coupling.
Furthermore, unlike in  Ref.~\cite{Benedetti:2019eyl}, in the present case the appearance of complex dimensions at some critical value of the marginal coupling seems to be compatible with the scenario conjectured in  Ref.~\cite{Kim:2019upg}, according to which it is a signal of an instability of the vacuum.

We have also encountered some of the recurring aspects of melonic theories (for rank 3, at least): for the short-range version, reality of the CFT constrains $\epsilon$ to stay very small; in the long-range version, we have instead the freedom to reach an integer dimension ($d=2$ in this case), by keeping the marginal coupling small, but at the price of loosing the energy-momentum tensor (as usual in long-range models \cite{Paulos:2015jfa}).
It would be interesting to get a better understanding of how general these features are.

One new feature that we found is that the fixed point of the short-range model has a non-diagonalizable stability matrix, even in the range of $\epsilon$ for which the exponents are real. This is an indication that the fixed-point theory is a logarithmic CFT, and thus it is non-unitary. We hope to explore this aspect more thoroughly in the near future.

Lastly, it would be important to understand the fate of our line of fixed points (in the long-range model) at higher orders in the $1/N$ expansion. 
At some order in the expansion we expect to find vertex corrections also to the wheel interaction, and therefore a non-zero beta function $\beta_1$.
A similar situation occurs in the vector $\phi^6$ model, where the leading-order beta function vanishes identically, but already at next-to-leading order in $1/N$ one finds a non-zero beta function \cite{Pisarski:1982vz}, thus reducing the leading-order line of fixed points to an isolated fixed point.

\section*{Acknowledgements}

We would like to thank Razvan Gurau and Guillaume Valette for useful  discussions.
The work of DB and SH is supported by the European Research Council (ERC) under the European Union's Horizon 2020 research and innovation program (grant agreement No818066).
RS was partially supported by the Spanish Research Agency  (Agencia
Estatal  de  Investigacion)  through  the  grants  IFT  Centro  de
Excelencia Severo Ochoa SEV-2016-0597, FPA2015-65480-P and
PGC2018-095976-B-C21. RS is currently partially supported by the
Israeli Science Foundation
Center of Excellence (grant No. 2289/18) and by the Quantum Universe
I-CORE program of the Israel Planning and Budgeting Committee (grant
No. 1937/12).

\newpage

\appendix

\section{Conventions for the interaction terms}
\label{ap:conventions}

We write here in an explicit form the interactions appearing in Eq.~\eqref{eq:int-action} and \eqref{eq:int-action-rank5}, as well as the quartic invariants, in terms of contraction operators built as linear combinations of products of Kronecker delta functions.

\subsection{Rank 3}
Using the compact notation $\mathbf{a}=(a_1a_2a_3)$,
the $U(N)^3$ quartic invariants, also known as pillow and double-trace invariants, respectively, are:
\begin{align}
I_p &=  \delta^p_{\mba\mbb; \mbc\mbd}\phi_{\mba}(x) \phib_{\mbb}(x)  \phi_{\mbc}(x) \phib_{\mbd }(x)\,,\\
I_d &= \delta^d_{\mba\mbb; \mbc\mbd }  \phi_{\mba}(x) \phib_{\mbb}(x)  \phi_{\mbc}(x) \phib_{\mbd }(x)\,,
\end{align}
with: 
\be
    \delta^p_{\mba\mbb; \mbc\mbd }=\frac{1}{3} \sum_{i=1}^3  \delta_{a_id_i} \delta_{b_ic_i} \prod_{j\neq i}  \delta_{a_jb_j}\delta_{c_j d_j} \; , \quad\quad  \delta^d_{\mba\mbb; \mbc\mbd }  = \delta_{\mba \mbb}  \delta_{\mbc \mbd}\,,
\ee
and $\delta_{\mba \mbb}  = \prod_{i=1}^3 \delta_{a_i b_i} $.

The sextic invariants depicted in Eq.~\eqref{eq:int-action} are instead:
\begin{align}
I_1 &=  \delta^{(1)}_{\mba\mbb \mbc\mbd\mbe\mbf} \phi_{\mba}(x) \phib_{\mbb}(x)  \phi_{\mbc}(x) \phib_{\mbd }(x)\phi_{\mbe }(x)\phib_{\mbf }(x)\,,\\
I_b &=\delta^{(b)}_{\mba\mbb; \mbc\mbd; \mbe\mbf }  \phi_{\mba}(x) \phib_{\mbb}(x)  \phi_{\mbc}(x) \phib_{\mbd }(x)\phi_{\mbe }(x)\phib_{\mbf }(x)\,,   \qquad b=2,\ldots,5\,,
\end{align}
with
%
\begin{align}
   & \delta^{(1)}_{\mba\mbb\mbc\mbd \mbe\mbf }= \d_{a_1b_1}\d_{a_2f_2}\d_{a_3d_3}\d_{c_1d_1}\d_{c_2b_2}\d_{c_3f_3}\d_{e_1f_1}\d_{e_2d_2}\d_{e_3b_3}\, ,\crcr
  & \delta^{(2)}_{\mba\mbb; \mbc\mbd; \mbe\mbf }= \frac{1}{9}\left( \sum_{i=1}^3 \sum_{j \neq i}  \delta_{a_if_i} \delta_{b_ic_i} \delta_{c_j d_j}\delta_{e_jf_j}\left(\prod_{k\neq i}  \delta_{a_kb_k} \right)\left(\prod_{l\neq j}  \delta_{e_l d_l} \right) \left( \prod_{m \neq i,j} \delta_{c_m f_m}\right)+ \mbc\mbd \leftrightarrow \mbe\mbf + \mbc\mbd \leftrightarrow \mba\mbb \right) \, ,\crcr
  & \delta^{(3)}_{\mba\mbb; \mbc\mbd; \mbe\mbf }= \frac{1}{3} \sum_{i=1}^3 \delta_{a_if_i} \delta_{b_ic_i} \delta_{d_i e_i}\prod_{j\neq i}  \delta_{a_j b_j}\delta_{c_j d_j}\delta_{e_j f_j} \, ,\crcr
  & \delta^{(4)}_{\mba\mbb; \mbc\mbd; \mbe\mbf }=\frac{1}{3}\left(\delta_{\mba\mbb}\delta^p_{\mbc\mbd;\mbe\mbf} + \delta_{\mbc\mbd}\delta^p_{\mba\mbb;\mbe\mbf}+ \delta_{\mbe\mbf}\delta^p_{\mba\mbb;\mbc\mbd}\right)\, ,\crcr
  & \delta^{(5)}_{\mba\mbb; \mbc\mbd; \mbe\mbf }=\delta_{\mba\mbb}\delta_{\mbc\mbd}\delta_{\mbe\mbf}\, .
 \end{align}
Besides the color symmetrization, to simplify the computation of the beta-functions, we have included a symmetrization with respect to exchange of pairs of black and white vertices.

\subsection{Rank 5}

Using the compact notation $\mathbf{a}=(a_1a_2a_3a_4a_5)$, the $O(N)^3$ melonic quartic invariants are:
\begin{align}
I_p &=  \delta^p_{\mba\mbb; \mbc\mbd}\phi_{\mba}(x) \phi_{\mbb}(x)  \phi_{\mbc}(x) \phi_{\mbd }(x)\,,\\
I_d &= \delta^d_{\mba\mbb; \mbc\mbd }  \phi_{\mba}(x) \phi_{\mbb}(x)  \phi_{\mbc}(x) \phi_{\mbd }(x)\,,
\end{align}
with: 
\be
   \delta^p_{\mba\mbb; \mbc\mbd }=\frac{1}{5} \sum_{i=1}^5  \delta_{a_id_i} \delta_{b_ic_i} \prod_{j\neq i}  \delta_{a_jb_j}\delta_{c_j d_j} \; , \quad\quad  \delta^d_{\mba\mbb; \mbc\mbd }  = \delta_{\mba \mbb}  \delta_{\mbc \mbd} \,,
\ee
and $\delta_{\mba \mbb}  = \prod_{i=1}^5 \delta_{a_i b_i} $.

The sextic invariants depicted in Eq.~\eqref{eq:int-action-rank5} are instead:

\begin{align}
J_1 &=  \delta^{(1)}_{\mba\mbb \mbc\mbd\mbe\mbf} \phi_{\mba}(x) \phi_{\mbb}(x)  \phi_{\mbc}(x) \phi_{\mbd }(x)\phi_{\mbe }(x)\phi_{\mbf }(x)\,,\\
J_b &=\delta^{(b)}_{\mba\mbb; \mbc\mbd; \mbe\mbf } \phi_{\mba}(x) \phi_{\mbb}(x)  \phi_{\mbc}(x) \phi_{\mbd }(x)\phi_{\mbe }(x)\phi_{\mbf }(x)\,,   \qquad b=2,\ldots,6\,,
\end{align}
with
\begin{align}
   & \delta^{(1)}_{\mba\mbb\mbc\mbd\mbe\mbf}  = \delta_{a_1 b_1}  \delta_{a_2f_2} \delta_{a_3 e_3}  \delta_{a_4 d_4 } \delta_{a_5 c_5}   \delta_{b_2 c_2}\delta_{b_3 d_3}\delta_{b_4 f_4} \delta_{b_5 e_5} \delta_{c_3 f_3}\delta_{c_4 e_4}\delta_{c_1 d_1}\delta_{e_1 f_1}\delta_{e_2 d_2}\delta_{d_5 f_5} \,  , \crcr
   & \delta^{(2)}_{\mba\mbb; \mbc\mbd; \mbe\mbf }= \frac{1}{60}\left( \sum_{i=1}^5 \sum_{j \neq i}  \delta_{a_ic_i} \delta_{b_id_i} \delta_{c_j e_j}\delta_{d_jf_j}\left(\prod_{k\neq i}  \delta_{a_kb_k} \right)\left(\prod_{l\neq j}  \delta_{e_l f_l} \right) \left( \prod_{m \neq i,j} \delta_{c_m d_m}\right)+ \mbc\mbd \leftrightarrow \mbe\mbf + \mbc\mbd \leftrightarrow \mba\mbb \right) \, ,\crcr
   & \delta^{(3)}_{\mba\mbb; \mbc\mbd; \mbe\mbf }= \frac{1}{5} \sum_{i=1}^5 \delta_{a_if_i} \delta_{b_ic_i} \delta_{d_i e_i}\prod_{j\neq i}  \delta_{a_j b_j}\delta_{c_j d_j}\delta_{e_j f_j} \, ,\crcr
  & \delta^{(4)}_{\mba\mbb; \mbc\mbd; \mbe\mbf }=\frac{1}{3}\left(\delta_{\mba\mbb}\delta^p_{\mbc\mbd;\mbe\mbf} + \delta_{\mbc\mbd}\delta^p_{\mba\mbb;\mbe\mbf}+ \delta_{\mbe\mbf}\delta^p_{\mba\mbb;\mbc\mbd}\right)\, ,\crcr
  & \delta^{(5)}_{\mba\mbb; \mbc\mbd; \mbe\mbf }=\delta_{\mba\mbb}\delta_{\mbc\mbd}\delta_{\mbe\mbf}\, , \\
  & \delta^{(6)}_{\mba\mbb; \mbc\mbd; \mbe\mbf }= \frac{1}{60} \sum_{i=1}^5 \sum_{j\neq i}\sum_{k\neq i,j}  \delta_{a_ic_i} \delta_{b_id_i} \delta_{c_j e_j}\delta_{d_jf_j}\delta_{a_k e_k}\delta_{b_k f_k}\left(\prod_{l\neq i,k}  \delta_{a_l b_l} \right)\left(\prod_{m\neq j,k}  \delta_{e_m f_m} \right) \left( \prod_{n \neq i,j} \delta_{c_n d_n}\right) \, . \nn
 \end{align}

\section{The melon integral}
\label{ap:melon}

In this section we compute the melon integral contributing to the wave function renormalization. 

We want to compute:
\begin{equation}
M_{\Delta}(p)=\int_{q_1,q_2,q_3,q_4}G_0(q_1)G_0(q_2)G_0(q_3)G_0(q_4)G_0(p+q_1+q_2+q_3+q_4) \,,
\end{equation}
with $G_0(p)=\frac{1}{p^{2\Delta}}$.

We will use the following formula to compute $M(p)$:
\begin{equation}
\int\f{\dd[d] k}{(2\pi)^d} \f{1}{k^{2\a}(k+p)^{2\b}}  = \f{1}{(4\pi)^{d/2}}\f{\G(d/2 - \a)\G(d/2 - \b)\G(\a +\b -d/2)}{\G(\a)\G(\b) \G(d - \a - \b)}\f{1}{|p|^{2(\a+\b - d/2)}} \,.
\label{eq:intG}
\end{equation}
We obtain:
\begin{equation}
M_{\Delta}(p)=\frac{p^{4d-10\Delta}}{(4\pi)^{2d}}\frac{\Gamma(d/2-\Delta)^5\Gamma(5\Delta-2d)}{\Gamma(\Delta)^5\Gamma(5d/2-5\Delta)} \,.
\end{equation}
For $\Delta=\frac{d}{3}$, this simplifies to:
\begin{equation} \label{eq:M-d/3}
M_{d/3}(p)=-\frac{p^{2d/3}}{(4\pi)^{2d}}\frac{3}{d}\frac{\Gamma(1-\frac{d}{3})\Gamma(\frac{d}{6})^5}{\Gamma(\frac{d}{3})^5\Gamma(\frac{5d}{6})} \,.
\end{equation}
We will also need the melon integral for $d=3-\epsilon$ and $\Delta=1$:
\begin{equation}  \label{eq:M-1}
M_1(p)=\frac{p^{2-4\epsilon}}{(4\pi)^{6-2\epsilon}}\frac{\Gamma(2\epsilon-1)\Gamma(\frac{1-\epsilon}{2})^5}{\Gamma(\frac{5}{2}(1-\epsilon)}\,.
\end{equation}
At first order in $\epsilon$, this gives:
\begin{equation}
M_1(p)=- \frac{p^{2-4\epsilon}}{(4\pi)^{6}}\frac{2\pi^2}{3\epsilon} + \mathcal{O}(1)\,.
\end{equation}

\section{Beta functions details}
\label{ap:betafun4}

\subsection{2-loop amplitude}

We want to compute the two-loop amputated Feynman integral (the candy) represented in the middle of Fig.~\ref{fig:bare3}.
We use the subtraction point defined in Sec.~\ref{sec:betas1}. Then, respecting the conservation of momenta, we can write the candy integral as:
\begin{equation}
D_{\Delta}(\mu)=\int_{q_1,q_2} G_0(q_1)G_0(q_2)G_0(-p_1-p_2-p_3-q_1-q_2).
\end{equation}
This gives with $G(q)=\frac{1}{q^{2\Delta}}$:
\begin{equation}
D_{\Delta}(\mu)=\int_{q_1,q_2} \frac{1}{q_1^{2\Delta}q_2^{2\Delta}(p_1+p_2+p_3+q_1+q_2)^{2\Delta}}\,.
\end{equation}
We use twice Eq.~\eqref{eq:intG} and obtain (using $|p_1+p_2+p_3|=\mu$):
\begin{equation}
D_{\Delta}(\mu)=\f{1}{(4\pi)^d}\f{\Gamma(d/2 - \Delta)^3\Gamma(3\Delta- d)}{\Gamma(\Delta)^3\Gamma(3d/2 - 3\Delta)}\f{1}{\mu^{2(3\Delta - d)}}\,.
\end{equation}
For $\zeta=1$, we set $\Delta=1$ and $d=3-\epsilon$. We obtain at first order in $\epsilon$:
\begin{equation}
D_1(\mu)=\mu^{-2\epsilon}\frac{1}{(4\pi)^3}\frac{2\pi}{\epsilon}+ \mathcal{O}(1)\,.
\end{equation}
For the modified propagator case, we set $\Delta=\frac{d+\epsilon}{3}$ and $d<3$. We obtain at first order in $\epsilon$:
\begin{equation}
D_{d/3}(\mu)=\mu^{-2\epsilon}\frac{1}{(4\pi)^d}\frac{\Gamma(\frac{d}{6})^3}{\Gamma(\frac{d}{3})^3\Gamma(\frac{d}{2})\epsilon}+ \mathcal{O}(1)\,.
\end{equation}

\subsection{4-loop amplitude}

We compute the following four-loop amputated Feynman integral: 
\begin{equation*}
\int \dd[d]x\dd[d]y ~ G(x-y)^4G(x-z)G(y-z)\,. 
\end{equation*}
Again we use the symmetric subtraction point and we can write the integral in momentum space as:
\begin{equation}
S_{\Delta}(\mu)=\int_{q_1,q_2,q_3,q_4} G_0(q_1)G_0(q_2)G_0(q_3)G_0(q_4)G_0(-p_1-p_2-q_4)G_0(-p_1-q_1-q_2-q_3-q_4) \,.
\end{equation}
With $G_0(q)=\frac{1}{q^{2\Delta}}$, this gives: 
\be
S_{\Delta}(\mu)=\int_{q_1,q_2,q_3,q_4} \f{1}{(p_1+q_1+q_2+q_3+q_4)^{2\Delta}}\f{1}{(q_4 + p_1 +p_2)^{2\Delta}}\f{1}{(q_1q_2q_3q_4)^{2\Delta}}\,.
\ee
We integrate loop by loop using Eq.~\eqref{eq:intG}, until we are left with a triangle-type one-loop integral:
\be
S_{\Delta}(\mu)=\f{1}{(4\pi)^{3d/2}}\left(\f{\G(d/2 - \Delta)}{\G(\Delta)}\right)^4 \f{\G(4\Delta - 3d/2)}{\G(2d - 4\Delta)}\int_{q_4} \f{1}{(q_4 + p_1 +p_2)^{2\Delta}} \f{1}{q_4^{2\Delta}}\f{1}{(p_1 +q_4)^{2(4\Delta - 3d/2)}}\,.
\ee
We use a Mellin-Barnes representation \cite{Davydychev:1995mq,ODwyer:2007brp} to rewrite the remaining integral as: 
\be
\begin{split}
\int_{q_4} \f{1}{(q_4 + p_1 +p_2)^{2\Delta}} \f{1}{q_4^{2\Delta}} \f{1}{(p_1 +q_4)^{2(4\Delta - 3d/2)}} & = \\
\f{\pi^{d/2} ((p_1+p_2)^2)^{d/2 - \sum_i \n_i}}{\Gamma(d - \sum_i \n_i) \prod_i \Gamma(\n_i)(2\pi i)^2}\int_{-i\infty}^{i\infty}&\frac{\dd s\dd t }{(2\pi)^d} x^s y^t\Gamma(-s)\Gamma(-t) \Gamma(d/2 - \n_2 - \n_3 - s)\\
&\times \Gamma(d/2 - \n_1 - \n_3 - t)\Gamma(\n_3 + s+t)\Gamma(\sum_i\n_i -d/2 + s+t)\,,
\end{split}
\ee
with $\n_1=\n_2=\Delta,\n_3=4\Delta - 3d/2$ and $x=p_1^2/(p_1+p_2)^2,y=p_2^2/(p_1+p_2)^2$. 

In the case $\z=1$, we set $\Delta=1$ and deforming the contour on the right and picking the residue at $s=t=0$,\footnote{The other contributions from poles at $(s=n\geq 1,t=m\geq 1)$, $(s=n-2\eps, t=m-2\eps)$ and $(s=n-2\eps,t=m)$, $(s=n,t=m-2\eps)$ (assuming $n,m \in \mathbb{N}$) cancel, as well as those at $(s=0,t=m\geq1)$ with $(s=0, t=m-2\eps)$ or $(s=n\geq1,t=0)$ with $(s=n-2\eps, t=0)$.} we find in the last $\Gamma$ function the only contribution to the pole in $1/\eps$ from a $d = 3-\eps$ expansion. Putting everything together, we find, at first order in $\epsilon$: 
\begin{equation}
S_1(\mu)=\f{\mu^{-4\epsilon}}{(4\pi)^{6}}\f{\Gamma(1/2)^4}{\Gamma(3/2)}\f{\Gamma(-1/2)}{2\eps} = \f{\mu^{-4\epsilon}}{(4\pi)^6}\f{-2\pi^2}{\eps}+ \mathcal{O}(1)\,.
\end{equation}

In the case $\z=(d+\epsilon)/3$, we set $\Delta=\z$ and again only the residue at $s=t=0$ gives a contribution to the pole in $1/\epsilon$. We find:

\begin{equation}
S_{d/3}(\mu)=\frac{\mu^{-4\epsilon}}{(4\pi)^{2d}}\frac{\Gamma(d/6)^4\Gamma(-d/6)}{2\epsilon\Gamma(d/2)\Gamma(d/3)^4\Gamma(2d/3)}+ \mathcal{O}(1) \,.
\end{equation}


\providecommand{\href}[2]{#2}\begingroup\raggedright\endgroup


\end{document}